\documentclass[article,fleqn]{cas-sc}
\pdfoutput=1
\usepackage{textcomp}
\usepackage{lastpage}
\usepackage{fancyhdr}
\usepackage{enumitem}
\usepackage{float}
\floatstyle{plaintop}
\restylefloat{table}
\usepackage{caption} 
\usepackage{placeins}
\usepackage[sort & compress]{natbib}
\setcitestyle{authoryear}
\newtheorem{theorem}{Theorem}

\newdefinition{rmk}{Remark}
\newproof{pf}{Proof}
\newproof{pot}{Proof of Theorem \ref{thm2}}
\usepackage[ruled,vlined]{algorithm2e}

\SetCommentSty{mycommfont}
\newcommand{\ra}[1]{\renewcommand{\arraystretch}{#1}}
\usepackage{empheq,multirow}
\usepackage{bbm}

\usepackage{amsmath}
\usepackage{enumitem}

\usepackage{url}
\usepackage{stfloats}
\usepackage[normalem]{ulem}
\usepackage{setspace}
\usepackage{tikz}

\def\tsc#1{\csdef{#1}{\textsc{\lowercase{#1}}\xspace}}
\tsc{WGM}
\tsc{QE}
\tsc{EP}
\tsc{PMS}
\tsc{BEC}
\tsc{DE}

\begin{document}

\let\WriteBookmarks\relax
\def\floatpagepagefraction{1}
\def\textpagefraction{.001}

\newtheorem{proposition}{Proposition}

\newtheorem{assumption}{A}

\title [mode = title]{Nested Vehicle Routing Problem: Optimizing Drone-Truck Surveillance Operations}

\author[]{Fanruiqi Zeng}
\fnmark[1]
\cormark[1]
\ead{fanruiqi.zeng@gatech.edu}
\credit{Conceptualization, Methodology, Software, Investigation, Data Curation, Visualization, Writing -- Original Draft}
\author[]{Zaiwei Chen}
\fnmark[1]
\credit{Conceptualization, Validation, Writing -- Review \& Editing}
\author[]{John-Paul Clarke}
\fnmark[2]
\credit{Conceptualization, Validation, Writing -- Review \& Editing, Supervision, Project administration}
\author[]{David Goldsman}
\fnmark[3]
\credit{Writing -- Review \& Editing, Supervision}

\cortext[cor1]{Corresponding author}
\fntext[fn1]{The author is affiliated with the Daniel Guggenheim School of Aerospace Engineering, Georgia Institute of Technology.}
\fntext[fn2]{The author is affiliated with the Department of Aerospace Engineering and Engineering Mechanics, The University of Texas at Austin.}
\fntext[fn3]{The author is affiliated with the H. Milton Stewart School of Industrial and Systems Engineering, Georgia Institute of Technology.}

\begin{abstract}
	Unmanned aerial vehicles or drones are becoming increasingly popular due to their low cost and high mobility. In this paper we address the routing and coordination of a drone-truck pairing where the drone travels to multiple locations to perform specified observation tasks and rendezvous periodically with the truck to swap its batteries. We refer to this as the Nested-Vehicle Routing Problem (Nested-VRP) and develop a Mixed Integer Quadratically Constrained Programming (MIQCP) formulation with critical operational constraints, including drone battery capacity and synchronization of both vehicles during scheduled rendezvous. An enhancement of the MIQCP model for the Nested-VRP is achieved by deriving the equivalent Mixed Integer Linear Programming (MILP) formulation as well as leveraging lifting and Reformulation-Linearization techniques to strengthen the subtour elimination constraints of the drone. Given the NP-hard nature of the Nested-VRP, we further propose an efficient neighborhood search (NS) heuristic where we generate and improve on a good initial solution by iteratively solving the Nested-VRP on a local scale. We provide comparisons of both the exact approaches based on MIQCP or its enhanced formulations and NS heuristic methods with a relaxation lower bound in the cases of small and large problem sizes, and present the results of a computational study to show the effectiveness of the MIQCP model and its variants as well as the efficiency of the NS heuristic, including for a real-life instance with 631 locations.  We envision that this framework will facilitate the planning and operations of combined drone-truck missions.
\end{abstract}

\begin{keywords}
aerial surveillance \sep cooperative vehicle routing \sep mixed integer programming \sep persistent operation \sep unmanned aerial vehicle
\end{keywords}

\maketitle

\section{Introduction}\label{sec:intro}

Unmanned aerial vehicles, commonly referred to as drones, are flight vehicles that can operate autonomously or be remotely controlled by a human operator or computer. Although drones were originally developed for military purposes, they have become increasingly popular in civilian applications such as logistics (see, e.g., \cite{7513397,tavana2017drone,dayarian2020same}); agriculture (see, e.g., \cite{tokekar2016sensor,valente2013aerial}), search and rescue (see, e.g., \cite{miao2017research,raap2017aerial,raap2017trajectory}), as well as aerial photography and surveillance (see, e.g., \cite{ccakici2016coordinated}).  

In this paper, we consider a nested vehicle routing problem (Nested-VRP) where: (a) a single drone is deployed to survey prescribed locations; (b) the locations to be surveyed are distributed across a large geographical area; (c) the duration of the surveillance at each location is unique to the type of survey to be conducted at that location; (d) the drone has limited flight endurance; (e) a single truck with an unlimited supply of fully charged batteries is used to recharge the drone; (f) the drone must rendezvous with the truck before running out of charge; (g) the time required to swap the batteries in the drone is a prescribed positive constant; and (h) perhaps most importantly the time required to complete the sequence of surveys must be minimized.

We seek to answer the following questions: (i) What is the optimal sequence of locations for the drone to visit? (ii) At which of these locations should the drone rendezvous with the truck? and (iii) What is the optimal routing for the truck? Further, because the information must be obtained frequently and in a timely fashion, we must do so via a computationally efficient heuristic algorithm that outperforms previous algorithms. This study contributes to the literature on cooperative vehicle routing in the following ways:

    \begin{itemize}
    \item Although it is not the main contribution of the paper, to the best of our knowledge, we are the first to provide answers to these questions via a single formulation that incorporates the following real-world considerations: (a) non-zero surveillance times at locations; (b) flight endurance limitations; (c) the requirement that the truck must arrive at the rendezvous location before the drone battery charge has expired; (d) the truck is allowed to perform a battery swap while shipping the drone from one location to the other; and (e) a non-zero battery swapping time.
    \item A comparison of the proposed Nested-VRP model to the state-of-the-art model regarding model compactness is present. Moreover, we apply linearization and constraint strengthening techniques to further enhance the model performance. We also analyze the complexity of the Nested-VRP model with and without prior information on the drone routing.
    \item We propose an effective Neighborhood Search (NS) heuristic to solve the Nested-VRP\@. Although the NS heuristic is widely studied in solving combinatorial optimization problems, the proposed heuristic includes innovations in evaluating the goodness of local geometry by measuring how efficiently the drone battery can be used in nested units.       
    \item An absolute lower bound on the mission makespan of the Nested-VRP is provided and serves as a benchmark for assessing the quality of heuristic solutions.  
	\item We conduct extensive computational experiments from which we empirically examine the improvement of the Nested-VRP model by applying linearization and constraint strengthening techniques, demonstrate the effectiveness and efficiency of the proposed NS heuristic, and extract valuable insights for practitioners.
	\end{itemize}

The reminder of the paper is organized as follows. In \S\ref{sec:literature}, we present a review of the literature on cooperative vehicle routing. Next, we describe and formulate the Nested-VRP in \S\ref{sec:problem} followed by a discussion of model compactness, complexity, and techniques to linearize as well as strengthen the MIQCP model. The proposed NS heuristic methodology is explained in \S\ref{sec:heuristic}, followed in \S\ref{sec:lowerbound} by a discussion of how we determine the lower bounding of its objective function value. In Section \S\ref{sec:computation}, we provide a detailed analysis of various numerical results as well as a case study. We share our conclusions and proposed directions for future research in \S\ref{sec:conclusion}.

\section{Literature review}\label{sec:literature}
There continues to be growing interest in the coordinated use of drones and trucks to increase  the efficiency of surveillance and transportation systems. The theoretical foundation for this work lies in the Traveling Salesman Problem (TSP) and its variant the Vehicle Routing Problem (VRP). Interested readers can consult surveys regarding solution methodologies for the TSP (see, e.g., \cite{rego2011traveling,dantzig1954solution,lin1973effective}) and VRP (see, e.g., \cite{toth2002vehicle,laporte1992vehicle,kulkarni1985integer}). The Nested-VRP we address, where we seek to optimize the routing for \textit{a single truck} and \textit{a single drone}, can be viewed as an extension of the VRP\@.

Several algorithms have been developed for the combined operation of a single-truck-single-drone system in a delivery context. \cite{murray2015flying} introduced the “Flying Sidekick Traveling Salesman Problem” (FSTSP), where (when appropriate) the delivery drone leaves the truck, completes a single delivery task and returns to the truck when it is at a subsequent customer location. The authors formulated the problem as a Mixed Integer Linear Program (MILP) and proposed two heuristic methodologies whose effectiveness were assessed and demonstrated via a series of computational experiments. The FSTSP heuristic starts by solving the TSP route\footnote[1]{The shortest tour for a person to visit a set of locations} for all customers.  Then, for each drone-eligible customer, the heuristic will decide whether to assign it to the drone tour or reinsert it into the truck tour at a different position in the TSP route. A similar problem, the Traveling Salesman Problem with Drone (TSP-D) was proposed by \cite{agatz2018optimization}. 

\cite{murray2020multiple} subsequently extended their single-truck-single-drone framework to allow the truck to cooperate with a team of drones. Substantial time savings are achieved at the expense of more-complex coordination between vehicles. Numerous other extensions have been proposed since then: (a) improving the formulation of FSTSP such as  \cite{daknama2017vehicle,ha2018min,de2018randomized,dell2019drone}; (b) extending the concept to $m$-truck-$m$-drone scenarios such as \cite{kitjacharoenchai2019multiple} and \cite{sacramento2019adaptive}; and (c) proposing new exact and heuristic methods such as  \cite{schermer2018algorithms,bouman2018dynamic,poikonen2019branch}. However, the aforementioned work addresses the limited battery capacity issue by restricting the drone to visit only one intermediate location between leaving and returning to the truck. While this assumption is reasonable in delivery problems, it is quite restrictive in the case of drones performing surveillance tasks.

\cite{poikonen2020mothership} introduced the ``Mothership and Drone Routing Problem (MDRP)” which considers the routing of a mothership and a drone to visit several designated locations. In the infinite-capacity drone routing problem (MDRP-IC) setting, in contrast to the models mentioned in the last paragraph, the drone is allowed to visit multiple targets consecutively before returning to the mothership for refueling. They devised an exact branch-and-bound solution approach and proposed two greedy heuristic approaches that were demonstrated to be competitive in achieving near-optimal solutions. But the model is fundamentally different from the Nested-VRP\@. While the mothership can move freely in 2D continuous space, the route of the truck in our problem is restricted to the road network which is idealized as straight lines between all pairs of locations. 

To date, the work most-similar in approach to ours is that of \cite{gonzalez2020truck}. They address the truck-drone team logistic (TDTL) problem via an MIP that generates the routes that the drone must follow to visit all the prescribed locations, and assigns rendezvous locations where the drone's batteries are replaced from the truck. Their overall goal is to minimize mission makespan. To deal with the inherent computational complexity, they propose a two-step heuristic approach and demonstrate its performance by comparing it to the exact solution obtained using the Gurobi Optimizer. TDTL departs from the general last mile delivery problem where the drone serves one location per operation. Instead, TDTL allows the drone to serve multiple locations per excursion from the truck. Each excursion involves a set of drone actions including launching from the truck, visiting multiple locations, and returning to the truck. Even though the characteristics of their problem are similar to our problem, their model cannot be adapted to the planning of a surveillance mission. In a typical surveillance mission, the drone battery is being used both when the drone travels between locations and when it executes its observation tasks. Taking into account the observation times at locations is crucial. In the extreme, one can imagine that if all the observation tasks require a full battery, then the truck would be required to visit every location to refuel the drone --- which reduces the surveillance problem to a TSP for the truck.  

Separately, efforts have also been made to address the battery limit issue. For example, \cite{7513397} derived an energy consumption model that can further be integrated into an MILP seeking to optimize routes of a fleet of drones to complete delivery tasks. \cite{cheng2018formulations} studied a multi-trip drone routing problem, where payload and traveling distance are accounted for in determining the drone’s energy consumption. However, they do not consider drones working in collaboration with trucks.  

All the work described above motivates the development of a Nested-VRP that takes into account the observation time at each location. In addition, the drone should be able to visit multiple locations between two consecutive battery swaps. Moreover, the Nested-VRP should also penalize the number of recharge stops by incurring a battery swap service time for each swap operation. And, for the sake of improving vehicle safety, the Nested-VRP should also include a restriction that the truck arrives at the rendezvous location before the drone battery is depleted.

\section{Problem definition}\label{sec:problem}

We describe the Nested-VRP in \S\ref{sec:des_def}, the associated mathematical formulation in \S\ref{sec:MM}, and prove some important results regarding the characteristics of our model in \S\ref{sec:characteristics}.

\subsection{Overview} \label{sec:des_def}
\begin{figure}
    \centering
    \includegraphics[width = \textwidth]{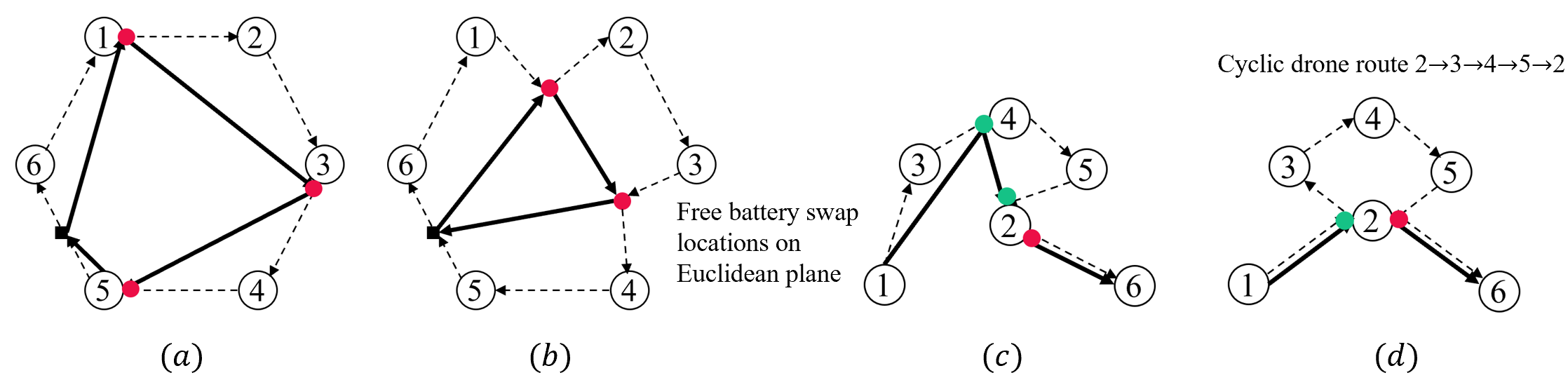}
    \caption{Potential improvements can be achieved by suspending some of the assumptions. (a)--(b) Allowing the battery swap locations to be anywhere on the 2D plane could potentially reduce the number of times the truck and drone rendezvous. (c)--(d) Allowing cyclic operation of the drone could potentially save mission makespan by taking advantage of geometry characteristics of hub-like locations.}
    \label{fig:assumtion}
\end{figure}

In the Nested-VRP, a set of geographically scattered locations is given. Each of these locations has an associated observation task with a prescribed duration. These observation tasks are completed by \textit{a single drone} with a limited battery life supported by \textit{a single truck} with an unlimited supply of fully charged batteries.  Due to limitations on the drone's battery capacity, the drone and truck must periodically meet so that an almost-discharged battery can be swapped with a completely-charged battery supplied by the truck. Specifically, the Nested-VRP considers the routing of the drone including traveling to and making an observation at each location. We make the following assumptions regarding the physical properties of the drone and the truck, as well as the principles guiding the collaboration of the two vehicles:

\begin{enumerate}[label=\textbf{A.\arabic*}]
    \item The drone is equipped with a replaceable battery that is fully charged before the start of the mission. The drone's flight endurance is limited in time due to the battery's limited capacity.
    \item Moreover, the drone's battery consumption is proportional to the active flight duration (i.e., the travel time between two locations or the observation time at the second location). The primary risk of using an oversimplified drone energy consumption model, as \cite{murray2020multiple} pointed out, is that the drone will be unable to reach the designated locations. Such issue raises serious concerns about the drone's safety in the event of an unplanned landing and potentially reduces the mission's efficiency.
    \item We further assume that the road segments between pairs of locations are straight lines. The drone and the truck move according to Euclidean distance between two locations with constant speeds. Therefore, the travel times of the drone and the truck both satisfy the triangle inequality.
    \item The drone, by default, moves no slower than the truck.
    \item The truck has a sufficient battery supply and/or the capability to recharge batteries en route. Thus, there are no constraints on the number of completely-charged batteries that are available when the truck rendezvous with the drone for battery swaps.
    \item The battery swap process can only occur during segments where the truck is carrying the drone or at survey locations either before or after the drone surveys the location. Suspending this assumption would significantly increase the flexibility for two vehicles to choose rendezvous locations, potentially saving time by reducing the coordination effort required by two vehicles, as shown in Figure \ref{fig:assumtion} (a)--(b).     
    \item Each swap operation, including collecting the drone, swapping the batteries, and re-positioning the drone for take-off, delays the mission by a predetermined amount of service time.
    \item Each location can be visited by the drone once and exactly once. This assumption eliminates solutions that contain cyclic operations (i.e., the vehicle starts and ends at the same location). As shown in Figure \ref{fig:assumtion}, (c) represents a Nested-VRP solution; in contrast, scenario (d) saves mission time by performing cyclic operations around location 2. In this case, location 2 delays the mission by an observation task at location 2 and a sequence of tasks associated with the cycle. To the best of the authors' knowledge, some most-promising research in studying the cyclic operation only consider a unit cycle of length two (see, e.g., \cite{wang2017vehicle}, \cite{poikonen2017vehicle}, \cite{bouman2018dynamic}, \cite{schermer2019matheuristic}) in a delivery context. In our case of a time-sensitive surveillance problem, it is an extremely challenging task to use a Mixed Integer Programming (MIP) model to describe the combinatorial choice of locations on a single cycle and describe the precedence relationship of two locations on two distinct cycles that are connected to the same hub. Further research on considering the cyclic operations in the Nested-VRP model is beyond the scope of this paper. \label{asu:cycle}
\end{enumerate}

The objective is to minimize the total mission time needed to complete all observation tasks including time spent traveling and conducting swapping services. Note that the total time the drone spends making observations is part of the mission but cannot be minimized because it is the sum of constant values.    

\begin{figure}[h!]
	\centering
	\includegraphics[width=0.7\textwidth]{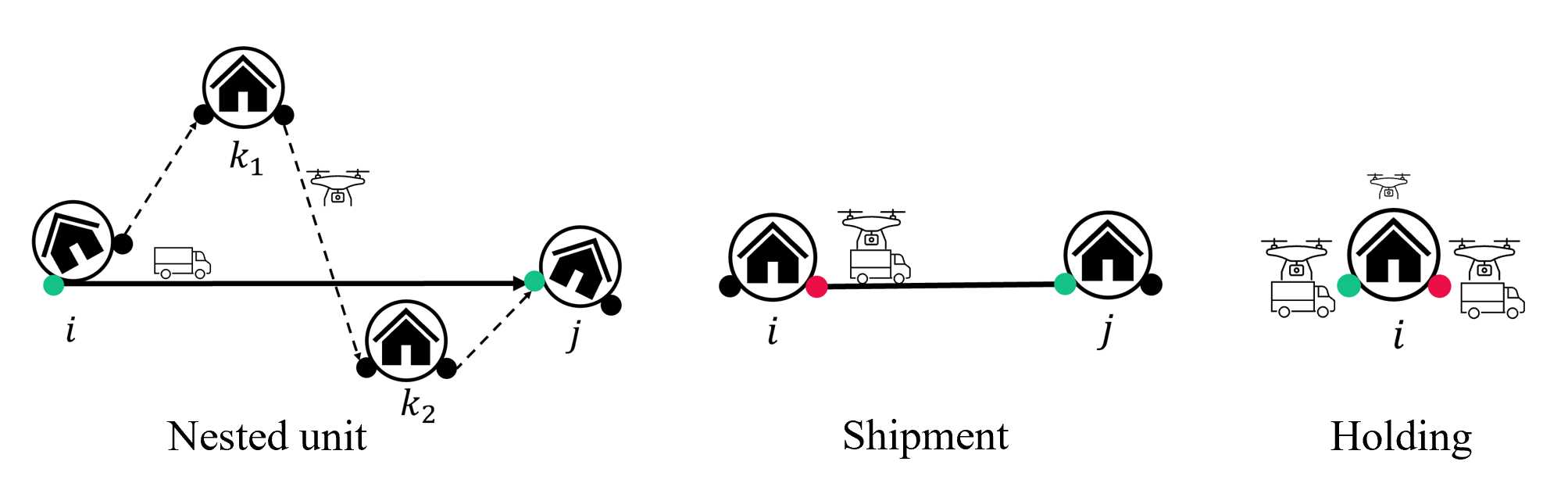}
	\caption{Illustration of a nested unit and its special formations}
	\label{fig:concept}
\end{figure}

To aid in our exposition of the problem and in the subsequent derivation of the mathematical formulation, we introduce the concept of a \textit{nested unit} as shown in Figure \ref{fig:concept}. In a nested unit, the truck travels from location $i$ to location $j$. Meanwhile, the drone departs from $i$, travels to and observes locations $\{k_1, k_2\}$, and finally meets up with the truck at location $j$. In summary, a nested unit consists of four components that are further explained as follows.

\begin{itemize}
	\item \textbf{Truck bridge} refers to the arc from $i$ to $j$. It connects two consecutive locations where a battery swap occurs.
	\item \textbf{Drone path} refers to the collection of arcs that deviate from the truck bridge, and guides the drone to observe a subset of locations assigned to the nested unit. The fact that the drone speed is usually greater than the truck speed makes it possible for the drone to observe multiple locations while the truck travels from $i$ to $j$.  
	\item Node $i$ is a \textbf{split location} from which the drone and the truck initiate their respective next tasks (i.e., taking observations, and delivery of a new battery to the next stop). Generally, the drone gets a fully charged battery and will take off right after the swapping process. Since the truck provides the swap service, it departs no earlier than the drone.
	\item Node $j$ is a \textbf{rendezvous location} at which the drone path and truck bridge merge together. A battery swap happens right after the two vehicles meet. 
	\item In special cases, the nested unit will be transformed and reduced to a \textbf{shipment} or \textbf{holding} pattern. 
	\begin{enumerate}
		\item A \textbf{shipment} pattern occurs when the truck and the drone both decide to traverse a relatively long arc without any observation task involved during the move. When a shipment happens, the truck ships the drone and executes a battery swap in transit. Depends on the required amount of time for completing a battery swap service, the shipment unit delays the mission by the maximum of the truck travel time and the battery swap service time. 
		\item A \textbf{holding} pattern can occur at a location due to a relatively long observation period. In this case, the truck is held at the location and provides batteries to the drone both before and after the drone observes that location. In this case, the truck bridge degrades to a trivial point. Since the holding pattern only considers one location, it differs fundamentally from a cyclic operation, which takes into account at least two locations, as discussed in Assumption \ref{asu:cycle}.
	\end{enumerate}	       
\end{itemize}

The Nested-VRP solution can be viewed as a collection of nested units without overlapping tasks. If the drone has unlimited battery capacity (i.e., infinite endurance time), the optimal solution is that the drone gets a battery at a depot once, follows a TSP route to visit and observe all locations, and returns to the depot without additional swaps along the tour. Referring to Figure \ref{fig:concept}, the optimal solution is a single large nested unit with a split location (the originating depot), a rendezvous location (the depot), and a single drone path (the TSP route). In this case, no truck bridge would be involved. However, if the drone has limited battery capacity, the drone can only operate over relatively short time intervals without a battery swap. It follows then that the single large nested unit must be decomposed into a sequence of smaller nested units in which excessively long arcs and prolonged observation tasks will be accommodated into shipment and holding patterns, respectively. Most importantly, in each nested unit, the drone should be able to complete the specified travel and observation tasks using only its battery capacity. Therefore, the total mission is reduced to having the drone complete the relatively few tasks associated with each nested unit with battery swaps undertaken at the split and rendezvous locations.

We further require that the truck should always arrive before the drone's battery is depleted. This restriction is referred to as a \textbf{sychronization constraint}. However, there is no preference regarding the order in which the two vehicles arrive at a rendezvous location as long as the truck arrives before the drone battery charge has expired. Note that the drone may occasionally have to hover at the rendezvous location if it arrives before the truck.  

To keep track of mission makespan, we define the interval between rendezvous (IBR) for each nested unit as the greater of the two vehicles' travel durations from the split location at the beginning of the nested unit to the rendezvous location at the end of the nested unit. Therefore, minimizing the mission makespan is equivalent to minimizing the summation of all the IBRs associated with nested units together with the total service times needed for battery swaps. Note that we regard a shipment as a special form of a  nested unit whose IBR is the maximum of the truck travel time and the battery swap service time. In the case where a nested unit reduces to a holding pattern, the truck remains stationary while the drone is observing a location. Therefore, the mission makespan increases by the amount of the drone's observation time at that location. The time increment is the IBR of the holding pattern. 

\subsection{Mathematical model} \label{sec:MM}

\begin{table}[H]
	\caption{Notations used in the Nested-VRP MIP formulation. \strut } 
	\centering
	\setlength{\tabcolsep}{4pt}
	\renewcommand{\arraystretch}{1.3}
	\begin{tabular}{ll} 
		\hline 
		\multicolumn{2}{l}{\textbf{Parameters}}\\
		\hline
		$T_{\text{bl}}$ & Battery capacity of drone.\\ 
		$T_{\text{s}}$ & Service time needed to swap a battery.\\
		$\mathcal{H}$ & Set of locations in graph $\mathcal{G} = (\mathcal{H}, \mathcal{A})$. \\
		$\mathcal{A}$ & Set of directed arcs in graph $\mathcal{G} = (\mathcal{H}, \mathcal{A})$.\\
		$\mathcal{S}$ & Set of arcs on the TSP route of a given graph $\mathcal{G}$.\\
		$o_i$,$i\in \mathcal{H} \setminus \{0,n+1\}$ & Observation time associated with location $i$.\\
		$\tau_{ij}^D$, $(i,j)\in \mathcal{A}$ & Drone's flight time from $i$ to $j$.  \\
		$\tau_{ij}^T$, $(i,j)\in \mathcal{A}$ & Truck's travel time from $i$ to $j$.\\
		\hline 
		\multicolumn{2}{l}{\textbf{Decision Variables}} \\
		\hline
		$x_{ij} \in \{0,1\}$, $(i,j)\in \mathcal{A}$  & If $x_{ij} = 1$, the drone flies from $i$ to $j$, 0 otherwise.\\
		$y_{ij}\in \{0,1\}$, $(i,j)\in \mathcal{A}$ & If $y_{ij} = 1$, the truck  travels from $i$ to $j$, 0 otherwise.\\
		$z_i^- \in \{0,1\}$, $i \in \mathcal{H} \setminus \{0\}$ & If $z_i^- = 1$, the drone swaps batteries at location $i$ immediately \it{before}\\
		& it observes location $i$,  $0$ otherwise.\\
		$z_i^+\in \{0,1\}$, $i \in \mathcal{H} \setminus \{n+1\}$ &  If $z_i^+ = 1$, the drone swaps batteries at location $i$ immediately \it{after}\\
		& it observes location $i$, $0$ otherwise.  \\
		$z_i\in \{0,1\}$, $i \in \mathcal{H}$ &  If $z_i = 1$, location $i$ is selected as a battery swap stop, $0$ otherwise.\\
		\hline   
	    \multicolumn{2}{l}{\textbf{Auxiliary Variables}} \\
		\hline
		$t_i^- \in [0, T_{\text{bl}}]$, $i \in \mathcal{H} \setminus \{0\}$ & Total travel and observation time from when drone departs the previous \\
		&  rendezvous location until it \textit{arrives} at location $i$.\\
		$t_i^+ \in [0, T_{\text{bl}}]$, $i \in \mathcal{H} \setminus \{n+1\}$ & Total travel and observation time from when drone departs the previous \\
		&  rendezvous location until it \textit{leaves} location $i$.\\  
		$u_i \in [0, n+1]$, $i \in \mathcal{H}$ & Order index of location $i$ in the solution of the drone route.\\
		$w_{ij}\in \{0,1\}$, $(i,j)\in \mathcal{A}$ & If $w_{ij}=1$, the truck ships the drone from location $i$ to $j$, $0$ otherwise.\\
		$l_i^-\in  \mathcal{R}_+$, $i \in \mathcal{H} \setminus \{0\}$  & IBR of a nested unit that terminates \textit{before} the drone observes location $i$.\\
		$l_i^+ \in  \mathcal{R}_+$, $i \in \mathcal{H} \setminus \{n+1\}$ & IBR of a nested unit that terminates \textit{after} the drone observes location $i$. \\  
		\hline
	\end{tabular}
	\label{tab:notation}
\end{table}
Given the comprehensive list of notations in Table \ref{tab:notation}, consider an undirected graph $\mathcal{G}=(\mathcal{H},\mathcal{A})$, where $\mathcal{H}=\{0,1,2,\ldots,n+1\}$ is the set of locations, and $\mathcal{A}=\{(i,j)\mid i\in \mathcal{H}\setminus \{n+1\},\;j\in \mathcal{H}\setminus \{0\}, \; i \neq j$\} is the set of arcs. Location $0$ is the origin, and location $n+1$ is the eventual destination, which we force to be the origin (so that the trip starts and stops at the same place).  For each location $ i \in \mathcal{H} \setminus \{0, n+1\}$, let $o_i$ be its non-negative observation time, which is assumed to be smaller than $T_{\text{bl}}$. If $o_i$ were to be greater than $T_{\text{bl}}$, then the location $i$ must be set as a swap stop, where at least one battery swap is required (depending on the observation time); and the ``final'' observation time (i.e., after the last swap) is the remainder of the quotient $o_i/T_{\text{bl}}$, that is, $o_i/T_{\text{bl}} - \lfloor o_i/T_{\text{bl}} \rfloor$, where $\lfloor \cdot \rfloor$ denotes the ``floor'' function. For each arc $(i,j) \in \mathcal{A}$, the time metrics $\tau_{ij}^T$ ($\tau_{ij}^D$) represent the truck (drone) travel times between pairs of locations.

A mission consists of planning the drone route $x_{ij}$, $(i,j)\in \mathcal{A}$, to visit all locations and designing the truck route $y_{ij}$, $(i,j)\in \mathcal{A}$, to delivery fully charged batteries. Specifically, the drone route and the truck route should intersect at a subset of locations $z_i$, $i \in \mathcal{H}$ where the truck performs battery swaps for the drone. The primary goal is to minimize the total mission time while operational constraints are satisfied. 

Next, we explain different sets of operational constraints in \S\S\ref{sec:DRC}--\ref{sec:IBR}. Constraints from the same set serve a particular function. In \S\ref{sec:OverallFormulation}, we put together all of the constraints and present the full Nested-VRP formulation.

\subsubsection{Drone route and truck route construction} \label{sec:DRC}
Both the truck and the drone depart from location $0$ and eventually return to location $n+1$. To ensure each location is observed exactly once by the drone, we require that every location in $\mathcal{H} \setminus \{0, n+1\}$ has one incoming arc and one outgoing arc as stated in constraints (\ref{model:drone_1}) -- (\ref{model:drone_2}). 

\begin{align}
& \sum_{j:(i,j) \in \mathcal{A}}x_{ij} = 1, \quad  \forall i \in \mathcal{H} \setminus \{n+1\} \label{model:drone_1}\\
&\sum_{i:(i,j) \in \mathcal{A} }x_{ij} = 1, \quad \forall j \in \mathcal{H} \setminus \{0\} \label{model:drone_2}
\end{align}

The above constraints are not sufficient to construct the drone route because they are also satisfied by subtours in the graph. We further introduce auxiliary variables $u_i$, $i \in \mathcal{H}$, which indicate the order of locations along the drone route. Any potential subtours in the graph will be eliminated by enforcing the following  Miller-Tucker-Zemlin (MTZ) subtour elimination constraints (SEC) proposed by \citep{miller1960integer}.

\begin{align}
    & u_0 = 0\\
    & 1 \le u_i \le n+1, \quad \forall\; i\in\mathcal{H} \setminus  \{0\} \label{model:ubound}\\
    & u_i - u_j  + 1 \le (n+1)(1-x_{ij}),\quad \forall (i,j) \in \mathcal{A}, \, i \neq 0 \label{model:drone_4}
\end{align}

The truck route is constructed in such a way that the truck only serves locations that are selected as battery swap stops (i.e., $z_i=z_i^- \vee z_i^+ = 1$). In the special case where the drone can finish the entire mission without any battery swaps, the truck travels from $0$ to $n+1$ which is equivalent to parks at the depot. The above requirements are captured by constraints (\ref{model:z_1})--(\ref{model:truck_2}). Most importantly, the truck visits the battery swap locations in the same order as that of the drone. The precedence relationship between battery swap locations is enforced by constraint (\ref{model:y_order}) which also eliminates truck subtours.

\begin{align}
    &z_i \le z_i^- + z_i^+, \quad \forall i \in \mathcal{H} \setminus \{0,n+1\} \label{model:z_1}\\
    &z_i \ge z_i^-, \quad \forall i \in \mathcal{H} \setminus \{0\} \label{model:z_2}\\
    &z_i \ge z_i^+, \quad \forall i \in \mathcal{H} \setminus \{n+1\}\label{model:z_3}\\
    & \sum_{j: (0,j) \in \mathcal{A}} y_{0, j} = 1 \label{model:truck_1} \\
    & \sum_{i:(i,j) \in \mathcal{A}}y_{ij} = z_j; \quad \sum_{k:(j,k) \in \mathcal{A}} y_{jk} = z_j, \quad \forall j \in \mathcal{H} \setminus \{0, n+1\} \label{model:truck_2}\\
    & u_i - u_j + 1 \le (n+1)(1-y_{ij}),\quad \forall (i,j) \in \mathcal{A}, \, i \neq 0 \label{model:y_order} 
\end{align}

In theory, the linear programming (LP) relaxation of the Nested-VRP model can be further tightened up by replacing the MTZ subtour elimination constraints by the classical formulation of the SEC constraints: $\sum_{(i,j) \in \mathcal{A},i \in \mathcal{S}, j \notin \mathcal{S}} x_{ij} \ge 2$, where $\mathcal{S} \subseteq \mathcal{H}$ \citep{dantzig1954solution}, denoted by DFJ\@. In the Nested-VRP model, however, substituting the MTZ subtour elimination constraints (\ref{model:ubound})--(\ref{model:drone_4}) with DFJ constraints for the drone and constraints  (\ref{model:y_order}) with DFJ constraints for the truck is not an equivalent transformation. First, the MTZ subtour elimination constraints (\ref{model:ubound})--(\ref{model:drone_4}) together with constraints (\ref{model:y_order}) serve not only to eliminate subtours in the drone route and the truck route respectively, but also to ensure that the truck travels in the same direction as the drone by visiting lower rank to higher rank locations encoded in variable $u_i$. Second, rather than visiting all locations, the truck only travels to and visits locations that are selected as battery swap locations. Only when the set of battery swap locations is determined can we explicitly formulate the DFJ constraints for the truck route. Given the above analysis, we pursue the goal of strengthening the MTZ subtour elimination constraints of the drone by examining lifting technique and Reformulation-Linearization technique (RLT) in \S\ref{subsec:F-Comp}.

Next, to ensure the synchronization constraint, the truck must spend no more than $T_{\text{bl}}$ time in transit between two consecutive rendezvous locations. In the special case where the truck ships the drone (i.e., $w_{ij} =1$), the synchronization constraint becomes redundant. This can be captured by the following constraint in which $M_1 = \max \limits_{(i,j) \in \mathcal{A}} \tau_{ij}^T$.

\begin{equation}
    \tau_{ij}^T y_{ij} \le T_{\text{bl}} + M_1 w_{ij}, \quad \forall (i,j) \in \mathcal{A} \label{model:waiting}
\end{equation}

\subsubsection{Time flow balance} \label{sec:TFB}

To keep track of the drone's battery consumption, we create an artificial timer. The timer records the current battery consumption of the drone by accumulating the total travel and observation time since leaving the previous rendezvous location. At each location, we introduce auxiliary variables ${t_j^-, t_j^+} \in [0, T_{\text{bl}}]$, which we refer to as the state of timer at location $j$. In particular, $t_j^-$ denotes the state of the timer when the drone arrives at location $j$ and $t_j^+$ denotes the state of the timer when the drone is about to leave location $j$. Once a timer is about to exceed the battery capacity $T_{\text{bl}}$, the timer is reset to $0$, corresponding to a battery swap. Constraints (\ref{model:time_1})--(\ref{model:time_2}) state the maximum value of the timer. 

Constraint (\ref{model:time_5}) defines variable $t_j^+$ which is the total travel and observation time from when the drone departs the previous rendezvous location until it leaves the location $j$. It depends on the state of the timer $t_j^-$ as well as the battery swap decision $z_j^-$ when the drone arrives at the location $j$. (a) If $z_j^-=0$, the drone continues the flight and observes location $j$. Thus, the timer keeps accumulating the drone flight time and increases by the amount of $o_j$, that is $t_j^+ = t_j^- + o_j$. (b) If $z_j^-=1$, the drone requires a battery swap service before observing location $j$. This is equivalent to end the previous nested unit and restart a new nested unit for the latter mission. In this case, the timer starts from 0 and becomes $o_j$ once the drone completes the observation task at the location $j$, that is $t_j^+ = o_j$.

Constraints (\ref{model:w_1})--(\ref{model:time_4}) are used to describe the relationship between variable $t_i^+$ and $t_j^-$ due to the drone travel from location $i$ to $j$. The battery consumption on any arc $(i,j) \in \mathcal{A}$ depends on both the drone route decision $x_{ij}$ and the truck shipment decision $w_{ij}$ associated with the arc $(i,j)$.  Constraints (\ref{model:w_1})--(\ref{model:w_5}) ensure that the truck ships the drone (i.e., $w_{ij}=1$) if and only if the two vehicles have decided to traverse the same arc $(i,j)$ (i.e., $x_{ij} = y_{ij} = z_i^+ = z_j^- = 1$). 

Typically, the drone travels from location $i$ to location $j$ alone, i.e., $x_{ij}=1$ and $w_{ij}=0$. Recall that the timer state is $t_i^+$ when the drone is about to leave location $i$ and $t_j^-$ when the drone arrives at location $j$. According to constraints (\ref{model:time_3})--(\ref{model:time_4}), if the arc $(i,j)$ is activated as part of the drone route, the states of the timer at both sides of the arc $(i,j)$ are regulated by the equation $t_j^- = t_i^+(1-z_i^+) + \tau_{ij}^D$. Specifically, if the drone has sufficient battery to cover the arc $(i,j)$, the drone requires no additional battery swaps (i.e., $z_i^+=0$) before leaving $i$. Therefore, the timer increases by the amount of traveling time $\tau^D_{ij}$, and $t_j^-$ becomes $ t_i^+ + \tau^D_{ij}$ when the drone arrives at location $j$. However, if the drone requires a battery swap to be able to cover arc $(i,j)$, a rendezvous location is added when the drone departs location $i$ (i.e., $z_i^+=1$). In this case, since the timer is set to $0$ at the beginning of the arc traveling, the timer accumulates the amount of the drone traveling time and becomes $\tau^D_{ij}$ when the drone reaches the endpoint of arc $(i,j)$.      

Constraints (\ref{model:time_3})--(\ref{model:time_4}) regulate the states of the timer at both sides of arc $(i,j)$. When the shipment happens, the drone's timer is set to $0$ at the time the drone arrives at location $j$ which corresponds to receiving a new battery. Note that the truck does not perform battery swaps at the rendezvous locations placed at both endpoints of the arc $(i,j)$. Therefore, these two rendezvous locations do not delay the mission (see objective function (\ref{model:obj})). 

\begin{figure}[h!]
	\centering
	\includegraphics[width =0.8 \textwidth]{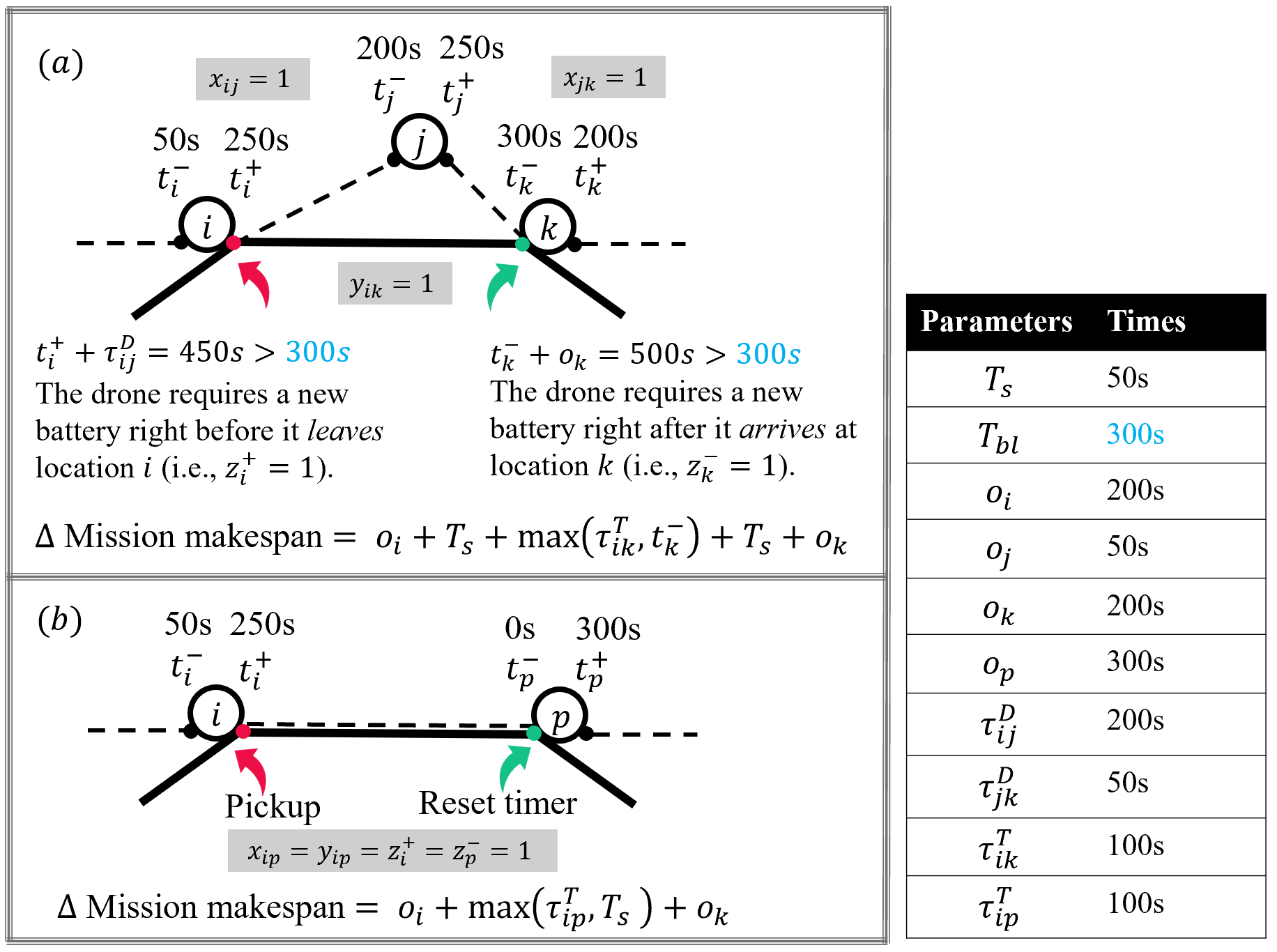}
	\caption{Time flow balance example. (a) Typically, a battery swap can occur when the drone is about to leave a location (red dot) or when it just arrives at a location (green dot). (b) In the special case where the truck ships the drone, a rendezvous location, corresponding to a pickup, is placed before they start traveling (red dot). A rendezvous location is also placed at the end of travel (green dot). Notice that the timer is in state $0$ when the two vehicles arrive at the designated location $p$. In addition, the stops at the two ends of arc $ip$ serve special purposes and do not delay the mission. \label{fig:timeflow}}
	\label{example}
\end{figure}

Arc $(i,j)$ has no impact on the state of the timer if it is not part of the drone route. This can be captured by constraints (\ref{model:time_3})--(\ref{model:time_4}) in which $M_2 = T_{\text{bl}}$. In Figure \ref{fig:timeflow}, we illustrate how  the state of the timer is updated as the mission continues.

\begin{align}
& t_j^- \le T_{\text{bl}}, \quad\forall j \in \mathcal{H} \setminus \{0\} \label{model:time_1}\\
& t_j^+ \le T_{\text{bl}}, \quad\forall j \in \mathcal{H} \setminus \{n+1\} \label{model:time_2}\\
& t_j^+ = t_j^-(1-z_j^-) + o_j,\quad \forall j \in \mathcal{H} \setminus \{0, n+1\} \label{model:time_5}\\  
& w_{ij} \le x_{ij}, \quad \forall (i,j) \in \mathcal{A} \label{model:w_1}\\
& w_{ij} \le y_{ij}, \quad \forall (i,j) \in \mathcal{A} \label{model:w_2}\\
& w_{ij} \le z_i^+, \quad \forall (i,j) \in \mathcal{A} \label{model:w_3}\\ 
& w_{ij} \le z_j^-, \quad \forall (i,j) \in \mathcal{A} \label{model:w_4}\\ 
& x_{ij} + y_{ij} + z_i^+ + z_j^- \le 3 + w_{ij}, \quad \forall (i,j) \in \mathcal{A} \label{model:w_5}\\
& t_j^- \le t_i^+(1-z_i^+) + \tau_{ij}^D(1-w_{ij}) + M_2(1-x_{ij}), \quad \forall (i,j) \in \mathcal{A} \label{model:time_3}\\
& t_j^- \ge t_i^+(1-z_i^+) + \tau_{ij}^D(1-w_{ij}) -M_2(1-x_{ij}), \quad \forall (i,j) \in \mathcal{A} \label{model:time_4}
\end{align}

\subsubsection{Interval between rendezvous} \label{sec:IBR}

 Let $l_j^-$ denote the IBR for the nested unit that terminates when the drone arrives at the location $j$. Likewise, let $l_j^+$ denote the IBR for the nested unit that ends after the drone observes location $j$.   

 Typically, the IBR of a nested unit should not exceed $T_{\text{bl}}$. In the special case when the nested unit reduces to a ``Shipment'' pattern, it must terminate at the arrival of a location $j$ (i.e., $z_j^- = 1, \sum_{i:(i,j) \in \mathcal{A}} w_{ij} = 1$ ). Thus, the IBR of such unit $l_j^-$ is unrestricted. Notice that for a location that is not a battery swap stop, the IBR of such a location does not exist and, therefore, is trivial. Constraints (\ref{model:lbound1})--(\ref{model:lbound2}) capture the above restrictions.  

\begin{align}
    & l_j^- \le T_{\text{bl}}  z_j^- + M_2 \sum_{i:(i,j) \in \mathcal{A}} w_{ij},  \quad \forall j \in \mathcal{H} \label{model:lbound1}\\
& l_j^+ \le T_{\text{bl}} z_j^+,  \quad \forall j \in \mathcal{H} \label{model:lbound2}
\end{align}

As indicated by constraints (\ref{model:late_1})--(\ref{model:late_2}), IBR $l_j^-$ is determined by comparing the drone's surveillance time $t_j^- $ and truck's travel time  $\sum_{i:(i,j) \in \mathcal{A}} \Big( \tau_{ij}^T y_{ij} + (\max(\tau_{ij}^T, T_{\text{s}}) - \tau_{ij}^T)w_{ij} \Big)$ since they both leave the previous rendezvous location. In the special case when the nested unit is a shipment unit connecting, for example, location $k$ and $j$, the drone timer $t_j^-$ at location $j$ becomes $0$ due to constraints (\ref{model:time_3})--(\ref{model:time_4}). In this case, the drone's surveillance time becomes trivial. As a result, the IBR $l_j^-$ should be no smaller than the amount of time the truck spends in shipping the drone from $k$ to $j$ while completing a battery swap service, that is $\max(\tau_{kj}^T, T_{\text{s}})$. 

As indicated by constraints (\ref{model:late_3})--(\ref{model:late_4}), IBR $l_j^+$ is the maximum of the drone's surveillance time $t_j^+$ and the truck's travel time $\sum_{i:(i,j) \in \mathcal{A}} \tau_{ij}^T y_{ij} (1 - z_j^-)$ for the nested unit that ends after the drone observes location $j$. Notably, if the nested unit reduces to a holding pattern (i.e., $z_j^- = z_j^+ = 1$), then the truck's travel time becomes $0$, and thus IBR $l_j^+$ should be no smaller than the drone's surveillance time $t_j^+$.

\begin{align}
& l_j^- \ge t_j^- - M_2(1-z_j^-), \quad \forall j \in \mathcal{H} \label{model:late_1}\\
&l_j^- \ge \sum_{i:(i,j) \in \mathcal{A}} \Big( \tau_{ij}^T y_{ij} + \big(\max(\tau_{ij}^T, T_{\text{s}}) - \tau_{ij}^T \big)w_{ij} \Big)  - M_1(1-z_j^-) , \quad \forall j \in \mathcal{H}  \label{model:late_2}\\
&l_j^+ \ge t_j^+ - M_2(1-z_j^+), \quad \forall j \in \mathcal{H} \label{model:late_3}\\
&l_j^+ \ge \sum_{i:(i,j) \in \mathcal{A}} \tau_{ij}^T y_{ij} (1 - z_j^-) - M_1(1-z_j^+), \quad \forall j \in \mathcal{H} \label{model:late_4}
\end{align}

\subsubsection{Overall formulation} \label{sec:OverallFormulation}

Our objective is to minimize the mission makespan, which consists of the IBRs of all nested units in the solution and the total service time for battery swaps. In the case where the truck ships the drone from location $i$ to $j$ (i.e., $w_{ij}=1$), the model enforces $z_i^+ = z_j^- =1$ for adding meetup stops at the two ends of arc $(i,j)$. Since no battery swaps happen at these two stops, the objective function adjusts for over counting  the battery swaps service time by adding $- \sum_{(i,j) \in \mathcal{A}} 2w_{ij}$. The complete Nested-VRP model is given as follows. 

\begin{align}
 (\text{Nested-VRP}) \quad \min_{} \quad &  \sum_{i\in \mathcal{H}} (l_i^- + l_i^+) + T_{\text{s}} \big( \sum_{i \in \mathcal{H}} ( z_i^- + z_i^+ )  -  \sum_{(i,j) \in \mathcal{A}}  2w_{ij}\big)  \label{model:obj}\\  
\textrm{s.t.} \quad & \mbox{constraints  (\ref{model:drone_1})--(\ref{model:late_4})} \nonumber \\
&z_{0}^+ = 1, \; t_{0}^+ = 0 \label{model:ini}\\ 
& x_{ij} \in \{0,1\}, \; y_{ij} \in \{0,1\}, \; w_{ij} \in \{0,1\}, \forall (i,j) \in \mathcal{A}\label{model:v1}\\
&z_i^-, \in \{0,1\},  \; t_i^-,  \in  \mathcal{R}_+, \; l_i^-,   \in  \mathcal{R}_+,\forall i \in \mathcal{H}\setminus \{n+1\} \; \label{model:v2}\\
&z_i^+, \in \{0,1\},  \; t_i^+,  \in  \mathcal{R}_+, \; l_i^+,   \in  \mathcal{R}_+,\forall i \in \mathcal{H}\setminus \{0\} \; \label{model:v3}\\
&z_i, u_i \in [0, n+1], \forall i \in \mathcal{H} \label{model:v4}
\end{align}

\subsection{Model characteristics} \label{sec:characteristics}
In \S\ref{sec:model_prop1}, we compare the Nested-VRP model to what we regard as the state-of-the-art model in terms of their relaxed polyhedra. Next, in \S\ref{sec:improve}, we derive a linearized Nested-VRP model and propose valid constraints to further strengthen the proposed formulation. Finally, in \S\ref{sec:complexity}, we investigate the complexity of the Nested-VRP model given prior drone route information.

\subsubsection{Relationship to the state-of-the-art model}\label{sec:model_prop1}
From a modeling perspective, it is of particular interest to compare the compactness of the proposed MIP model to other state-of-the-art models. To the best of our knowledge, the most-similar model concerning truck-drone coordinated routing is that of \cite{gonzalez2020truck} --- the Truck Drone Team Logistics (TDTL) model. Since the TDTL model does not consider the observation times associated with each location nor the service time in swapping the battery, we will first derive a special version of the Nested-VRP model where we neglect the battery swap service time and set the observation time to zero for every location. Our special version of the Nested-VRP is called Zero Observation Nested-VRP (ZONVRP)\@. Since the two models to be compared apply different notations with different physical meanings, a linear transformation $\Phi$ that maps the ZONVRP variables to that of TDTL is needed. We will show that the LP relaxation of the Nested-VRP model is tighter than the LP relaxation of the TDTL with respect to a linear transformation. This is noteworthy because the majority of commercial MIP Optimizers have a branch-and-bound component that leverages the associated LP to iteratively search for the optimal solution. Thus, a tighter formulation usually requires the evaluation of fewer branching nodes thereby reducing computation time. Further, even if the optimal solution can not be obtained within the time limit, the MIP Optimizer can provide a better bound on the optimal value of the problem at termination when using a tighter formulation.

\begin{theorem} \label{thm:compact}
 Denote the feasible set of ZONVRP under a linear transformation $\Phi$ as $P_1$, and denote the feasible set of TDTL as $P_2$. Then $P_1$ is a proper subset of $P_2$. (See Appendix \ref{app:proof} for $\Phi$ construction and a detailed proof.)
\end{theorem}

\subsubsection{Model linearization \& formulation strengthening} \label{sec:improve}
One possible way to reduce model complexity is to transform the structure of the model into a more-amiable one. Specifically, the mathematical formulation of the Nested-VRP model consists of a handful of quadratic terms in the constraints, which leads to a Mixed Integer Quadratically Constrained Program (MIQCP)\@. An efficient methodology for directly tackling the MIQCP model requires careful design and relies on a combination of special techniques for decreasing the "upper bound" (e.g., heuristics to obtain integral solutions) or increasing the lower bound (e.g., valid inequality) which is challenging for commercial optimizers. Tracing back to the seventies, \cite{jeroslow1973there} showed that solving an MIQCP is deemed as a formidable task due to its notorious computational intractability. This motivates us to linearize the MIQCP model to its MILP equivalent. Solving an MILP model is one of the most-successful achievements in computational optimization. Theoretical and computational research progress over the last six decades has shown that MILPs can be solved in many if not all practical settings. We further propose an MILP equivalent of the Nested-VRP model in Proposition \ref{prop:MILP}.
\begin{proposition} \label{prop:MILP}
The following pairs of constraints are equivalent: (i) linear constraints (\ref{eq:15}) and quadratic constraints (\ref{model:time_5});  (ii) linear constraints (\ref{eq:2122}) and quadratic constraints (\ref{model:time_3}) -- (\ref{model:time_4}); (iii) linear constraints (\ref{eq:28}) and quadratic constraints (\ref{model:late_4})

\begin{equation}\label{eq:15}
\begin{cases}
    & P_j \in [0, T_{bl}]\\
    & P_j \le T_{bl}z_j^-\\
    & P_j \le t_j^-\\
    & P_j \ge t_j^- - T_{bl} (1-z_j^-)\\
    &t_j^+ = t_j^- -  P_j +o_j, \quad \forall j \in H \setminus \{0, n+1\} \tag{$15^\star$} 
\end{cases}
\end{equation}

\begin{equation}\label{eq:2122}
\begin{cases}
    & F_i \in [0, T_{bl}]\\
    & F_i \le T_{bl}z_i^+\\
    & F_i \le t_i^+\\
    & F_i \ge t_i^+ - T_{bl} (1-z_i^+)\\
    & t_j^- \le t_i^+ - F_i + \tau_{ij}^D(1-w_{ij}) + M_2(1-x_{ij}),\forall (i,j) \in A \\
    &t_j^- \ge t_i^+ - F_i + \tau_{ij}^D(1-w_{ij}) - M_2(1-x_{ij}),\forall (i,j) \in A \tag{$21^\star)-(22^\star$}
\end{cases}
\end{equation}

\begin{equation}\label{eq:28}
\begin{cases}
    & Q_{ij} \le y_{ij}, \quad \forall (i,j) \in A \\
    & Q_{ij} \le z_j,\quad \forall (i,j) \in A \\
    & Y_{ij} + z_j \le Q_{ij}+1,\quad \forall (i,j) \in A \\
    & l_j^+ \ge \sum_{i:(i,j) \in \mathcal{A}} \tau_{ij}^T\big(  y_{ij} - Q_{ij} \big) - M_1(1-z_j^+), \quad \forall j \in \mathcal{H} \tag{$28^\star$}
\end{cases}
\end{equation}
where auxiliary variables $P_j = t_j^-z_j^-, \forall j \in H \setminus \{0, n+1\}$, $F_i = t_i^+z_i^+, \forall i \in H \setminus \{n+1\}$, and $Q_{ij} = y_{ij} z_j^-, \forall (i,j) \in A$.
\end{proposition}

Another way to possibly speed up the problem-solving process is to strengthen the Nested-VRP formulation. When solving the Nested-VRP model, as we stated earlier, the commercial optimizer typically adopts the spirit of branch-and-bound algorithm at its heart which involves repeatedly solving the LP relaxation and leveraging the generated bounds to guide branching decisions. Such an exact approach favors two strategies to resolve the need for integrality for variables. The first strategy is to identify effective cutting planes with the goal of iteratively reducing the search space by introducing linear inequalities. The second strategy aims at tightening up the polyhedral representation of the mathematical model at the root node by reformulating the model. In this paper, we adopt the second strategy and employ two different techniques to tighten up the MTZ subtour elimination constraints (\ref{model:ubound})--(\ref{model:drone_4}) that describe the routing decisions of the drone $x_{ij} , \forall (i,j) \in A$. \\

\cite{desrochers1991improvements} use a lifting technique to strengthen the MTZ subtour elimination constraints for the Traveling Salesman Problem. For the Nested-VRP, we apply their ideas with minor variations.

\begin{proposition}\label{prop:dl}
The constraints
\begin{align}
    & 1 + (1 - x_{0i}) + (n-2)x_{i,n+1} \le u_i \le (n+1) - (n-1)x_{0i} - (1 - x_{i,n+1}), \quad \forall i \in H \setminus \{0, n+1\} \tag{DL-1} \label{dl1}\\
    & u_i - u_j + (n+1)x_{ij} + (n-1)x_{ji} \le n,  \quad \forall (i,j) \in A, i \neq 0 \tag{DL-2} \label{dl2}
\end{align} 
are valid inequalities for the Nested-VRP\@. (See Appendix \ref{app:prop1} for a detailed proof.)
\end{proposition}

\cite{sherali2002tightening} proposed a novel RLT to derive even tighter relaxations for the Asymmetric Traveling Salesman Problem (ATSP) that is based on the MTZ formulation. The resulting new formulation of the ATSP has been theoretically and computationally shown to outperform the lifted-MTZ formulation proposed by \cite{desrochers1991improvements}. We apply the proposed RLT on strengthening the MTZ subtour elimination constraints in the Nested-VRP model.

\begin{proposition} \label{prop:sd}
The constraints
\begin{align}
     & \sum_{j:(i,j) \in A, j \neq n+1} y_{ij} + nx_{i, n+1} - u_i = 0, \quad \forall i \in H \setminus \{n+1\} \tag{SD-1} \label{sd_1}\\
    & \sum_{i:(i,j) \in A} y_{ij} + 1 = u_j,  \quad \forall j \in H \setminus \{0\} \tag{SD-2} \label{sd_2}\\
    & x_{ij} \le y_{ij} \le nx_{ij}, \quad \forall (i,j) \in A, i \neq 0 \tag{SD-3} \label{sd_3}\\
    & u_i + (n+1)(x_{ij}-1) + nx_{ji} \le y_{ij} + y_{ji} \le u_i - (1-x_{ij}), \quad \forall (i,j) \in A, i \neq 0 \tag{SD-4} \label{sd_4}\\
    & 1 + (1-x_{0i}) + (n-2)x_{i,n+1} \le u_i \le n +1 - (n-1)x_{0i} - (1 - x_{i, n+1}), \quad \forall i \in H \setminus \{0, n+1\} \tag{SD-5} \label{sd_5}
\end{align}
are valid inequalities for the Nested-VRP\@. (See Appendix \ref{app:prop2} for a detailed proof.)
\end{proposition}

Table \ref{tab:compareF} presents the original Nested-VRP model and three new models with different choices of linearization and strengthening techniques. For simplicity, the original Nested-VRP model is referred to as MIQCP and the linearized Nested-VRP model is referred to as MILP\@. Similarly, we refer to the MILP as MILP+DL or MILP+SD when the drone's MTZ subtour elimination constraints are replaced by constraints (\ref{dl1})\mbox{--}(\ref{dl2}) or (\ref{sd_1})\mbox{--}(\ref{sd_5}). In addition, Table \ref{tab:compareF} compares the size of the models in terms of the numbers of variables and constraints. In spite of the fact that the linearization process inevitably introduces $O(n^2+2n)$ additional variables and $O(3n^2+12n)$ additional constraints, it does convert the MIQCP model into one for which a large number of solution algorithms have been developed. The formulation strengthening techniques, as stated in Proposition \ref{prop:dl} and Proposition \ref{prop:sd}, do not bring more variables into either the MILP+DL or the MILP+SD model. However, with the use of the RLT, we introduce $O(n^2 + 3n)$ additional constraints into the MILP+SD model as compared to the MILP model.

\begin{table}[h!] 
\resizebox{\textwidth}{!}{
\begin{tabular}{lllll}
\toprule
\multicolumn{1}{l}{}  & \multicolumn{1}{l}{MIQCP}  & \multicolumn{1}{l}{MILP} & \multicolumn{1}{l}{MILP+DL} &  \multicolumn{1}{l}{MILP+SD} \\ 
\cmidrule{2-5}
& (\ref{model:drone_1})--(\ref{model:v4}) & (\ref{model:drone_1})--(\ref{model:v4}) & (\ref{model:drone_1})--(\ref{model:v4}) & (\ref{model:drone_1})--(\ref{model:v4})  \\
& & with (\ref{model:time_5}) substituted by (\ref{eq:15})  & with (\ref{model:time_5}) substituted by (\ref{eq:15}) & with (\ref{model:time_5}) substituted by (\ref{eq:15})\\
\multicolumn{1}{r}{Formulation}  &  & with (\ref{model:time_3}) -- (\ref{model:time_4}) substituted by (\ref{eq:2122})  & with (\ref{model:time_3}) -- (\ref{model:time_4}) substituted by (\ref{eq:2122}) & with (\ref{model:time_3}) -- (\ref{model:time_4}) substituted by (\ref{eq:2122})    \\
& & with (\ref{model:late_4}) substituted by (\ref{eq:28})  & with (\ref{model:late_4}) substituted by (\ref{eq:28}) & with (\ref{model:late_4}) substituted by (\ref{eq:28}) \\
& & & with (\ref{model:ubound}) -- (\ref{model:drone_4}) substituted by (\ref{dl1}) -- (\ref{dl2}) & with (\ref{model:ubound}) -- (\ref{model:drone_4}) substituted by (\ref{sd_1}) -- (\ref{sd_5})  \\ 
\cmidrule{2-5}
\multicolumn{1}{r}{Binary variables}  & $O(3n^2 + 9n)$ & $O(4n^2 + 11n)$  & $O(4n^2 + 11n)$ & $O(4n^2 + 11n)$    \\
\cmidrule{2-5}
\multicolumn{1}{r}{Continuous variables} & $O(5n)$ & $O(7n)$  & $O(7n)$ & $O(7n)$    \\
\cmidrule{2-5}
\multicolumn{1}{r}{Constraints}  & $O(10n^2 + 35n)$ & $O(13n^2 + 47n)$ & $O(13n^2 + 47n)$ &  $O(14n^2 + 50n)$\\
\bottomrule
\end{tabular}
}
\caption{Compare MIQCP, MILP, MILP+DL, and MILP+SD formulations and corresponding size as a function of number of locations $n$.\label{tab:compareF}}
\end{table}

\subsubsection{Model complexity}\label{sec:complexity}
To understand the complexity of the Nested-VRP, we will evaluate the computational effort required throughout the decision-making process. (i) First, the drone route is constructed by sequencing $n$ locations. Together with the origin $0$ and eventual destination $n+1$, there exists $n!$ different drone routes. Each possible drone route has a unique battery consumption pattern along the tour. (ii) Second, given a specific possible drone route, the swap stops assignment is a problem of finding a subset of locations that naturally split the drone route into path segments. In particular, the drone's flight duration and the truck's ground travel time for each of these segments should  both be within the battery limit. We can see that the choices of the set of charging locations is highly sensitive to the battery usage corresponding to the drone route. Thus, if implementing a drone route requires excessive battery swaps to ensure flight continuity, an adjustment to the drone route may reduce the coordination effort for the truck to serve batteries. As the name of the model suggests, the decisions regarding the drone route and truck route are intertwined dynamically and tied by the decisions on swap locations. To solve the Nested-VRP, we should navigate through the tasks of planning a good drone route and scheduling battery swaps in a versatile manner. Neither component seems to dominate the other, and that question merits further investigation---if partial information about the Nested-VRP solution is given, how much effort is needed to obtain the complete Nested-VRP solution? 

In the following, we will demonstrate that, given a fixed order for the drone to visit all locations, we can solve within polynomial time the remaining Nested-VRP solution---including the truck route and the placement of swap stops---that minimizes the mission makespan.

\begin{theorem}
	\label{thm:ESC}
	Given a fixed order of a set of locations representing a known drone route, the partial Nested-VRP solution ---including a subset of locations as swap stops and the truck route---can be solved in polynomial time.
\end{theorem}

\begin{pf}
With the drone route specified, the optimal Nested-VRP solution can obtained by finding the cheapest collection of non-overlapping nested units such that the union of the drone paths from each unit aligns with the predetermined drone route. In the following, we will first construct the set of all feasible nested units. Each of these nested units is associated with a cost (i.e., IBR plus battery service time)\@. Then, the collection of nested units (CNU) with the smallest mission time is obtained by solving an integer program efficiently. We name it the CNU problem.

Let $(s_0,s_1,\ldots,s_{n+1})$ be a permutation of the node set $\mathcal{H}$, where $s_0 = 0$ and $s_{n+1} = n+1 $. In following this order, the drone departs from a location $s_i$, travels to the next location $s_{i+1}$ which takes time $\tau^D_{s_is_{i+1}}$, and spends $o_{s_{i+1}}$ time for surveying location $s_{i+1}$. Let $W=(w_0, w_1, w_2, \ldots,w_{2n+1})$ denote the consecutive tasks to be completed by the drone in the mission, where $w_0 = o_{s_0} = 0,w_1 = \tau^D_{s_0 s_1}, w_{2k} = o_{s_k}, w_{2k+1} = \tau^D_{s_k s_{k+1}}, \forall k =1, \ldots,n$. According to the assumptions of the problem, a rendezvous location is always placed at the beginning of a mission which corresponds to equipping the drone with a full battery. In addition, for simplicity, we assume that any other rendezvous location $i \in \{0, 1, \ldots, 2n+1\}$ should be limited to where the task $w_i$ is completed (i.e., the drone just arrives at a location or just completes an observation task). 
 
For a nested unit with the first rendezvous location at the end of completing task $w_i$ and the second rendezvous location at the end of completing task $w_j$, we denote the nested unit as $(i,j, l_{ij}), i, j \in \{0, 1, \ldots, 2n+1\}, \, i < j$, where $l_{ij}$ is the cost of the nested unit measured in time. Specifically, variables $l_{ij}$ is the sum of the battery swap service time for obtaining a new battery at the start of the nested unit and the interval between rendezvous of the unit. Recall that when the truck ships the drone, the nested unit is in a special form of ``Shipment" which delays the mission by the maximum of the truck travel time and  the battery swap service time. To determine $l_{ij}$, we further define $D_{ij} = \sum_{k=i+1}^{j} w_k$ as the total drone travel and surveillance time for completing the ($i+1$)th task and all others up to and including the $j$th task. Likewise, define $T_{ij}$ as the truck travel time moving from where the $i$ task is completed to where the $j$th task will be completed by the drone. A nested unit is feasible if both $D_{ij}$ and $T_{ij}$ are within the battery limit $T_{\text{bl}}$ unless the nested unit is in a form of ``Shipment". The nested unit, if picked, will delay the mission by $l_{ij} \in \mathcal{R_+}$, which is defined as follows.

\begin{equation}
l_{ij} = \begin{cases}
 \max (T_{ij}, T_s)&  \text{if}\; i \; \text{is even}, \; j = i+1,  \text{``Shipment''}  \\
 \max(D_{ij}, T_{ij}) + T_s & \text{if } j > i+1,  D_{ij} \le T_{\text{bl}} \ \text{or} \  T_{ij} \le T_{\text{bl}}, \text{``Nested unit''}\\
 \infty  & \text{otherwise}
\end{cases}\nonumber 
\end{equation}

Let $U = \{(i, j, l_{ij})\ : \forall i, j \in \{0, 1, \ldots, 2n+1\}, \, i < j\}$ denote the set of all possible formations of nested units, the CNU problem is to find the least-delayed subset of nested units such that all tasks are completed/covered exactly once. We now formulate the CNU problem. Define the binary matrix $A \in \{0,1\}^{|U||W|}$. For each nested unit $u \in U$, we have $A_{uk} = 1$ if task $w_k$ is covered by nested unit $u$, and $0$ otherwise. For simplicity, let $l_u$ represents the time cost of nested unit $u$. Define decision variables $x_u = \{0, 1\}, \forall u \in \{1, \ldots, |U|\}$. If $x_u=1$, then nested unit $u$ is selected, $0$ otherwise. 

\begin{align}
\label{model:CNU}
(\text{CNU}) \ \ \min \quad & \sum_{u=1}^{ |U|}l_ux_u\\
\textrm{s.t.} \quad & \sum_{u=1}^{|U|}A_{uk}x_u =1,\quad \forall k \in \{1, 2, \ldots, 2n+1\} \nonumber \\
& x_u \in \{0, 1\},\quad \forall i \in \{1, 2, \ldots, |U|\}  \nonumber
\end{align}

The matrix $A$ is total unimodular (TU) since each row of matrix $A$ consists of consecutive ones \citep{schrijver1998theory}. Then since $A$ is TU, the non-empty polyhedron $P(b) = \{ Ax = b, x \ge 0 \}$ has integral vertices for the all-integral vector $b$ \citep{wolsey1999integer}; in our case,  $b$ is a vector of all $1$s. Therefore, we can solve the CNU model by solving its linear relaxation and still achieve integer solutions. The time effort in solving a linear program is polynomially bounded by the total number of variables $|U|$ \citep{karmarkar1984new}.
\end{pf}

\section{Heuristic methodology via neighborhood search} \label{sec:heuristic}

The Nested-VRP, being an extension of the Traveling Salesman Problem, is difficult from a theoretical perspective. In particular, as the size of the problem increases, the complexity of the MIQCP model and limited computation time do not allow for an exact solution. This motivates us to develop a heuristic methodology that is expected to produce good solutions given a computational time budget.

 The neighborhood search (NS) framework, first introduced in \cite{shaw1998using}, has been demonstrated as a powerful tool for solving difficult combinatorial optimization problems. A generic NS starts with an initial feasible solution to the problem of interest. Each NS iteration involves destructing the current best-known solution and reconstructing a better candidate by modifying local decisions. The best solution obtained by the time of termination is recorded as the final result.
 
 As we have discussed, the optimal Nested-VRP solution can be characterized as the time-minimizing collection of non-overlapping nested units each of which obeys the battery capacity limit and the synchronization constraint. Given a Nested-VRP instance, we denote its optimal solution as $I_{\text{opt}}$ which consists of the optimal configuration of nested units. In our NS framework, we start with a feasible solution $I$ of the same instance. Initially, the feasible solution $I$ forms the nested units in a suboptimal way (i.e., at least one nested unit does not match that in the optimal solution $sol$). Next, $I$ can be further improved by iteratively destructing mismatched nested units and regrouping them with their neighboring units. When the search process terminates, the heuristic returns the best known solution $I^*$. Therefore, the key to success of an NS heuristic approach to solving the Nested-VRP boils down to: (i) Finding a good initial feasible solution that contains the least possible mismatched nested units in comparison to the true (unknown) optimal solution as a starting point. (ii) Identifying the set of undesirable nested units that are more likely mismatched compared to the (unknown) optimal solution. (iii) An effective destruction and reconstruction process to fix the local mismatches.
 
\begin{figure}[h!]
	\centering
	\makebox[\textwidth][c]{
	\includegraphics[width=\textwidth]{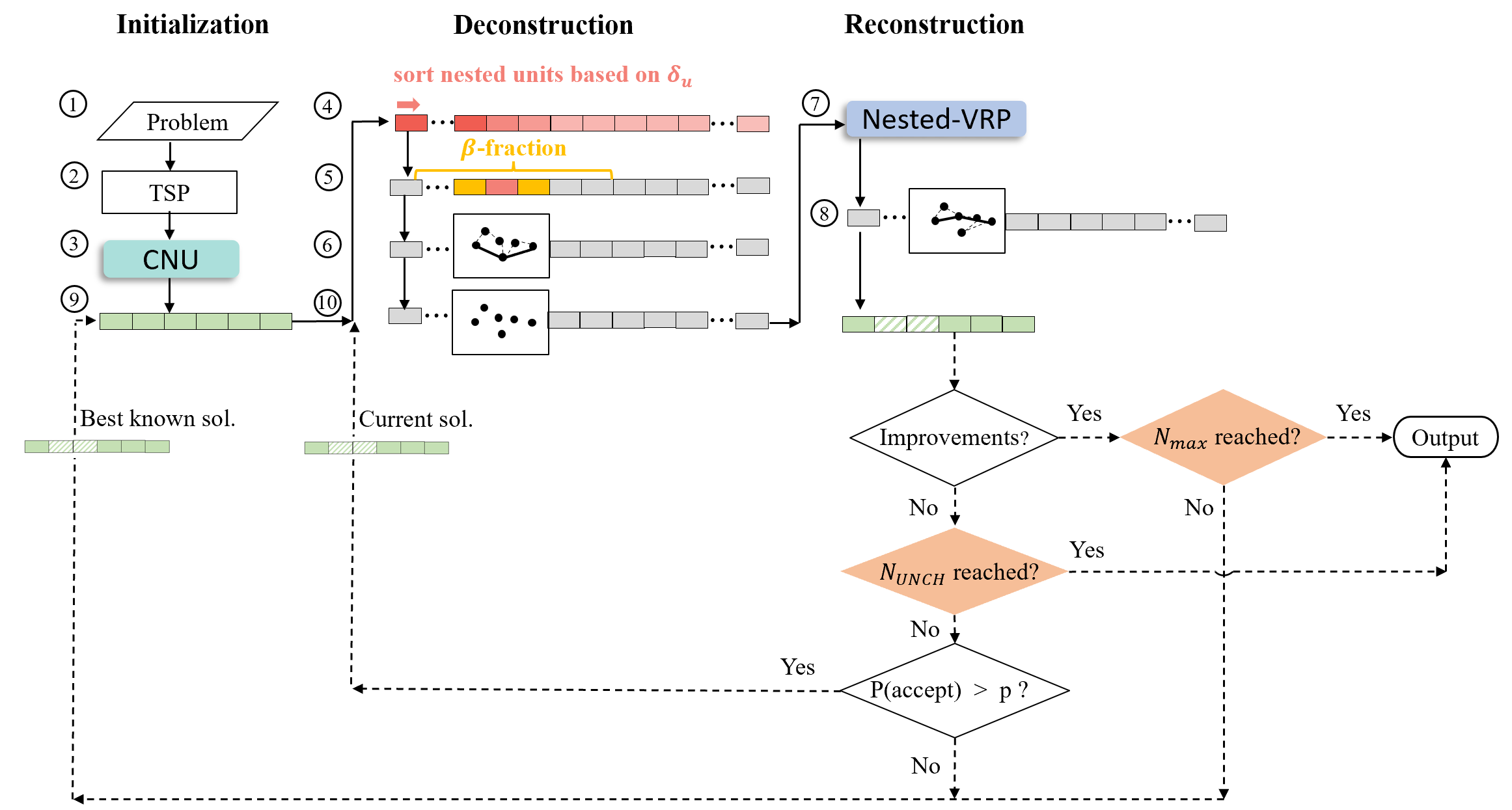}}
	\caption{Flow chart for the NS heuristic, including Initialization, Destruction, and Reconstruction phases.}
	\label{fig:LNS}
\end{figure}

 We summarize the overall NS heuristic in Figure \ref{fig:LNS}. The initialization, destruction, reconstruction, and termination components are further discussed in \S\S\ref{Initialization}, \ref{sec:Destruction}, \ref{sec:Reconstruction}, and \ref{sec:Termination}, respectively. In addition, a formal description of the NS heuristic is stated at the end of this section.

\subsection{Initialization}\label{Initialization}
Suppose that the order to visit all locations (i.e., the drone route) is given.  By Theorem \ref{thm:ESC}, the full set of Nested-VRP solutions can be determined by further solving the CNU problem (\ref{model:CNU}). Therefore, finding a good initial drone route is critical in obtaining a feasible Nested-VRP solution.   

We expect that the initial feasible solution has as few mismatched nested units compared to the unknown optimal solution. Empirically, we have observed from the solutions of small-size problems (whose exact optimal solutions are actually obtainable) that even if the TSP order is not always guaranteed to be the best drone route, it at least aligns with the optimal drone routing most of the time. In fact, in the situation where the true optimal drone route is not the TSP route, we have still empirically observed that a large portion of nested units, both in the heuristic solution and the optimal solution, share the same configurations. Moreover, \cite{murray2015flying} has conducted computational experiments to investigate how the optimality of the TSP tour, which is used as a starting point for the proposed heuristic method, affects the solution quality. The results show that starting with the optimal TSP tour, the proposed heuristic finds higher-quality solutions than starting with sub-optimal TSP tours generated by other TSP heuristics (e.g., nearest neighbor method).

Thus, in the initial Nested-VRP solution, we enforce that the drone route is the same as the TSP route, i.e., the shortest-time tour to visit every location exactly once.  With the known drone route, the decisions on the truck route and the assignment of battery stops along the tour can be obtained by further solving the CNU problem (see Figure \ref{fig:LNS}, steps 1--3). 

\subsection{Destruction}\label{sec:Destruction}
Given a feasible solution to be improved, the destruction process is to remove the nested units that are likely misaligned with those in the true optimal solution. Since the Nested-VRP favors a solution with a minimum makespan, each nested unit is expected to pack in as many observation tasks and be as efficient with battery energy as possible. Therefore, if one nested unit only consumes a small fraction of battery capacity and leaves a great amount of energy wasted when a new battery swap occurs, the nested unit is identified as undesirable. A nested unit $u$ is deemed to be less desirable if the amount of wasted battery capacity, named battery slackness $\delta_u$ is relative large. Motivated by this, the destruction process involves sorting all nested units in \textit{non-increasing} order of $\delta_u$ and randomly choosing a nested unit from the top $\beta$-fraction of the list as the bad unit at the current iteration. Intuitively, an NS heuristic parameterized by a larger $\beta$ is less tolerant of inefficient battery energy usage. 

With the bad nested unit identified, the NS heuristic proceeds to destroy the bad nested unit itself as well as its neighbors. Regarding the severity of the destruction process, if the impacted neighborhood is limited, there is not much freedom for the reconstruction process to identify a better choice. On the contrary, once a large portion of the initial solution is destructed, the reconstruction process is equivalent to resolving a Nested-VRP of a relatively larger size. To alleviate the brunt of computational complexity per iteration, we currently restrict the destruction process to the bad unit plus a single neighbor located either immediately before or after the bad unit. At the end of the destruction process (see Figure \ref{fig:LNS}, steps 4--6), we obtain a set of locations that were previously covered by the bad unit and its neighbor. To complete the observation tasks at these locations, we still need to determine the drone route, truck route, and assignment of the battery swap stops for coordinating the truck and the drone. We present the destruction function in Algorithm \ref{alg:destruction}.  

\vspace{.1in}

\begin{algorithm}[H]
	\label{alg:destruction}
	\SetAlgoLined
	\KwData{$I$: The current feasible solution\;}
	\KwResult{$U$: bad nested units to be destroyed\; $L$: the set of locations that are included in set $U$\;}
	$pool \leftarrow$ decompose the current Nested-VRP solution to a set of nested units \;
	\For{$\text{nested unit } u$ in $pool$}{compute battery slackness $\delta_u$ \;}
	$pool \leftarrow$ sort all nested units in an non-increasing order of $\delta_u$\; 
	$l \leftarrow$ size of the pool\;
	$U$ $\leftarrow$ randomly choose a unit from the first to $\beta l$-th units from $pool$ and pick one of its neighboring units \;
	$L$ $\leftarrow$ locations that are covered by nested units in $U$ \;
	\caption{Destruction Function}
\end{algorithm}

\subsection{Reconstruction}\label{sec:Reconstruction}
Nested units that have been destroyed at the end of the destruction process are in the form of free locations in the graph. In the reconstruction process, our goal is to sequence the free locations and pick a subset of the free locations as swap stops to minimize the total time for the drone to receive battery replacements, travel, and complete observation tasks associated with these free locations. This can be carried out by solving the Nested-VRP on the free locations locally and exactly. 

Even though the Nested-VRP model suffers from computational complexity as the size of the problem increases, in the reconstruction process, the number of locations that need to be solved in each iteration is relatively small. This enables us to take advantage of the Nested-VRP model that produces a local operation plan with the smallest makespan (see Figure \ref{fig:LNS}, steps 7--8). Interestingly, during the local reconstruction, the order of the free locations will be altered to explore the portion of time-saving benefits lost due to pre-fixing the drone route. 

At the end of the reconstruction process, we obtain a new set of nested units. Compared to the grouping before being destructed, if the new sets have a smaller makespan, we accept the solution and replace the old grouping with the new one (see Figure \ref{fig:LNS}, step 9).   Otherwise, we will accept it with a probability $1/2$. Accepting a worse solution allows the heuristic to step out of a local minimum and explore for the global minimum. If the solution is rejected, the heuristic proceeds to the next iteration with the same best-known solution (Figure \ref{fig:LNS}, step 10). After that, the destruction process picks a different bad unit and continues the search process. The reconstruction function is detailed in Algorithm \ref{alg:reconstruction}.

\vspace{.1in}

\begin{algorithm}[H]
	\label{alg:reconstruction}
	\SetAlgoLined
	\KwData{$I$: The current feasible solution\; $U$: The set of bad units\; $L$: The set of locations that are covered by bad units\;}
	\KwResult{$I'$: a candidate solution\;}
	$L'$ = Nested-VRP($L$) \tcp*[l]{$L'$ is a collection of nested units obtained by solving the Nested-VRP model on locations in $L$}
	$I'\leftarrow (I \backslash U) \cup L'$ \tcp*[l]{merge good nested units with the newly formed nested units} 
	\caption{Reconstruction Function}
\end{algorithm}

\subsection{Termination criteria}\label{sec:Termination}
The destruction and reconstruction process is repeated until the makespan savings between iterations become marginal. Specifically, if there is no improvement achieved for $N_{\text{UNCH}}$ consecutive iterations, then the loop is terminated. Also, to safeguard the run-times, we restrict the total number of iterations to $N_{\text{max}}$, which is instance dependent. 

As a reflection, the art of performing this heuristic involves starting with a good feasible solution, exploring possible nested units for potential improvements, and optimizing Nested-VRP exactly on a local scale. At a higher level, the iterations between destruction and reconstruction can be viewed as a negotiation between the drone and truck routing decisions. Given a drone route, the truck can accept part of the workload for good nested units and reject the leftover workload required by bad nested units. In return, the drone will change its route with the hope that both parties are satisfied. The overall heuristic methodology is formally stated in Algorithm \ref{alg:main}.

\vspace{.1in}

\begin{algorithm}[H]
	\label{alg:main}
	\SetAlgoLined
	\KwData{Nested-VRP\;}
	\KwResult{$I^*$: the best known collection of nested units\;}
	$S$ = TSP(Problem)\;
	$I$ = CNU($S$)\;
	\While{Stopping condition is not satisfied}{
		$U, L$ = Destruction($I$)\;
		$L'$ = Nested-VRP($L$)\;
		$I'$ = Reconstruction($I\backslash U$, $L'$)\; 
		\eIf{$I'$ shows improvement}{
			$I \leftarrow I'$\;			
			Update stopping criteria\;
		}{$I \leftarrow I'$ with probability $1/2$\;Update stopping criteria\;
		}
	}
    \textbf{return} $I^* = I$
	\caption{NS Heuristic Overview}
\end{algorithm}

\section{Lower bounding method} \label{sec:lowerbound}

For an optimization problem, a lower bound is a value that is known to be less than or equal to the optimum. Generally, a lower bound is used for evaluating the quality of a solution solved by a heuristic when the optimal solution is unattainable. Ideally, a tighter lower bound gives a more-qualified guarantee of a near-optimal solution. In this section, we focus on deriving a lower bound on the mission makespan of the Nested-VRP\@. We compare the solutions obtained by the NS heuristic to the lower bound value. The tightness of the proposed lower bound will be evaluated in future work.

\begin{figure}[h!]
	\centering
	\includegraphics[width=\linewidth]{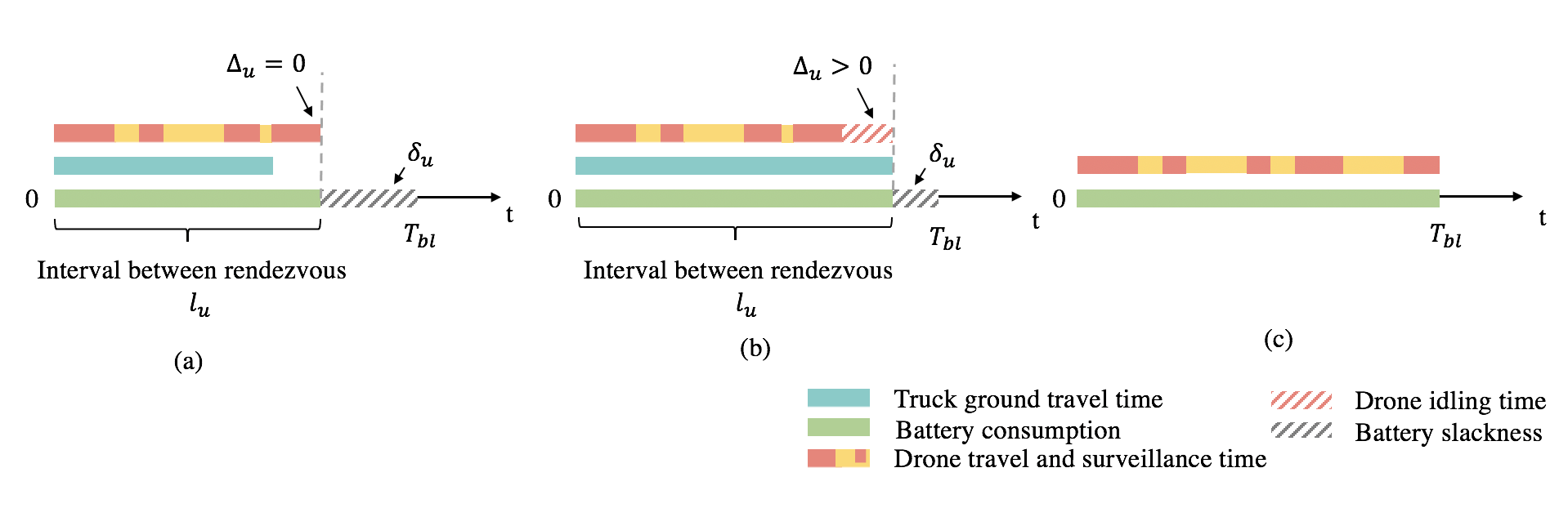}
	\caption{Battery usage in a nested unit. (a) The drone arrives later than the truck and thus the IBR $l_u$ of the nested unit is determined by the drone's surveillance time (i.e., traveling and observing). The drone's idling time $\Delta_u$ is $0$. The wasted battery energy $\delta_u$ is the difference between $T_{\text{bl}}$ and $l_u$. (b) The truck arrives later than the drone and thus the IBR $l_u$ of the nested unit is determined by the truck's traveling time. The drone arrives at the rendezvous location and idles for time $\Delta_u$. The wasted battery energy $\delta_u$ is the difference between $T_{\text{bl}}$ and $l_u$. (c) The drone battery can be swapped at anytime, anywhere. Thus, the drone's idling time $\delta_u$ and the wasted battery time $T_{\text{bl}}-l_u$ are trivial. \label{fig:batteryuse}}
\end{figure}

We start by investigating the battery usage in each of the nested units in the solution. Given a nested unit $u$, the drone surveillance time includes time spent on traveling between locations and completing observing tasks. As depicted in Figure \ref{fig:batteryuse}(a), if the truck arrives at a rendezvous location first, then once the drone arrives at the rendezvous location, the drone relinquishes all remaining battery life before it obtains a new battery. The battery slackness is denoted as $\delta_u$. However, in Figure \ref{fig:batteryuse}(b), if the drone arrives at the rendezvous location first, the drone will idle for time $\Delta_u$ while waiting for the truck. Once the truck arrives, the drone lands on the truck and releases all remaining battery life $\delta_u$. 

\vspace{.1in} 

 A lower bounding technique is based on relaxing some of the constraints in the original Nested-VRP model. First, instead of restricting the swap stops to take place at locations, the drone is allowed to replace its battery either en route from one location to the other or while observing at a location. Second, we simplify the synchronization constraint that ordinarily requires the truck to meet up the drone before the drone battery charge has expired; now we allow the drone to replace the battery itself without the truck being involved. This relaxed Nested-VRP is illustrated in Figure \ref{fig:batteryuse}(c). In this case, the drone idling time $\Delta_u$ and battery slackness $\delta_u$ are trivial. It is clear that the total number of swap stops is purely proportional to the battery consumption that occurs only during drone travel and surveillance activities.
 
Given the above relaxations, an optimistic estimate of the mission makespan consists of the following three components: (i) the minimum time spent in drone routing (i.e., the TSP route); (ii) the constant time spent in observing locations; and (iii) the smallest number of swap stops multiplied by the battery swap service time. We formally establish the lower bound on the objective value of the Nested-VRP model in Theorem \ref{thm:LB}.

 \begin{theorem}\label{thm:LB}
     Given a Nested-VRP instance described in the graph $\mathcal{G} = (\mathcal{H}, \mathcal{A})$, let $\mathcal{S}$ denote the set of arcs on the TSP route where $\mathcal{S}= \{(i,j): (i,j)\in \mathcal{A}, (i,j)\ \text{is on the TSP route}.\}$. The lower bound on the value of an optimal solution of the Nested-VRP, $LB$, can be computed by Equation (\ref{eq:lb}). (See Appendix \ref{app:proof_lb} for the detailed proof.)
 \end{theorem}
   \begin{equation}
   \text{LB}=\sum_{(i,j) \in \mathcal{S}} \tau_{ij}^D + \sum_{k \in \mathcal{H}} o_k + \left\lfloor \frac{1}{T_{\text{bl}}} \Big( \sum_{(i,j) \in \mathcal{S}} \tau_{ij}^D  + \sum_{k \in \mathcal{H}} o_k \Big) \right \rfloor T_{\text{s}}  \label{eq:lb}  
 \end{equation}
 
 \begin{algorithm}
	\SetAlgoLined
	\KwData{$G=(H, A)$: Nested-VRP instance with locations in set $H$ and arc set $A$\;
	$\tau^D/o/T_s/T_{bl}$: Drone travel time/Observation time/Battery service time/Battery lifetime\;
	}
	\KwResult{LB: the lower bounding value of the mission makespan of the Nested-VRP instance \;}
	$S \leftarrow $ TSP(H) \tcp*[l]{Generate the shortest drone route to visit all location in $H$} 
	$t_1=0$\;
	\For{each arc (i,j) on route $S$}{
	 $t_1 = t_1 + \tau_{ij}^D$\;  
	}
	$t_2=0$\;
	\For{each location $i$}{
	 $t_2 = t_2 + o_i$\;  
	}
	Return $t_1 + t_2+ \left \lfloor \frac{1}{T_{\text{bl}}} \Big( t_1  + t_2 \Big) \right\rfloor T_{\text{s}}$ 
	\caption{Compute the lower bound of Nested-VRP mission makespan}
\end{algorithm}

\section{Computational experiments}\label{sec:computation}

\subsection{Experimental setup}

In this section, we conduct a series of experiments to investigate the performance of two different approaches: the exact approaches, such as MIQCP and its variants, and the NS heuristic. Our goals are three-fold: (i) From a modeling perspective, we will examine and compare the strength of Nested-VRP model and its variants that apply linearization and constraint strengthening techniques;  (ii) From an algorithm design standpoint, we will demonstrate that the proposed heuristic is adequate to support the drone-truck surveillance mission; (iii) From an operational standpoint, we will further examine how to achieve the most-economic solution by carefully analyzing the model parameters.

All experiments are performed on a set of benchmark instances from \cite{agatz2018optimization} where the authors randomly generate locations on a 2D plane following different patterns, which are labeled: uniform, single-center, and double-center. The uniform pattern consists of locations whose $x$ and $y$ coordinates are uniformly sampled from $\{0, \ldots, 100 \}$ independently. The single-center represents a circular city in which locations are closer to the center with higher probability. The double-center pattern mimics a city with two centers that are 200 distance units away from each other. Around each center, locations are distributed in the same way as the single-center pattern. The benchmark dataset is naturally split into small cases and large cases according to the number of locations, $N$. The ``small'' set considers possible location numbers $\{5, 6, 7, 8, 9, 10\}$ while the ``large'' set considers numbers in the pool of $\{20, 50, 75, 100, 175, 250\}$. By default, the drone's cruising speed is 1 unit distance per unit time. By varying the truck speed among $\{1, 0.5, 0.3333\}$, while keeping the drone speed as 1, we achieve the speed ratios $\alpha = \{1,2,3\}$ between the two vehicles. To give practical sense to the data, we treat 1 unit distance as 100 meters and constant drone cruising speed 1 unt distance per unit time as 30 meters per second. For clarity, a scenario is referred to as a subset of data sharing the same $(\text{pattern}, N, \alpha)$ characteristics. One scenario includes $10$ instances. For example, the scenario $(\text{pattern=uniform}, N = 5, \alpha = 1)$ consists of $10$ instances, where each instance corresponds to a Nested-VRP in which the drone, with the same speed as the truck, observes $5$ locations that are randomly generated by following a uniform pattern.   

Additional information is needed for solving the Nested-VRP\@. First, the battery capacity is set to 900 seconds by default, which is enough for the drone to complete a round trip along the diagonal lines of the largest 2D plane across all instances from the small data set. Second, a single battery swap service takes 100 seconds. Since the benchmark data does not provide any information about the observation times associated with locations, we randomly generate the observation times at each location by drawing uniformly from $[0, 250]$ seconds. In short, given a scenario characterized as $(\text{pattern}, N, \alpha)$, one of its instances is further characterized by observation times $O$ for $N$ locations, battery capacity $T_{\text{bl}}$, battery swap service time $T_{\text{s}}$; and this is described as $(\text{pattern}, N, \alpha, O, T_{\text{bl}}, T_{\text{s}})$. All the input data features are listed in Table \ref{tab:features}.

\begin{table}[h!]
	\caption{\label{tab:features} Summary of data features \strut}
	\centering
	\ra{0.9}
	\begin{tabular}{@{}lll@{}}
		\toprule
		\textbf{Notations} & \textbf{Features} & \textbf{Values} \\
		\midrule
		$P$ & patterns & \{uniform, single-center, double-center\}\\
		$N$ & number of locations &  \{ 5, 6, 7, 8, 9, 10 \}\\
		& &  \{ 20, 50, 75, 100, 175, 250 \}  \\		
		$ T_{\text{bl}} $ & battery capacity &  900 seconds \\
		$T_{\text{s}}$  & battery swap service time & 100 seconds\\
		$ O $    &observation time  & \textit{Uniform}[0, 250] seconds \\
		$ \alpha $ & speed ratio of truck and drone & \{ 1, 2, 3 \}\\	
		\bottomrule
	\end{tabular}    
\end{table}

By varying the shapes of the location pattern, the number of locations, and the speed ratios of the two vehicles, a total of $6\times3\times3\times10$ computational experiments were conducted on a computer with an $\text{Intel}^{\text{\textregistered}}$ $\text{Xeon}^{\text{\textregistered}}$ CPU E5-2687W v4 3.00 GHz processor and 64.0 GB installed RAM\@. All the algorithms and models are coded in Python 3.7.8. The MIQCP, MILP, MILP+DL, MILP+SD models are solved via Gurobi 9.1.2. We limit Gurobi run-times to 15 minutes.  

\subsection{Formulation comparison results} \label{subsec:F-Comp}
    The first set of experiments aims at empirically assessing the impact of pure linearization process on the model performance and examine whether the linearized Nested-VRP model with the original MTZ subtour elimination constraints being strengthened per Proposition \ref{prop:dl} or Proposition \ref{prop:sd} would exhibit improvements in terms of runtime, number of nodes explored to achieve optimality, and tightness of the lower bound at the root node. Therefore, we solve data instances parameterized in the pattern = \{uniform, singlecenter, doublecenter\}, $N = 8$, $\alpha$ = \{1, 2, 3\} using the MIQCP, MILP, MILP+DL, and MILP+SD formulations presented in Table \ref{tab:compareF}, respectively. When solving each one of the mentioned formulations, Table \ref{tab:comparePerf} presents the average runtime $T_{\text{sol}}$ in seconds, the average number of nodes explored to achieve optimality $N_{\text{node}}$, and the average gap at the root node $\gamma_0$ in percentage. Moreover, we compare the relative performance of the MILP, MILP+DL, MILP+SD models to that of the MIQCP model in the second row of the table (i.e., changes with respect to the MIQCP model). 
    
    This study reveals that the pure linearization process is able to reduce runtime by 35.67 seconds on average --- which is equivalent to approximately a 25.16\% runtime reduction with respect to the MIQCP model. This observation can be further explained by the fact that when the Gurobi Optimizer solves the MILP model, the Optimizer explores on average $451195$ fewer nodes in the branch-and-bound search process as compared to solving the MIQCP model. Compared to the MILP model, MILP+DL makes slightly further progress in both reducing the runtime and the number of nodes explored. Most fortuitously, the MILP+SD outperforms all previous models by offering 92.29 seconds runtime reduction, which is 64.88\% less than that needed by the MIQCP model; and by requiring 639639 fewer node explorations, which is 84.15\% less than that required by the MIQCP model. However, it is somewhat surprising that all MILP, MILP+DL, and MILP+SD models exhibit slight but not obvious improvements in tightening up the lower bound produced by the LP relaxation at the root node. These computational results clearly demonstrate that the enhancement of the Nested-VRP model by linearization and constraints tightening techniques improves the model performance and reduces runtime to produce provably optimal solutions. In addition, the observed speedup is not caused by tightening the polyhedral representation of the original Nested-VRP model at the root node. To further enhance the Nested-VRP model, future research could shift the focus to identifying effective cutting planes according to the data structure of the model.

Given the superior performance of the MILP+SD model, we are interested in extending the computational experiments and further assessing and comparing the MILP+SD's capability in terms of closing gap during optimizing process with respect to the MIQCP model in the following sections.

\begin{table}[h!] 
\caption{Performance attained by employing MIQCP, MILP, MILP+DL, and MILP+SD in solving Nested-VRP\@.   \label{tab:comparePerf}}
\resizebox{\textwidth}{!}{%
\begin{tabular}{lcrrrcrrrcrrrcrrl}
\toprule
        & \multicolumn{3}{c}{MIQCP} & \multicolumn{1}{l}{} &\multicolumn{3}{c}{MILP}  & \multicolumn{1}{l}{} & \multicolumn{3}{c}{MILP+DL}  & \multicolumn{1}{l}{} & \multicolumn{3}{c}{MILP+SD}                                            &  \\ \cline{2-4} \cline{6-8} \cline{10-12} \cline{14-16}
        & $T_{\text{sol}}$ (seconds) & \multicolumn{1}{r}{$N_{\text{node}}$} & \multicolumn{1}{r}{$\gamma_0$ (\%) } &\multicolumn{1}{r}{}
        &$T_{\text{sol}}$ (seconds) & \multicolumn{1}{r}{$N_{\text{node}}$} & \multicolumn{1}{r}{$\gamma_0$ (\%) } & \multicolumn{1}{r}{} 
        &$T_{\text{sol}}$ (seconds)  & \multicolumn{1}{r}{$N_{\text{node}}$} & \multicolumn{1}{r}{$\gamma_0$ (\%) } & \multicolumn{1}{r}{} 
        & $T_{\text{sol}}$ (seconds) & \multicolumn{1}{r}{$N_{\text{node}}$} & \multicolumn{1}{r}{$\gamma_0$ (\%) } &  \\
Mean & \multicolumn{1}{r}{141.79} & 760138.10   & 119.82  &  & \multicolumn{1}{r}{106.13} & 308942.80 & 118.89  &   & \multicolumn{1}{r}{99.75} & 296040  & 118.70 & & \multicolumn{1}{r}{49.50} & 120498.60  & 118.71  &  \\
$\text{Change wrt MIQCP} $    & \multicolumn{1}{r}{}    - &       -  &        - &                   
        & \multicolumn{1}{r}{-35.67} & -451195   & -0.93 &           
        & \multicolumn{1}{r}{-42.04} & -464098   & -1.12  &
        & \multicolumn{1}{r}{-92.29} & -639639   & -1.12   &  \\ 
        \bottomrule
\end{tabular}
}
\end{table} 

\subsection{Small data set results} \label{subsec:small}

\begin{figure}[h!]
	\resizebox{\textwidth}{!} {
		\begin{tabular}{ccc}
			\includegraphics[width=100mm]{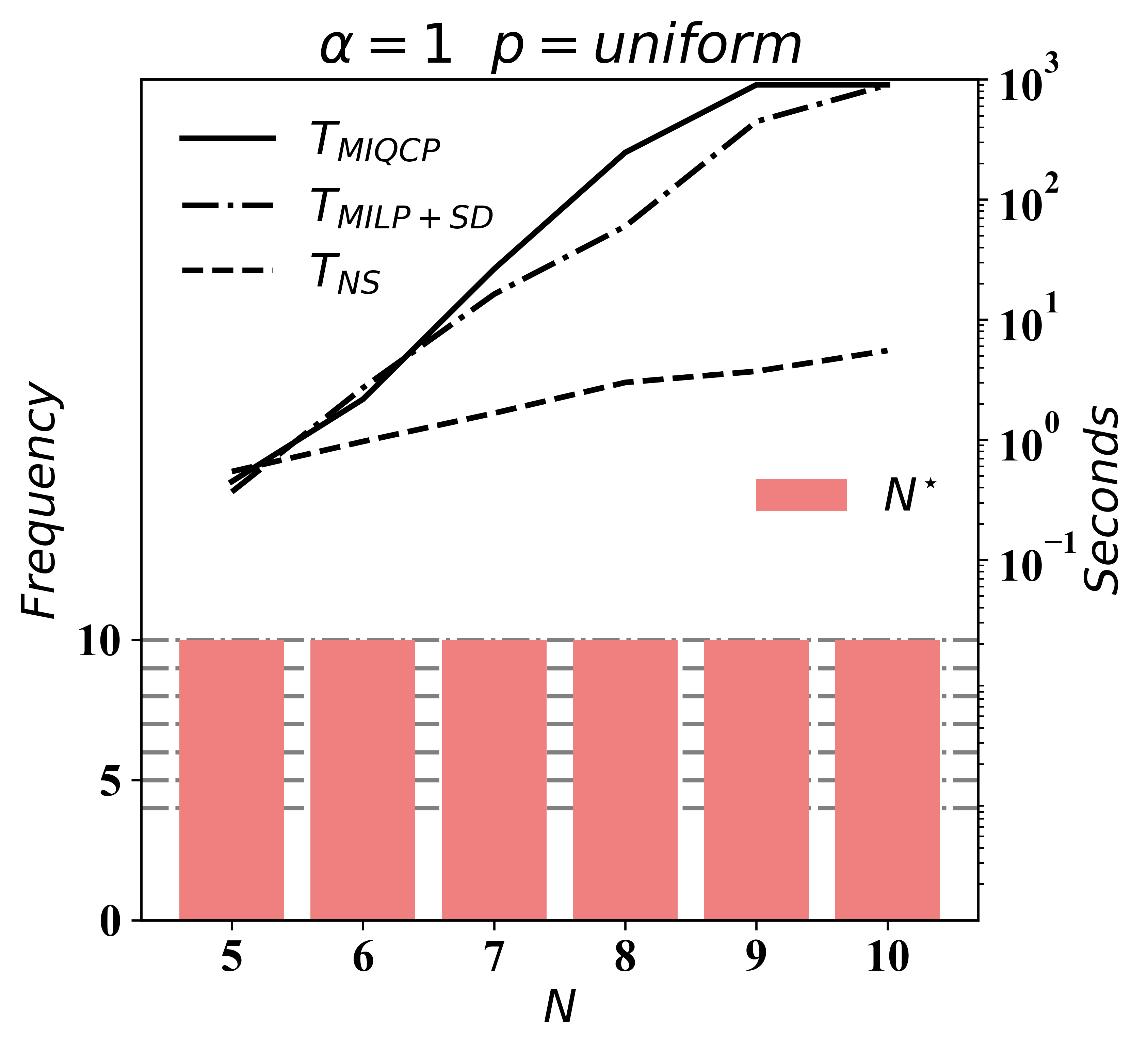} &   \includegraphics[width=100mm]{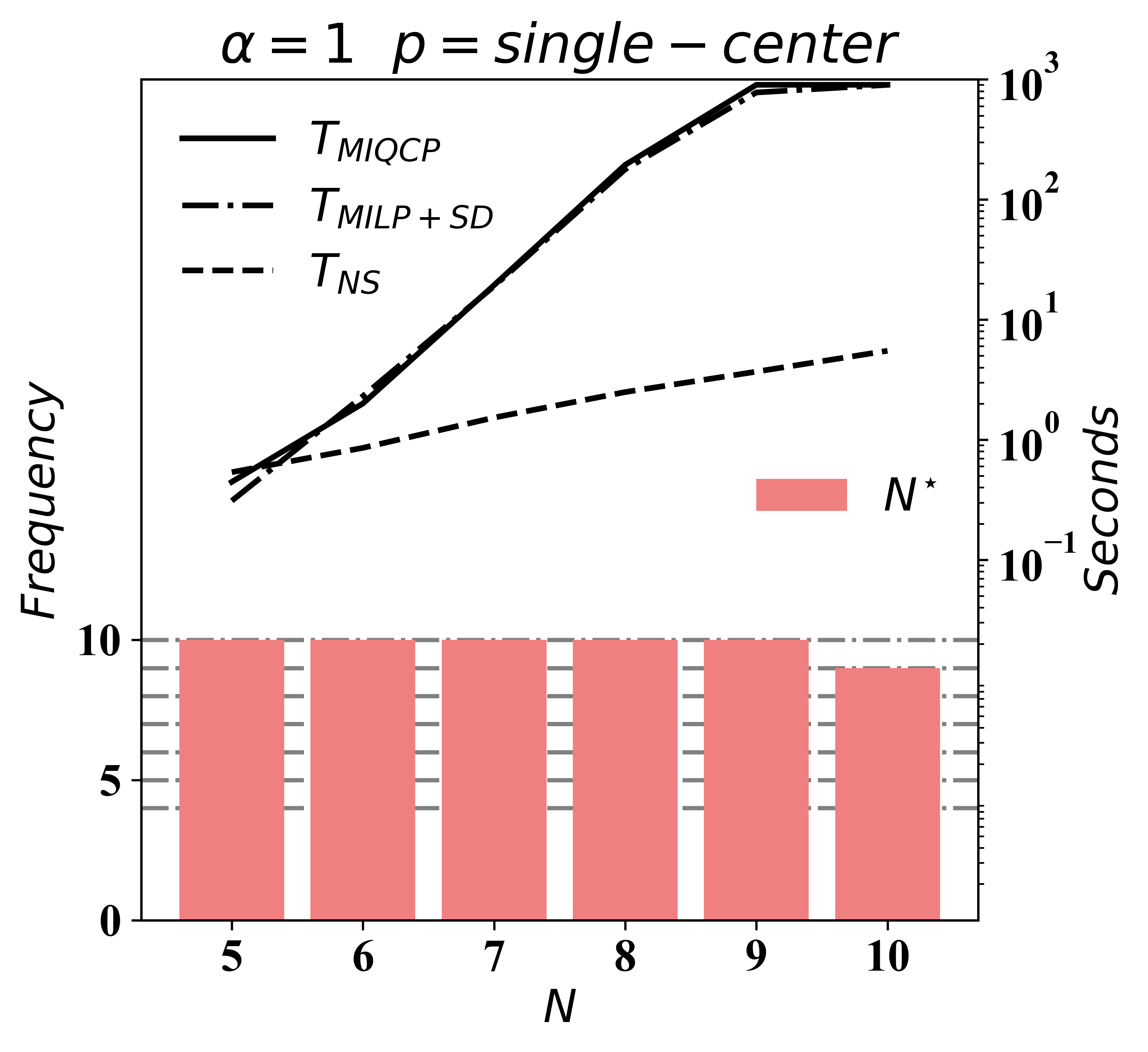} & \includegraphics[width=100mm]{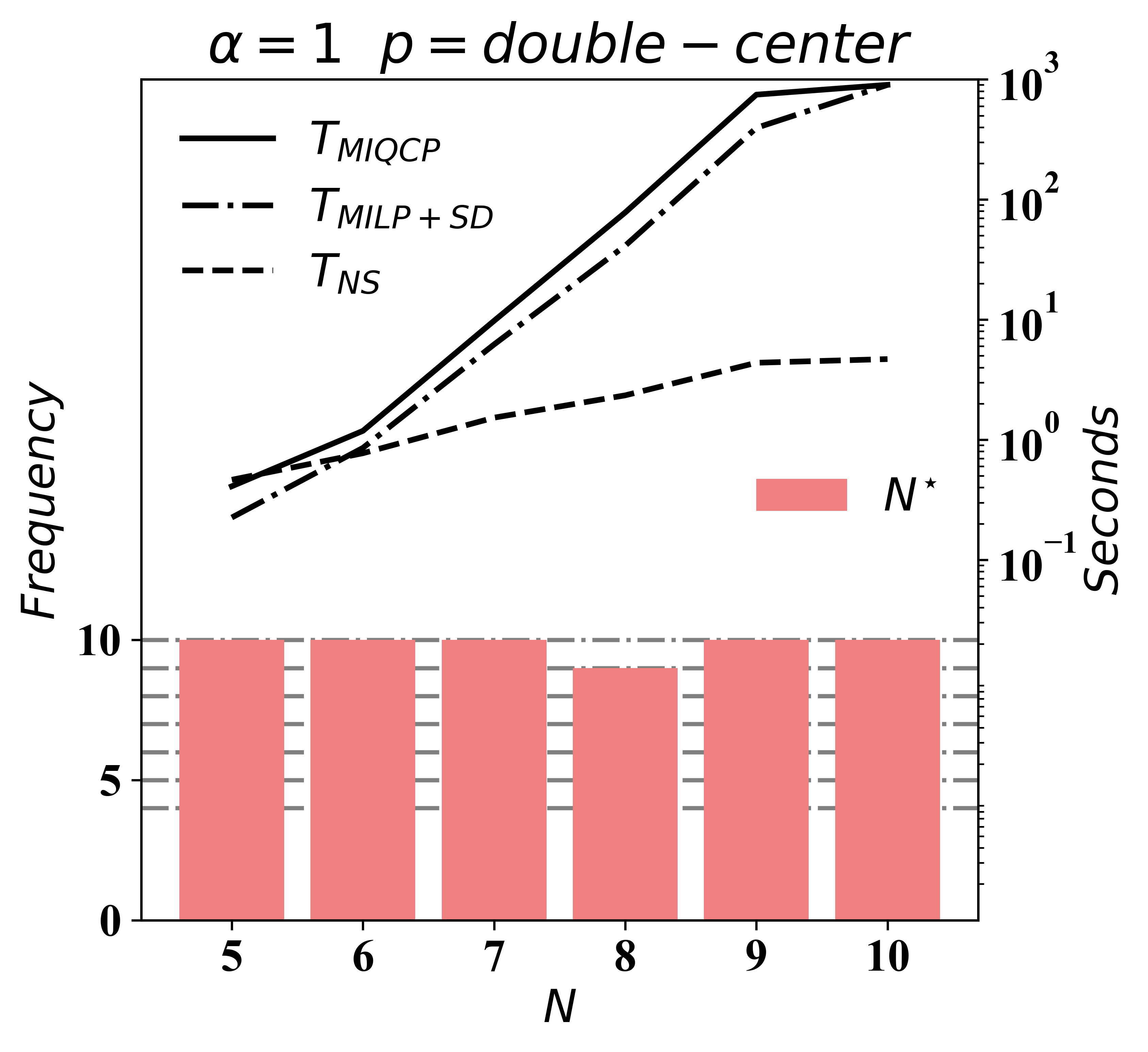}\\
			\includegraphics[width=100mm]{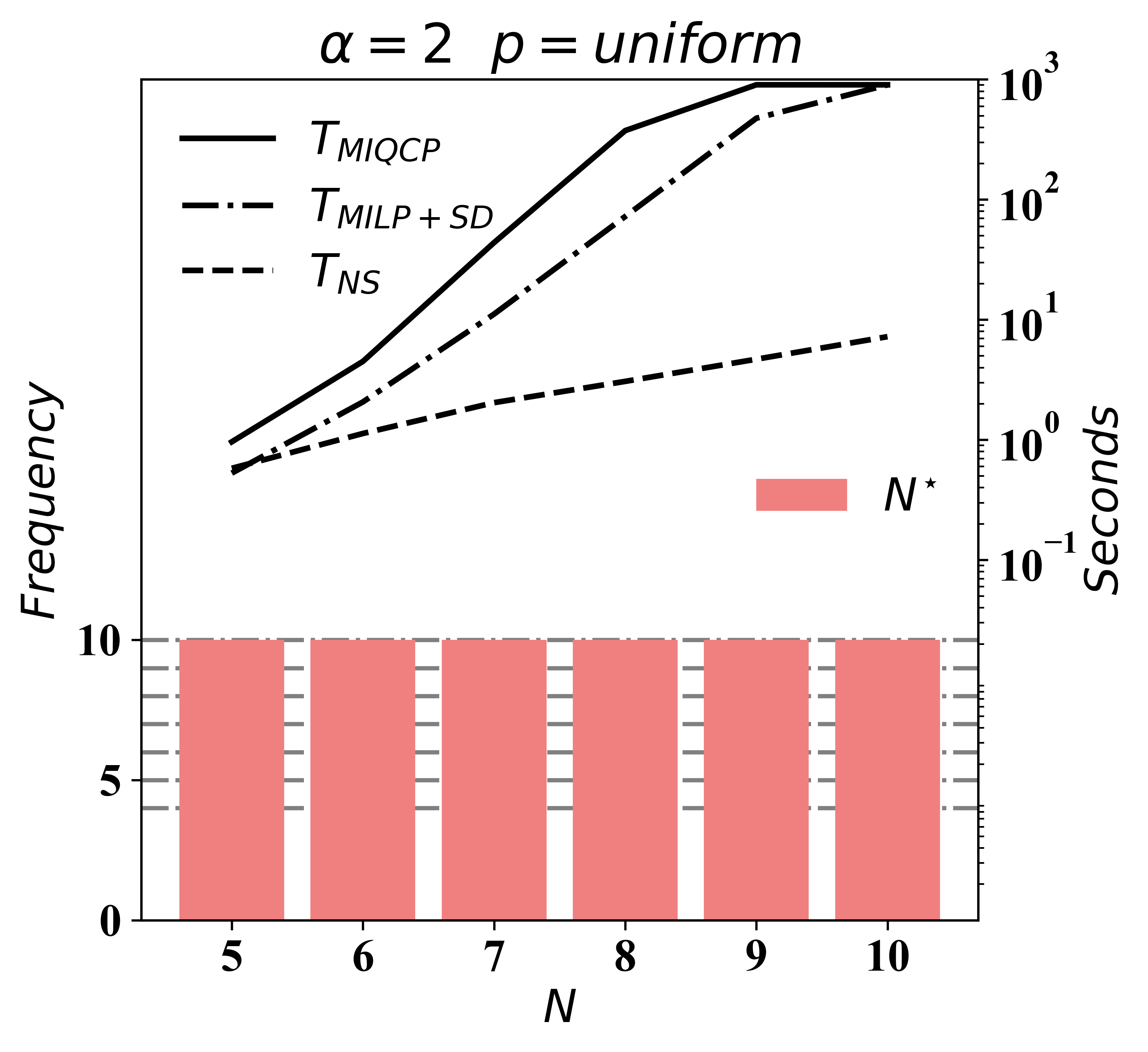} &   \includegraphics[width=100mm]{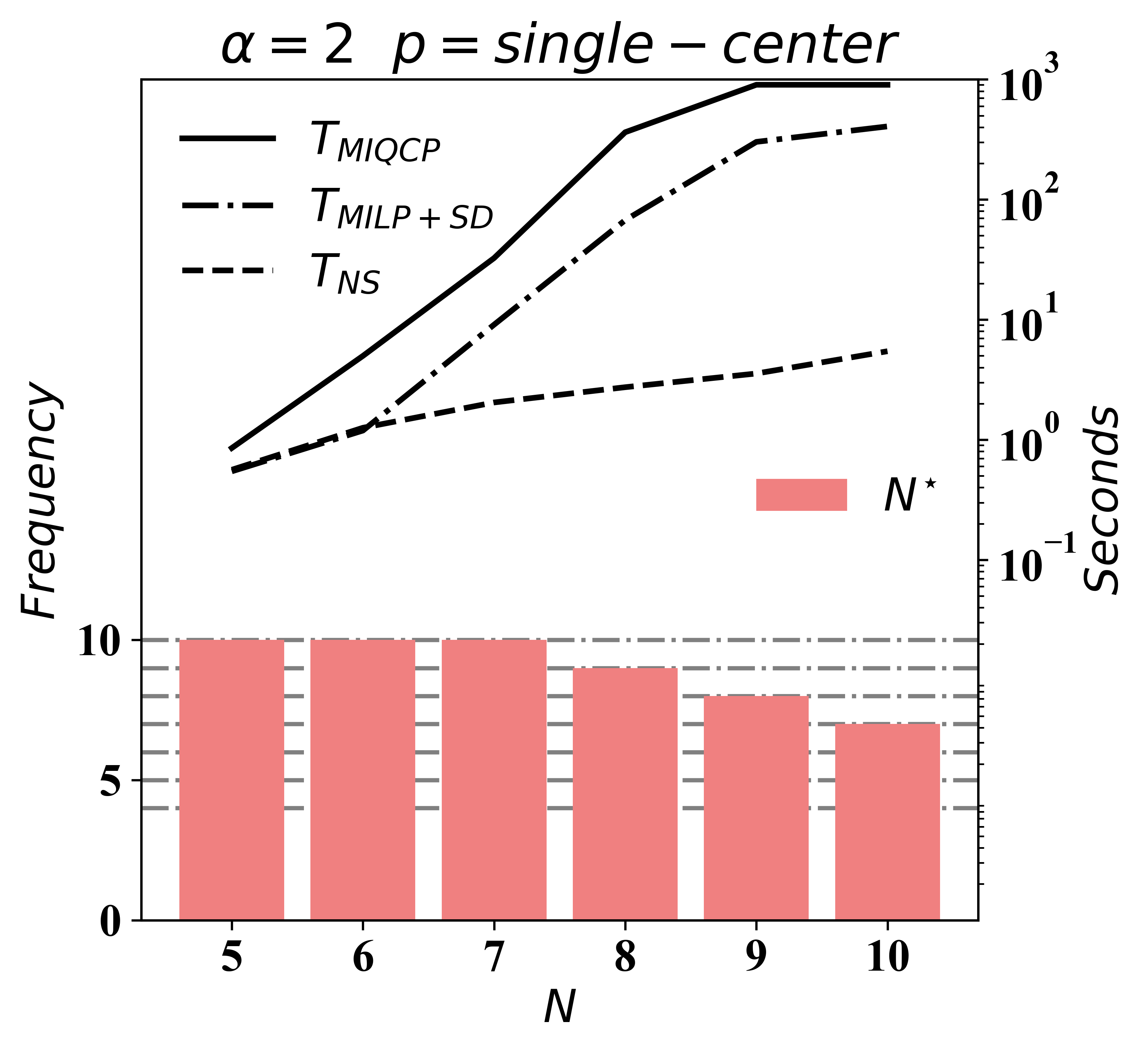} & \includegraphics[width=100mm]{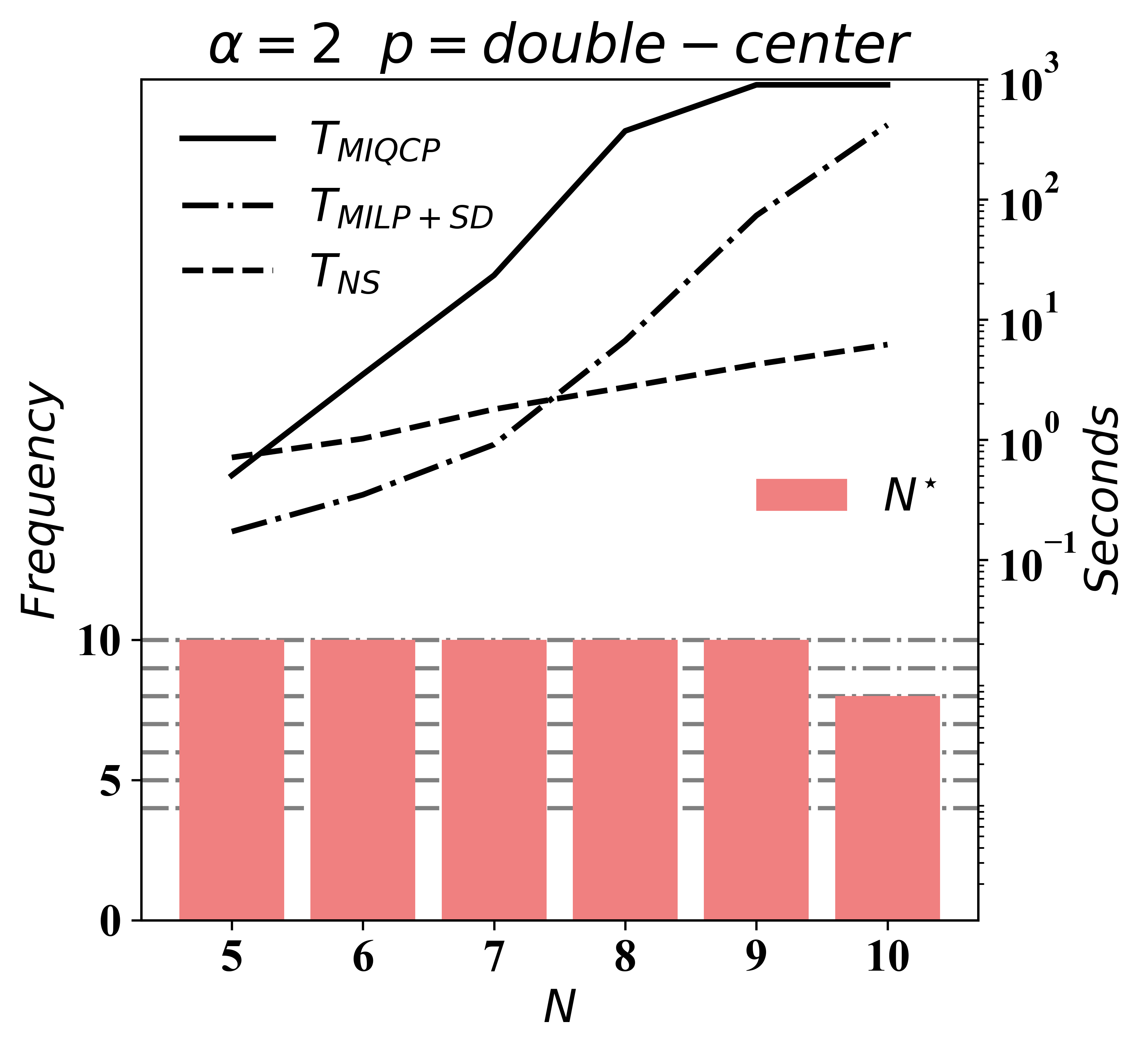} \\
			\includegraphics[width=100mm]{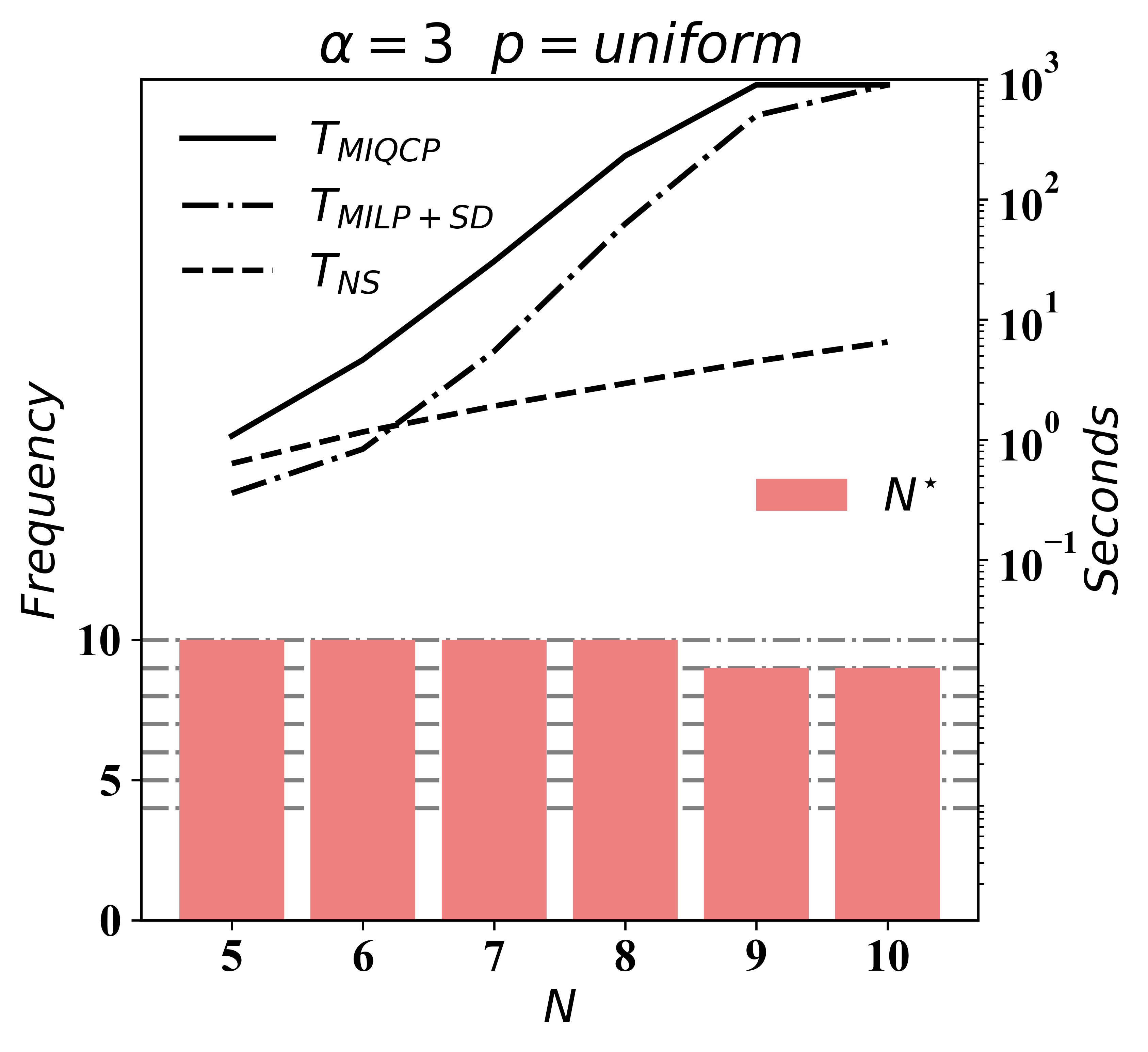} &   \includegraphics[width=100mm]{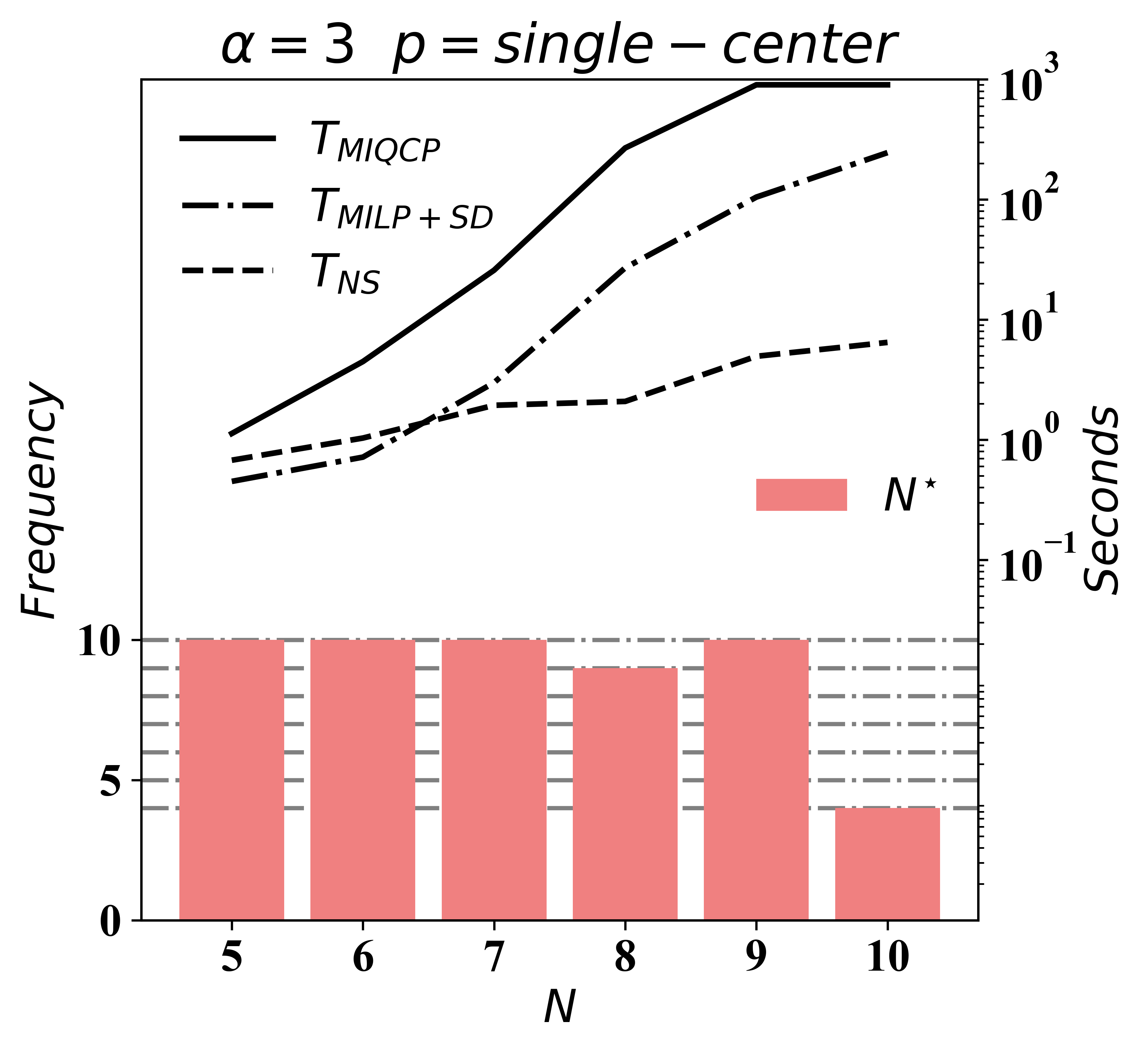} & \includegraphics[width=100mm]{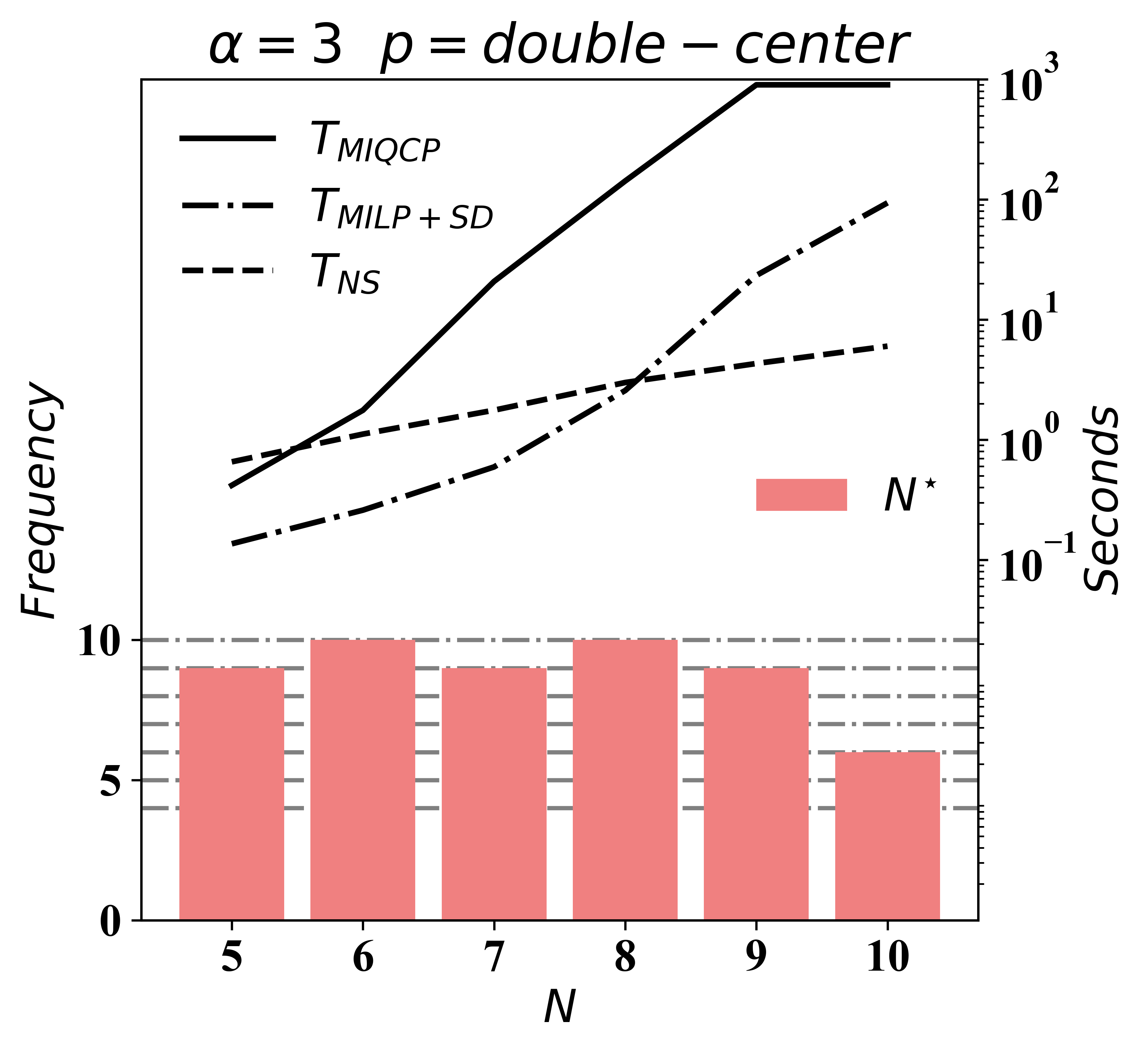} \\
		\end{tabular}
	}
    	\caption{Frequency of optimal solutions found and comparison of run-times of the MIQCP and MILP+SD exact approaches as well as the NS heuristic. In a particular row, we consider the uniform, single-center, and double-center pattern shapes. In the vertical direction, we arrange the figures by increasing the speed ratio of the two vehicles from $\alpha = 1$ to $\alpha = 3$. In each subfigure, a red column corresponds to a scenario characterized by $(\text{pattern}, N, \alpha)$. Each scenario is based on the $10$ instances provided by the benchmark dataset. Recall that for each scenario, $N^\star$ counts the total number of instances where the NS heuristic finds a better solution than the exact approach. For example, the NS heuristic does better in 9 out of 10 instances in scenario $(\text{single-center}, N=10, \alpha = 1)$. In addition, the solid (dashed) line depicts the trend of computational time growth of the MIP approach (NS heuristic) as $N$ increases. \label{fig:small}}
\end{figure}

In this section, we solve in three ways the Nested-VRP for each instance contained in the small data set via the MIQCP model, the MILP+SD model, and the NS heuristic. In Figure \ref{fig:small}, we present the computational results obtained from solving instances belonging to uniform, single-center, and double-center scenarios. Detailed statistics are documented in Tables \ref{tab:s_uniform}, \ref{tab:s_singlecenter}, \ref{tab:s_doublecenter}, respectively, in Appendix \ref{app:small_data}. Within each table, we report the results in three subgroups by differentiating $\alpha = \{ 1, 2, 3 \}$. In particular, when solving the MIQCP and MILP+SD models, we record the average optimal mission makespan $C_{\text{MIQCP}}/C_{\text{MILP+SD}}$ for instances belonging to the same scenario. Additionally, we track the optimality gap $\gamma_{\text{MIQCP}}/\gamma_{\text{MILP+SD}}$ and runtime $T_{\text{MIQCP}}/T_{\text{MILP+SD}}$ reported from the Gurobi Optimizer. In the following experiments, the NS heuristic is parameterized by setting $\beta = 0.25$, $N_{\text{UNCH}} = 5$, and $N_{\text{max}} = 20$ as the termination criteria. Plus, the reconstruction process of the NS heuristic employs the MILP+SD model whose superior performance has been demonstrated in Section \S\ref{subsec:F-Comp}. In Figure \ref{fig:small}, we compare the performance of the NS heuristic to the exact approaches by looking at the total number of instances, $N^\star$, where the NS heuristic achieves a solution with lower mission makespan as compared to the best solution provided by the Gurobi Optimizer from solving either the MIQCP or MILP+SD models, for each scenario. The average run-times $T_{\text{NS}}$ are recorded as well.

First, we evaluate the performance of the MIQCP and MILP+SD exact approaches. As a highlight, the MILP+SD model outperforms the MIQCP model in that it can provably solve 474/540 instances to optimality, whereas the MIQCP can only certify the optimality for 363/540 instances given the 15-minute cutoff time. In particular, both exact approaches achieve a $0$ gap for all data instances with no more than 8 locations. When the number of locations $n$ is increased to 9, the MIQCP model fails to produce optimal solutions for any instance and achieves an average optimality gap at 34.03\%. As a contrast, the MILP+SD model is able to aggressively close the gap to 4.30\% on average. Moreover, when $N = 10$, the MIQCP model produces solutions with an average optimality gap of 53.73\%  which is more than double the gap obtained by the MILP+SD model, that is 21.87\%. The loss of optimality due to the increased complexity of the problem has also been observed by \cite{gonzalez2020truck}, who solved an MIP model sharing similar features. These observations align with those found in \S\ref{subsec:F-Comp} and further confirm that the MILP+SD model tends to solve small-sized instances to optimality significantly faster than the original MIQCP model by leveraging the linearization scheme and RLT\@. More precisely, the MILP+SD model achieves aggressive speedups by drastically reducing the number of nodes explored during the branch-and-bound process.

Regarding the computation efficiency of the proposed exact approaches, we further compare our original MIQCP model without performing the linearization and constraints strengthening scheme to the TDTL model introduced by \cite{gonzalez2020truck}. Recall that the Nested-VRP considers a more-complex operational scenario where each location associates with a non-zero observation time.  In addition, the truck and drone are allowed to travel simultaneously from one location to another and complete battery swaps in transit. We observe that the MIQCP model is able to solve data instances (single-center, $N = 10$, $\alpha=3$) and obtains a solution within a $62.6\%$ average optimality gap in around 900 seconds. However, the TDTL model provides a sub-optimal solution with a $345.2\%$ optimality gap after spending 1322.2 seconds in solving the same data instances. The same reasoning can be applied to data instances (pattern = \{single-center, double-center\}, $N = 10$, $\alpha$ = \{1, 2, 3\}). The observed superior performance of the MIQCP model is a byproduct of Theorem 1. From the above observations, we can conclude that the proposed MIQCP model as well as its improved variants have good performance when the problem size is relatively small.

Second, we investigate the relationship between vehicle speed ratios $\alpha$ and the mission makespan $C_{\text{MIP}}$. The purpose is to provide practical guidelines to pair a drone and a truck, with different speed settings, to achieve more operational efficiency. We do so by comparing the average percentage time savings (APTS) due to the change of vehicle speed ratio from $\alpha_1$ to $\alpha_2$ for a specific geometrical pattern $p$ of interest. Given the pattern of interest $p \in \{\text{uniform, single-center, double-center}\}$, vehicle speed ratios of interest $\alpha_1, \alpha_2 \in \{1, 2, 3\}$, and possible number of locations $n \in \{5, 6, 7, 8, 9, 10 \}$, let $C_{\text{MILP+SD}}(p, n, \alpha_{k}), k \in \{1, 2\}$ denote the average mission makespan of scenario $(\text{pattern} = p, N = n, \alpha = \alpha_k)$. Then, we can define $\text{APTS}(p,\alpha_1, \alpha_2)$ as follows. 

\begin{equation}
\text{APTS}(p, \alpha_1, \alpha_2) \; \equiv \; 
\sum_{n \in \{5, 6, 7, 8, 9, 10\}} \frac{C_{\text{MILP+SD}}(p, n, \alpha_2) - C_{\text{MILP+SD}}(p, n, \alpha_1)}{6\,C_{\text{MILP+SD}}(p, n, \alpha_1)} \nonumber
\end{equation}

A more-negative $\text{APTS}(p,\alpha_1, \alpha_2)$ indicates that more time savings can be achieved by changing the speed ratio from $\alpha_1$ to $\alpha_2$ while the pattern $p$ remains the same. Take the set of data from the uniform pattern as an example, in which case $\text{APTS}(\text{uniform}, 1, 2) = -13.32
\%$ and $\text{APTS}(\text{uniform}, 1, 3) = -14.96
\%$. In other words, increasing the vehicle speed ratio from $1$ to $2$ only provides additional time savings by $1.57\%$. However, in the double-center scenario, increasing the vehicle speed ratio from $1$ to $2$ further reduces the Nested-VRP mission makespan by $|(-30.91\%) - (-19.80\%)| = 11.17\%$. This observation has significant implications for how to invest in a truck-drone team to survey the local region. A drone model with a higher maximum cruising speed is typically more expensive than a standard one. As a result, understanding how the observation tasks are distributed geometrically in the local region and what is the appropriate, but not necessarily maximum, speed ratio between the two vehicles can achieve satisfactory operation efficiency within budget.

Third, with regard to the performance of the NS heuristic, we observe that the NS heuristic is able to produce better solutions compared to the exact approach for \textbf{514} out of \textbf{540} total small data instances. This is supported by the evidence in Figure \ref{fig:small}. A closer look at instances where the NS heuristic fails to compete over the exact technique reveals that the NS heuristic has difficulty in searching for the optimal Nested-VRP solution when dealing with complex geometry, double-center pattern. However, there is no conclusive results on how the total number of locations and the vehicle speed ratio affect the performance of the proposed NS heuristic.

In terms of computation efficiency, the NS heuristic can obtain Nested-VRP solutions of high quality with significantly less computation time than the exact technique with a 15-minute cutoff time for 514 out of 540 small data instances. In particular, the run times for the NS heuristic to solve a small data instance is on average 3.04 seconds with a standard deviation of 1.90 seconds. In the next section, we further examine how the NS heuristic performs when dealing with problems of larger size. 

\subsection{Large data set results} \label{subsec:large}

For solving larger-scale problems, we first perform the MIQCP and the MILP+SD exact approaches via Gurobi Optimizer with a 15-minute cutoff time. In particular, we warm start the exact approach using the initial feasible Nested-VRP solution provided at the end of initialization phase of the NS heuristic as discussed in \S \ref{Initialization}. However, these two exact approaches fail to provide satisfactory solutions. Therefore, we apply the NS heuristic to solve Nested-VRP involving large data sets. In the following experiments, the heuristic is parameterized by setting $\beta = 0.25$, $N_{\text{UNCH}} = 5$ and $N_{\text{max}} = 50$ as the termination criteria. By default, the reconstruction process of the NS heuristic employs the MILP+SD formulation. To evaluate the quality of the solutions, we leverage the lower bound value introduced in \S\ref{sec:lowerbound}.  

Recall that the large data set considers locations of sizes $\{20, 50, 75, 100, 175, 250\}$ from uniform, single-center, and double-center geometrical patterns. Additionally, the speed ratio between the drone and the truck varies from $\{1, 2, 3\}$. Each scenario, characterized by $(\text{pattern}, N, \alpha )$, includes 10 instances.

Tables \ref{tab:l_uniform}, \ref{tab:l_singlecenter}, and \ref{tab:l_doublecenter} provide a comparison of the computational performance resulting from the application of various approaches on data from uniform, single-center, double-center patterns, respectively. Within a table, each row summarizes the statistics of interest per scenario. For each scenario, the statistics considered are: 
\begin{itemize}
\item The average mission makespan of the solutions obtained via Gurobi with warm-start (15-minute cutoff time) using the MIQCP model $C_{\text{15mins}}^{\text{MIQCP}}$ or the MILP+SD model $C_{\text{15mins}}^{\text{MILP+SD}}$; via solving the CNU problem $C_{\text{CNU}}$ (initialization of the NS Heuristic); and the NS heuristic $C_{\text{NS}}$.

\item The estimate of the lower bound on mission makespan $C_{\text{lb}}$.

\item The relative optimality gaps $\gamma_{\text{\text{15mins}}}^{\text{MIQCP}} $, $\gamma_{\text{\text{15mins}}}^{\text{MILP+SD}} $, $\gamma_{\text{CNU}} $, $\gamma_{\text{NS}}$ with respect to the estimate of the lower bound.

\item For the NS heuristic, we also report the average run-times $T_{\text{NS}}$, number of iterations $\#_{\text{iter}}$, and number of recharge stops $N_{\text{s}}$. 
\end{itemize}

\begin{table}[h!]
	\centering
	\caption{Results from solving instances from a uniform pattern in the large data set.}
	\ra{1.1}
	\resizebox{\textwidth}{!} {
		\begin{tabular}{@{}rrrrrrrrrrrrrrrrrr@{}}
			\toprule
			\textbf{uniform} && \textbf{MIQCP} && \textbf{MILP+SD} &&  \multicolumn{5}{c}{\textbf{NS Heuristic}} && \textbf{LB}  && \multicolumn{4}{c}{\textbf{Gap Comparison (\%)}}\\
			\cmidrule{3-3} \cmidrule{5-5}  \cmidrule{7-11} \cmidrule{13-13} \cmidrule{15-18}   
			$N$ && $C_{\text{15mins}}^{\text{MIQCP}}$  && $C_{\text{15mins}}^{\text{MILP+SD}}$ && $T_{\text{NS}}\ (s)$ & $\#_{\text{iter}}$ & $N_{\text{s}}$ & $C_{\text{CNU}}$ & $C_{\text{NS}}$ && $C_{\text{lb}}$ && $\gamma_{\text{CNU}} $ & $\gamma_{\text{\text{15mins}}}^{\text{MIQCP}} $  &$\gamma_{\text{\text{15mins}}}^{\text{MILP+SD}} $& $\gamma_{\text{NS}} $   \vspace{1 pt}\\  
			\hline
			$\alpha =1$\\
			\hline
20  &  & 5630.7  &  & 5630.7  &  & 4.9   & 3.8  & 17.1  & 5630.7  & 5242.7  &  & 5029.5  &  & 12.05 & 12.05 & 12.05 & 4.18 \\
50  &  & 13505.8 &  & 13505.8 &  & 13.3  & 6.2  & 30.3  & 13505.8 & 12976.1 &  & 12637.2 &  & 6.88  & 6.88  & 6.88  & 2.66 \\
75  &  & 20597.7 &  & 20597.7 &  & 25.2  & 10.6 & 50.2  & 20597.7 & 20040.7 &  & 18995.5 &  & 8.43  & 8.43  & 8.43  & 5.50 \\
100 &  & 26979.9 &  & 26979.9 &  & 40.1  & 14.8 & 64.7  & 26979.9 & 26430.4 &  & 24841.1 &  & 8.63  & 8.63  & 8.63  & 6.42 \\
175 &  & 46891.6 &  & 46891.6 &  & 99.8  & 24.0   & 104.6 & 46891.6 & 46249.2 &  & 43882.2 &  & 6.87  & 6.87  & 6.87  & 5.40 \\
250 &  & 66213   &  & 66213   &  & 186.9 & 34.1 & 143.7 & 66213.0 & 65514.7 &  & 62437.5 &  & 6.06  & 6.06  & 6.06  & 4.94 \\
			\hline
			$\alpha =2$\\
			\hline
20  &  & 5265.7  &  & 5265.7  &  & 5.3   & 2.9  & 11.9  & 5265.7  & 5094.0  &  & 4953.8  &  & 6.42  & 6.42  & 6.42  & 2.91 \\
50  &  & 13123.6 &  & 13123.6 &  & 17.4  & 6.7  & 27.4  & 13123.6 & 12557.4 &  & 12399.0 &  & 5.94  & 5.94  & 5.94  & 1.33 \\
75  &  & 19915   &  & 19915   &  & 27.8  & 9.7  & 44.1  & 19915.0 & 19254.2 &  & 18885.7 &  & 5.45  & 5.45  & 5.45  & 1.95 \\
100 &  & 26300.2 &  & 26300.2 &  & 47.8  & 11.4 & 57.2  & 26300.2 & 25618.8 &  & 24957.4 &  & 5.38  & 5.38  & 5.38  & 2.65 \\
175 &  & 45664.3 &  & 45664.3 &  & 107.1 & 22.7 & 92.4  & 45664.3 & 44951.3 &  & 43731.9 &  & 4.42  & 4.42  & 4.42  & 2.79 \\
250 &  & 64342.6 &  & 64342.6 &  & 204.1 & 34.3 & 134.3 & 64342.6 & 63605.2 &  & 61824.4 &  & 4.08  & 4.08  & 4.08  & 2.88 \\
			\hline
			$\alpha =3$\\
			\hline
20  &  & 5194.5  &  & 5194.5  &  & 12.5  & 3.3  & 9.5   & 5194.5  & 5036.0  &  & 4885.1  &  & 6.29  & 6.29  & 6.29  & 3.13 \\
50  &  & 13427.2 &  & 13427.2 &  & 22.1  & 6.8  & 29.7  & 13427.2 & 12814.7 &  & 12680.9 &  & 5.93  & 5.93  & 5.93  & 1.05 \\
75  &  & 18847.2 &  & 18847.2 &  & 42.5  & 10.9 & 41.8  & 18847.2 & 18213.5 &  & 18001.5 &  & 4.71  & 4.71  & 4.71  & 1.17 \\
100 &  & 26282.7 &  & 26282.7 &  & 51.6  & 14.3 & 57.6  & 26282.7 & 25589.0 &  & 25186.9 &  & 4.36  & 4.36  & 4.36  & 1.60 \\
175 &  & 45199.6 &  & 45199.6 &  & 140.3 & 27.6 & 94.8  & 45199.6 & 44466.9 &  & 43584.7 &  & 3.71  & 3.71  & 3.71  & 2.02 \\
250 &  & 64445.2 &  & 64445.2 &  & 229.9 & 35.7 & 136.7 & 64445.2 & 63678.9 &  & 62349.8 &  & 3.36  & 3.36  & 3.36  & 2.13\\
\hline
 Summary &&\multicolumn{10}{l}{The NS heuristic solves 146/180 instances to within 5\% of the $C_{\text{lb}}$}\\
  && \multicolumn{10}{l}{The NS heuristic solves 180/180 instances to within 10\% of the $C_{\text{lb}}$}  \\
\bottomrule
		\end{tabular}
	}
\label{tab:l_uniform}
\end{table}

\begin{table}[h!]
	\centering
	\caption{Results from solving instances from a single-center pattern in the large data set.}
	\ra{1.1}
	\resizebox{\textwidth}{!} {
		\begin{tabular}{@{}rrrrrrrrrrrrrrrrrr@{}}
			\toprule
			\textbf{single-center} && \textbf{MIQCP} && \textbf{MILP+SD} &&  \multicolumn{5}{c}{\textbf{NS Heuristic}} && \textbf{LB}  && \multicolumn{4}{c}{\textbf{Gap Comparison (\%)}}\\
			\cmidrule{3-3} \cmidrule{5-5}  \cmidrule{7-11} \cmidrule{13-13} \cmidrule{15-18}   
			$N$ && $C_{\text{15mins}}^{\text{MIQCP}}$  && $C_{\text{15mins}}^{\text{MILP+SD}}$ && $T_{\text{NS}}\ (s)$ & $\#_{\text{iter}}$ & $N_{\text{s}}$ & $C_{\text{CNU}}$ & $C_{\text{NS}}$ && $C_{\text{lb}}$ && $\gamma_{\text{CNU}} $ & $\gamma_{\text{\text{15mins}}}^{\text{MIQCP}} $  &$\gamma_{\text{\text{15mins}}}^{\text{MILP+SD}} $& $\gamma_{\text{NS}} $   \vspace{1pt}\\  
			\hline
			$\alpha =1$\\
			\hline
20  &  & 6107.7  &  & 6053.8  &  & 2.3   & 4.5  & 21.2  & 6107.7  & 5730.6  &  & 5594.5  &  & 9.18 & 9.18 & $8.26^\star$ & 2.43 \\
50  &  & 14435.1 &  & 14435.1 &  & 15.7  & 11.8 & 43.1  & 14435.1 & 13938.7 &  & 13633.9 &  & 5.87 & 5.87 & 5.87 & 2.23 \\			
75  &  & 22012.6 &  & 22012.6 &  & 24.0    & 14.1 & 65.4  & 22012.6 & 21431.1 &  & 20665.6 &  & 6.51 & 6.51 & 6.51 & 3.69 \\
100 &  & 29020.6 &  & 29020.6 &  & 30.9  & 16.8 & 75.7  & 29020.6 & 28402.6 &  & 27087.9 &  & 7.13 & 7.13 & 7.13 & 4.85 \\
175 &  & 47608.6 &  & 47608.6 &  & 67.0    & 21.2 & 116.5 & 47608.6 & 46952.1 &  & 43953.1 &  & 8.31 & 8.31 & 8.31 & 6.81 \\
250 &  & 67988.1 &  & 67988.1 &  & 311.7 & 35.7 & 160.8 & 67988.1 & 67288.9 &  & 62497.5 &  & 8.79 & 8.79 & 8.79 & 7.67 \\
			\hline
			$\alpha =2$\\
			\hline
20  &  & 5923.1  &  & 5923.1  &  & 7.3   & 3.6  & 11.1  & 5923.1  & 5814.8  &  & 5670.4  &  & 4.52 & 4.52 & 4.52 & 2.63 \\
50  &  & 13247.3 &  & 13247.3 &  & 19.6  & 7.8  & 30.1  & 13247.3 & 12746.3 &  & 12586.6 &  & 5.26 & 5.26 & 5.26 & 1.29 \\
75  &  & 19940.3 &  & 19940.3 &  & 30.5  & 9.8  & 44.1  & 19940.3 & 19328.5 &  & 19035.8 &  & 4.75 & 4.75 & 4.75 & 1.54 \\
100 &  & 26604.5 &  & 26604.5 &  & 47.1  & 15.8 & 58.8  & 26604.5 & 26037.7 &  & 25333.0 &  & 5.01 & 5.01 & 5.01 & 2.77 \\
175 &  & 45876   &  & 45876   &  & 115.3 & 29.9 & 103.3 & 45876.0 & 45166.6 &  & 43461.6 &  & 5.55 & 5.55 & 5.55 & 3.92 \\
250 &  & 65259.4 &  & 65259.4 &  & 167.5 & 31.8 & 133.6 & 65259.4 & 64567.7 &  & 61760.9 &  & 5.66 & 5.66 & 5.66 & 4.54 \\
			\hline
			$\alpha =3$\\
			\hline
20  &  & 5642.4  &  & 5642.4  &  & 7.3   & 1.0    & 7.9   & 5642.4  & 5498.8  &  & 5352.5  &  & 5.44 & 5.44 & 5.44 & 2.69 \\
50  &  & 13334.9 &  & 13334.9 &  & 19.0    & 7.8  & 30.0    & 13334.9 & 13092.7 &  & 12803.1 &  & 4.17 & 4.17 & 4.17 & 2.21 \\
75  &  & 20214   &  & 20214   &  & 28.6  & 8.8  & 46.4  & 20214.0 & 19552.3 &  & 19316.6 &  & 4.67 & 4.67 & 4.67 & 1.21 \\
100 &  & 26984.7 &  & 26984.7 &  & 45.2  & 14.5 & 55.3  & 26984.7 & 26310.0 &  & 25924.0 &  & 4.09 & 4.09 & 4.09 & 1.49 \\
175 &  & 46005.2 &  & 46005.2 &  & 88.3  & 20.6 & 99.1  & 46005.2 & 45303.4 &  & 44111.9 &  & 4.29 & 4.29 & 4.29 & 2.69 \\
250 &  & 65213   &  & 65213   &  & 273.2 & 34.2 & 135.9 & 65213.0 & 64472.3 &  & 62415.8 &  & 4.48 & 4.48 & 4.48 & 3.30 \\
\hline
 Summary  &\multicolumn{10}{l}{The NS heuristic solves 157/180 instances to within 5\% of the $C_{\text{lb}}$}\\
 & \multicolumn{10}{l}{The NS heuristic solves 180/180 instances to within 10\% of the $C_{\text{lb}}$}  \\
			\bottomrule
		\end{tabular}
	}
\label{tab:l_singlecenter}
\end{table}

\begin{table}[h!]
	\centering
	\caption{Results from solving instances from a double-center pattern in the large data set.}
	\ra{1.1}
	\resizebox{\textwidth}{!} {
		\begin{tabular}{@{}rrrrrrrrrrrrrrrrrr@{}}
			\toprule
			\textbf{double-center} && \textbf{MIQCP} && \textbf{MILP+SD} &&  \multicolumn{5}{c}{\textbf{NS Heuristic}} && \textbf{LB}  && \multicolumn{4}{c}{\textbf{Gap Comparison (\%)}}\\
			\cmidrule{3-3} \cmidrule{5-5}  \cmidrule{7-11} \cmidrule{13-13} \cmidrule{15-18}   
			$N$ && $C_{\text{15mins}}^{\text{MIQCP}}$  && $C_{\text{15mins}}^{\text{MILP+SD}}$ && $T_{\text{NS}}\ (s)$ & $\#_{\text{iter}}$ & $N_{\text{s}}$ & $C_{\text{CNU}}$ & $C_{\text{NS}}$ && $C_{\text{lb}}$ && $\gamma_{\text{CNU}} $ & $\gamma_{\text{\text{15mins}}}^{\text{MIQCP}} $  &$\gamma_{\text{\text{15mins}}}^{\text{MILP+SD}} $& $\gamma_{\text{NS}} $   \vspace{1pt}\\  
			\hline
			$\alpha =1$\\
			\hline
20  &  & 7424.6  &  & 7424.6  &  & 2.4   & 3.5  & 21.9  & 7424.6  & 7089.6  &  & 6933.5  &  & 7.13  & 7.13  & 7.13 & 2.25 \\
50  &  & 15512.9 &  & 15512.9 &  & 10.4  & 8.6  & 47.0    & 15512.9 & 15060.0 &  & 14913.1 &  & 4.06  & 4.06  & 4.06 & 1.00 \\
75  &  & 22348   &  & 22348   &  & 23.1  & 14.0   & 68.0    & 22348.0 & 21810.7 &  & 21347.0 &  & 4.70  & 4.70  & 4.70 & 2.17 \\
100 &  & 29956.4 &  & 29956.4 &  & 39.3  & 22.7 & 92.7  & 29956.4 & 29421.4 &  & 28390.7 &  & 5.51  & 5.51  & 5.51 & 3.63 \\
175 &  & 50159.1 &  & 50159.1 &  & 100.3 & 29.1 & 121.5 & 50159.1 & 49555.0 &  & 46999.4 &  & 6.72  & 6.72  & 6.72 & 5.44 \\
250 &  & 69377.8 &  & 69377.8 &  & 260.0   & 40.5 & 178.8 & 69377.8 & 68733.0 &  & 64420.0 &  & 7.70  & 7.70  & 7.70 & 6.69 \\
			\hline
			$\alpha =2$\\
			\hline
20  &  & 6405.8  &  & 6405.8  &  & 5.5   & 1.7  & 11.6  & 6405.8  & 6015.4  &  & 5891.2  &  & 9.09  & 9.09  & 9.09 & 2.21 \\
50  &  & 13840.9 &  & 13840.9 &  & 17.7  & 5.5  & 28.7  & 13840.9 & 13308.6 &  & 13145.2 &  & 5.31  & 5.31  & 5.31 & 1.25 \\
75  &  & 20762.9 &  & 20762.9 &  & 30.7  & 9.8  & 47.3  & 20762.9 & 20154.9 &  & 19899.4 &  & 4.35  & 4.35  & 4.35 & 1.28 \\
100 &  & 26963.8 &  & 26963.8 &  & 47.4  & 12.9 & 57.0    & 26963.8 & 26323.6 &  & 25805.4 &  & 4.49  & 4.49  & 4.49 & 2.00 \\
175 &  & 46667.6 &  & 46667.6 &  & 102.5 & 19.6 & 105.6 & 46667.6 & 46011.9 &  & 44461.9 &  & 4.96  & 4.96  & 4.96 & 3.49 \\
250 &  & 66497.8 &  & 66497.8 &  & 297.9 & 36.6 & 146.5 & 66497.8 & 65816.6 &  & 63210.3 &  & 5.20  & 5.20  & 5.20 & 4.12 \\
			\hline
			$\alpha =3$\\
			\hline
20  &  & 5701    &  & 5587.7  &  & 6.8   & 3.2  & 9.8   & 5701.0  & 5282.0  &  & 5111.8  &  & 11.65 & 11.65 & $9.44^\star$ & 3.38 \\
50  &  & 13400.3 &  & 13400.3 &  & 25.9  & 6.9  & 26.8  & 13400.3 & 12863.3 &  & 12736.8 &  & 5.32  & 5.32  & 5.32 & 1.11 \\
75  &  & 20255.2 &  & 20255.2 &  & 40.1  & 10.4 & 39.3  & 20255.2 & 19670.6 &  & 19512.0 &  & 3.82  & 3.82  & 3.82 & 0.80 \\
100 &  & 26478.9 &  & 26478.9 &  & 52.8  & 14.3 & 54.1  & 26478.9 & 25825.0 &  & 25472.4 &  & 3.95  & 3.95  & 3.95 & 1.38 \\
175 &  & 45429.1 &  & 45429.1 &  & 108.2 & 17.4 & 97.7  & 45429.1 & 44716.7 &  & 43657.1 &  & 4.06  & 4.06  & 4.06 & 2.43 \\
250 &  & 65414   &  & 65414   &  & 298.4 & 31.6 & 139.3 & 65414.0 & 64697.7 &  & 62760.9 &  & 4.23  & 4.23  & 4.23 & 3.08\\
\hline
 Summary  &\multicolumn{10}{l}{The NS heuristic solves 155/180 instances to within 5\% of the $C_{\text{lb}}$}\\
 & \multicolumn{10}{l}{The NS heuristic solves 180/180 instances to within 10\% of the $C_{\text{lb}}$}  \\
			\bottomrule
		\end{tabular}
	}
\label{tab:l_doublecenter}
\end{table}

First, we examine the performance of the two exact approaches in solving large data instances. Recall that for solving each large instance, we warm-start the exact approaches by using the initial feasible solution obtained by the CNU model. As can be observed in columns $\gamma_{\text{15mins}}^{\text{MIQCP}}$, $\gamma_{\text{15mins}}^{\text{MILP+SD}}$, and $\gamma_{\text{CNU}}$ in each of Tables \ref{tab:l_uniform}--\ref{tab:l_doublecenter}, both exact approaches experience difficulty in further reducing the mission makespan upon the initial feasible solution when the searching process is curtailed due to the limited run-time constraint. However, the MILP+SD model shows slight improvement in solving data scenarios characterized as (single-center, N = 20, $\alpha$ = 1) and (double-center, N = 20, $\alpha$ = 3) as highlighted by $\star$. These results indicate the failure in leveraging exact approaches for solving large data instances, which aligns with the conclusions in \cite{gonzalez2020truck}. On the one hand, the difficulties in closing the gap come from the effects of the number of locations, the geometrical distribution of the locations, and the vehicles' speed ratio. We hypothesize that as the distribution of the locations becomes more skewed, the speed ratio between the two vehicles becomes more incompatible, and the size of the problem increases, so that the drone experiences more difficulties in collaborating with the truck to replace batteries. From a computational complexity perspective, when solving the Nested-VRP, finding the optimal solution requires an intensive search for the best combination of drone route, truck route, and battery swap locations. Therefore, the exact approach can only achieve a less-than-optimal solution given the run-time limit of 15 minutes. On the other hand, when using the Gurobi optimizer and given a near-optimal feasible solution, the failure to achieve optimality could be partially explained by the slow converging behavior of the lower bound in the branch-and-bound process. 

Second, we explore the advantages of applying the proposed NS heuristic. Compared to the exact approaches, the NS heuristic improves the solution quality by combining effective initialization strategy and local improvement scheme. To assess the effectiveness of the initialization strategy, when solving each Nested-VRP instance, we evaluate the relative optimality gap $\gamma_{\text{CNU}}$ of the solution obtained by solving the CNU problem with respect to the lower bound value. The column $\gamma_{\text{CNU}}$ suggests that solving the CNU problem produces robust feasible Nested-VRP solutions with a 5.83\% average gap and a 1.91\% standard deviation relative to the lower bound value $C_{\text{lb}}$ across all large instances. We further examine the effectiveness of the pure local search process of the NS heuristic by assessing its contribution to optimality. By referring to columns $\gamma_{\text{CNU}}$ and $\gamma_{\text{NS}}$, across all scenarios, the local search process contributes to a further 2.86\% gap reduction on average given the initial feasible solution provided by solving the CNU problem. 

It is also important to highlight that further reduction, obtained by the local search process, comes from exploring a relatively small number of nested units specified by the initial feasible solution. To see this, the number of battery swaps $N_{\text{s}}$ is an indicator of the total number of nested units that are included in the final solution, while the number of iterations $\#_{\text{iter}}$ tells how many bad nested units are reorganized before reaching the final decision. As an example, in Table \ref{tab:l_singlecenter}, in the scenario $(\text{single-center},  N = 250, \, \alpha = 1)$, the NS heuristic takes on average 35.7 iterations to finalize the Nested-VRP solution, which consists of 160.8 nested units. Approximately 22.20\% of the units are reorganized before the NS heuristic terminates. A similar analysis can be applied to other scenarios.              

Third, we assess the quality of the solutions provided by the NS heuristic. The NS heuristic is able to produce solutions with objective values that are reasonably close to the lower bound. Specifically,according to the summary of the Tables \ref{tab:l_uniform}--\ref{tab:l_doublecenter}, $(146+151+155)/(180+180+180) = 84.81\%$ of instances achieve results that deviate from the lower bound solution by less than 5\%. Moreover, all instances can be solved within a 10\% gap with respect to the proposed lower bound value $C_{\text{lb}}$. It is worth noticing that the LB produces an over-optimistic estimation of the optimal value. Further tightening the LB will potentially gain more-accurate insights into the performance of the NS heuristic method. In terms of the NS's computation efficiency, the column $T_{\text{NS}}$ in Tables \ref{tab:l_uniform}--\ref{tab:l_doublecenter} show that the time needed for NS to solve a Nested-VRP instance scales up linearly as a function of total number of locations. To summarize the above discussion, the proposed NS heuristic is effective and efficient in producing high-quality solutions to the Nested-VRP\@.

\subsection{Real-world case study}\label{subsec:real}
In this section, we focus on a realistic application of the Nested-VRP to surveillance in the aftermath of the 2017 Santa Rosa Wildfire. Our goal is to assess the effectiveness and efficiency of the NS heuristic in solving practical applications. Most importantly, we will further demonstrate the robustness of the proposed heuristic by empirically investigating its convergence behavior.

\begin{figure} [h!]
	\centering
	\includegraphics[width =0.8\linewidth]{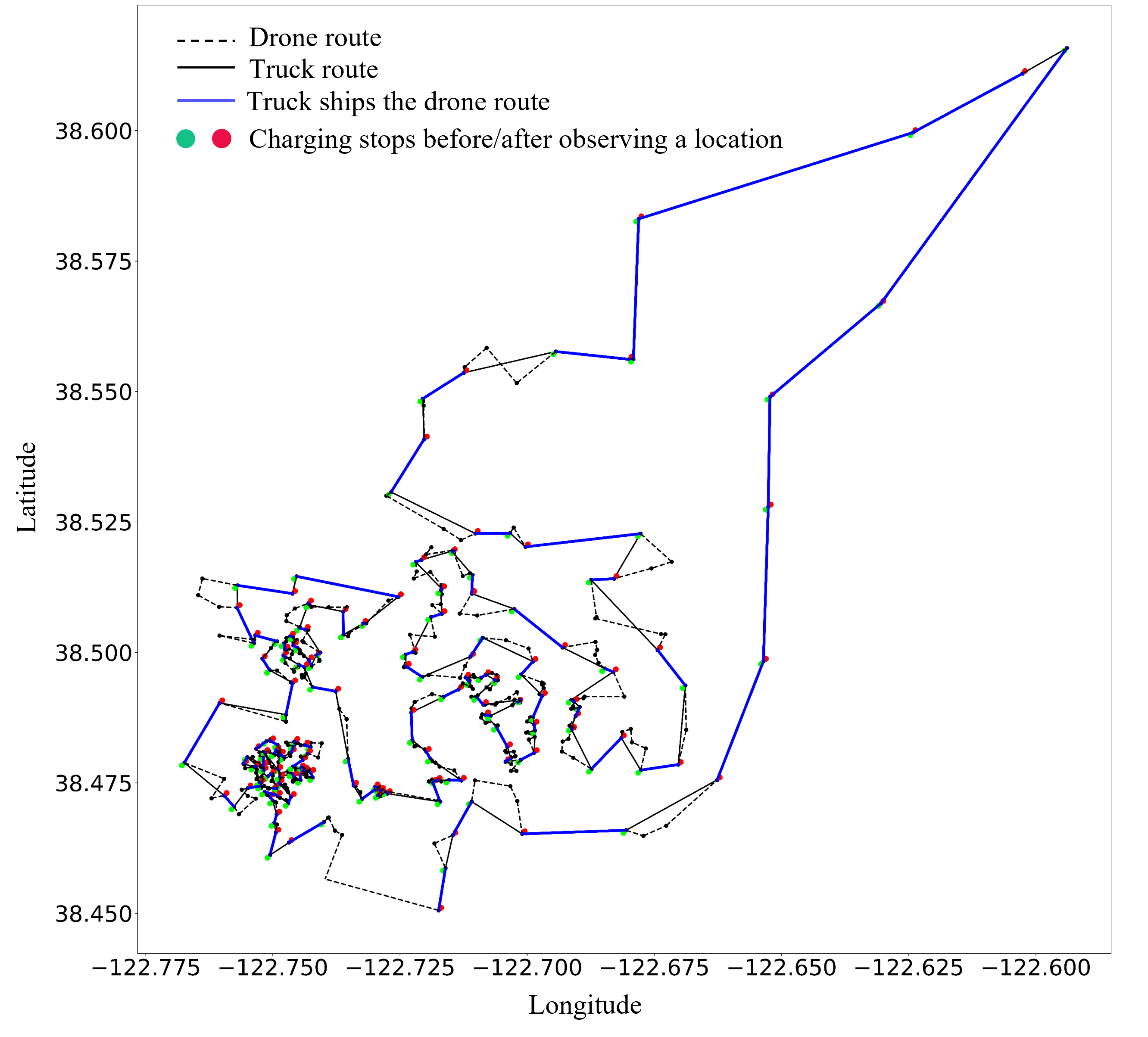}
	\caption{Nested-VRP solution for the practical instance. The black solid line is the truck route, while the black dashed line represents the drone route. The blue line corresponds to the truck ships the drone. Circles depict the subset of locations serving as swap stops. Green and red colors differentiate the battery swaps that happen before or after the drone observes a location. \label{fig:case_sol}}
\end{figure}

After the devastating fire disaster, an insurance company decided to send out a truck and a drone to inspect and collect home damage evidence from 631 clients' properties. The information regarding all house locations is given. We set the truck speed as 5 m/s (considering the difficult ground conditions) and the drone speed as 10 m/s; thus, $\alpha = 2$. The battery capacity is set to 10 minutes. The data cleaning process includes bundling observation tasks for apartments in the same building as a single observation task at the building's location. Correspondingly, the observation time is the sum of that for each apartment. We perform the NS heuristic on this specific instance for 100 independent runs to account for the randomness in the searching process. In each run, the NS heuristic search bad nested units on the top $\beta = 0.3$ fraction list per iteration and will terminate if the program observes no improvements for $N_{\text{UNCH}} =5$ consecutive iterations or the total number of iterations exceeds $N_{\text{max}}=50$. We depict the best-known solution in Figure \ref{fig:case_sol}. 

\vspace{.1in} 

\begin{figure}[h!]
	\centering
	\includegraphics[width =\linewidth]{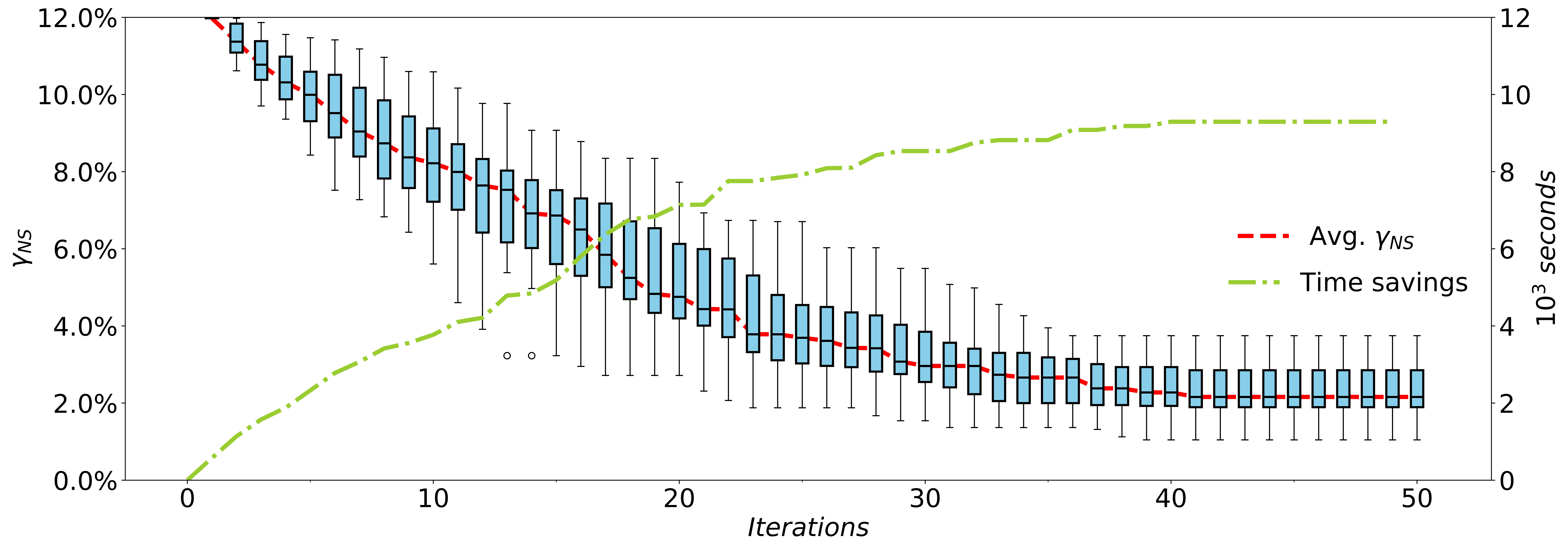}
	\caption{NS heuristic performance when solving the practical instance 100 times. The red line shows that the average optimality gap $\gamma_{\text{NS}}$ decreases as the iteration number increases across the $100$ tries. The green line shows that, compared to the mission makespan of the initial Nested-VRP solution obtained by solving CNU program, the NS heuristic finds a better solution via an effective local search scheme iteration by iteration. A better solution corresponds to a solution with a shorter mission makespan and thus corresponds to a larger time savings.  \label{fig:case_performance}}
\end{figure}

The best-known solution suggests that at least 26 hours 51 minutes are required to complete the entire mission, which consists of vising 218 meet-up stops along the tour. We summarize the performance of the NS heuristic in Figure \ref{fig:case_performance}. In this figure, the $i$th blue box describes the distribution of $\gamma_{\text{NS}}$ at the $i$th iteration across 100 runs. The average trend of $\gamma_{\text{NS}}$ as the number of iterations increases is presented in the red curve. In addition, the green curve reports the average amount of time savings obtained per iteration. 

\vspace{.1in} 

\textbf{Run-times:} As the NS heuristic suggests, the average run-times for solving this Nested-VRP instance of size 631 is 1032.6 seconds (17 minutes 13 seconds), which can be considered as efficient for planning a full-day mission. 

\textbf{Convergence:} Referring to Figure \ref{fig:case_performance}, the red line summarizes its average trend across 100 runs. The red line suggests that the NS heuristic produces a Nested-VRP solution of smaller mission makespan as the local search process proceeds. Specifically, after solving the CNU model in the initialization stage, the NS heuristic produces a feasible Nested-VRP solution of 11.98\% average optimality gap away from the estimate of lower bound value. Even though the red line has some ``almost flat'' segments (i.e., the 23th to 26th iterations), which indicate a marginal improvement between iterations in the destruction and reconstruction process, these stationary stages were transient and eventually progressed to a better solution. This is evidence to show NS's ability to escape local minima. Promisingly, the NS heuristic saves on average 9292.5 seconds (2 hour 35 minutes) from an initial CNU feasible solution and produces final results that are 2.16\% closer to the lower bound on average.    

It is worth noting that for a realistic Nested-VRP instance, implementing a drone-truck surveillance team to complete observation tasks may occasionally provide no time-saving benefits. Referring to Figure \ref{fig:case_sol}'s upper-right corner, we observe that the truck travels to multiple sites with the drone on-board. This is due to the geometrical configuration constituting the locations of the sites: relatively far away from each other and out of the drone's reach from one location to the other. The solution is almost reduced to having a truck complete the surveillance task alone by following a partial TSP tour. In this case, the drone, even though it has the advantage of high cruising speed, does not contribute to speeding up the mission. Therefore, we lose the potential time-saving benefit of hiring a drone-truck surveillance team. This observation suggests that while matching the speed ratio between the truck and drone is crucial, picking the appropriate drone model with sufficient flight duration to support the mission is also important to the overall mission performance. 

The results from solving small, large, and real-world Nested-VRP instances shed light on how to use the proposed methodology to solve a broader problem involving the coordination of multiple trucks and drones to complete a surveillance mission. Consider a large area with thousands of locations to observe. One strategy is to divide the area into dense, sub-dense, sparse, and truck-only sub-areas. Locations in the truck-only sub-area are typically located far apart from one another in the same sub-area and from the rest. In this case, using a drone and truck team to perform surveillance tasks is not beneficial. Then, for each sub-area except for the truck-only sub-area, we employ a single truck or a team of a single truck and a single drone. The proposed strategy has three benefits: (i) Excluding locations in the truck-only sub-area reduces the size of the Nested-VRP; (ii) Pairing a single-truck-single-drone team with a sub-area allows the practitioner to take a finer look at the geometrical distribution of the locations within the sub-area and adjust the speed ratio of the truck and the drone so as to obtain the most time savings; (iii) Solving the Nested-VRP for each sub-area allows the Nested-VRP model to be solved in parallel, potentially resulting in significant run-time savings.

\section{Conclusion}\label{sec:conclusion}
In this paper, the Nested-VRP is formally defined and formulated as a MIQCP program. Given the operational assumptions described in \S\ref{sec:des_def}, our model is capable of finding the best trade-off between the drone's routing plan and swap assignments along the route such that the total mission duration is minimized. An enhancement of the original Nested-VRP model by linearization and constraints strengthening techniques is examined and has been demonstrated effective in shortening computational time to obtain optimal solution. In situations where the Gurobi Optimizer fails to solve the Nested-VRP exactly, we further propose an NS heuristic approach that is based on destruction and local reconstruction principles. Our extensive experiments on small, large, and realistic instances have demonstrated the proposed heuristic's ability to obtain reasonably good results while requiring substantially lower run-times compared to the MIP exact approach when the size of the problem is large.

Our empirical study has the following implications for future practitioners. (i) The geometrical distribution of the set of locations should be evaluated first to see the potential time-savings benefits that are achievable by implementing the drone-truck surveillance team. (ii) The speed of the two vehicles should be matched at the right ratio to maximize the overall time savings.

The model developed in this paper could enable a spectrum of applications in the field of aerial surveillance. Future work could include the ground traffic as a stochastic element into the model such that the ground delays will factor in the planning. Regarding the drone's flight performance, further research into how the drone's energy consumption in observation and flying mode affects routing decisions is critical for real-world application. In terms of algorithmic design, one might propose a sophisticated exact approach inspired by branch-and-bound or a more-efficient heuristic that produces near-optimal solutions with reduced run-times. From a modeling perspective, future research should include advanced modeling techniques to better represent the unique features of the Drone-Truck surveillance problem (e.g., cyclic vehicle operation) as well as more-realistic operational scenarios (e.g., the drone can depart or arrive at any point along arcs that form the road network). While we only consider the case with a single truck and a single drone, one possible direction could be coordinating a team of drones together with multiple trucks to accomplish the goals and assessing the benefits of introducing more agents.

\vspace{.1in}

\noindent

\printcredits
\bibliographystyle{apalike}
\bibliography{bib}

\appendix
\setcounter{figure}{0} \renewcommand{\thefigure}{A.\arabic{figure}}
\setcounter{table}{0} \renewcommand{\thetable}{A.\arabic{table}}
\section{Proof of Theorem \ref{thm:compact}}\label{app:proof}
\begin{pf}
First, we set forth the ZONVRP formulation. Since there are no observations, we do not distinguish battery swaps that happen before or after observing a location. Therefore, for each location, we define $ z_j  \in \{0, 1 \}, \forall j \in \mathcal{H}$. If $z_j=1$, location $j$ is a battery swap location, 0 otherwise. Likewise, each location only requires a variable $l_j \in \mathcal{R}_+, \forall j \in \mathcal{H}$ to record the IBR time. The ZONVRP model is presented as follows:

\begin{align}
(\text{ZONVRP})\quad  \min_{} \quad  &  \sum_{i\in \mathcal{H}} l_i + T_s \sum_{i \in H} z_i - T_s\sum_{(i,j) \in A} 2w_{ij}  \nonumber\\
\textrm{s.t.} \quad  & \sum_{j:(i,j) \in \mathcal{A}}x_{ij} = 1, \quad  \forall i \in \mathcal{H} \setminus \{n+1\} \label{nmodel:drone_1}\\
&\sum_{i:(i,j) \in \mathcal{A} }x_{ij} = 1, \quad \forall j \in \mathcal{H} \setminus \{0\} \label{nmodel:drone_2}\\
& u_0 = 0\\
& 1 \le u_i \le n+1, \quad \forall\; i\in\mathcal{H} \setminus  \{0\} \label{nmodel:ubound}\\
& u_i - u_j  + 1 \le (n+1)(1-x_{ij}),\quad \forall (i,j) \in \mathcal{A}, \, i \neq 0 \label{nmodel:drone_4}\\
& \sum_{j: (0,j) \in \mathcal{A}} y_{0, j} \le 1 \label{nmodel:truck_1} \\
& \sum_{i:(i,j) \in \mathcal{A}}y_{ij} = z_j; \quad \sum_{k:(j,k) \in \mathcal{A}} y_{jk} = z_j, \quad \forall j \in \mathcal{H} \setminus \{0, n+1\} \label{nmodel:truck_2}\\
& u_i - u_j + 1 \le (n+1)(1-y_{ij}),\quad \forall (i,j) \in \mathcal{A}, \, i \neq 0 \label{nmodel:y_order} \\
&\tau_{ij}^T y_{ij} \le T_{\text{bl}} + M_1 w_{ij}, \quad \forall (i,j) \in \mathcal{A} \label{nmodel:waiting}\\
& t_j^- \le T_{\text{bl}}, \quad\forall j \in \mathcal{H} \setminus \{0\} \label{nmodel:time_1}\\
& t_j^+ \le T_{\text{bl}}, \quad\forall j \in \mathcal{H} \setminus \{n+1\} \label{nmodel:time_2}\\
& t_j^+ = t_j^-(1-z_j),\quad \forall j \in \mathcal{H} \setminus \{0, n+1\} \label{nmodel:time_5}\\  
& w_{ij} \le x_{ij}, \quad \forall (i,j) \in \mathcal{A} \label{nmodel:w_1}\\
& w_{ij} \le y_{ij}, \quad \forall (i,j) \in \mathcal{A} \label{nmodel:w_2}\\
& w_{ij} \le z_i, \quad \forall (i,j) \in \mathcal{A} \label{nmodel:w_3}\\ 
& w_{ij} \le z_j, \quad \forall (i,j) \in \mathcal{A} \label{nmodel:w_4}\\ 
& x_{ij} + y_{ij} + z_i + z_j \le 3 + w_{ij}, \quad \forall (i,j) \in \mathcal{A} \label{nmodel:w_5}\\
& t_j^- \le t_i^+(1-z_i) + \tau_{ij}^D(1-w_{ij}) + M_2(1-x_{ij}), \quad \forall (i,j) \in \mathcal{A} \label{nmodel:time_3}\\
& t_j^- \ge t_i^+(1-z_i) + \tau_{ij}^D(1-w_{ij}) -M_2(1-x_{ij}), \quad \forall (i,j) \in \mathcal{A} \label{nmodel:time_4}\\
& l_j \le T_{\text{bl}}  z_j + M_2 \sum_{i:(i,j) \in \mathcal{A}} w_{ij},  \quad \forall j \in \mathcal{H} \label{nmodel:lbound1}\\
& l_j \ge t_j^- - M_2(1-z_j), \quad \forall j \in \mathcal{H} \label{nmodel:late_1}\\
&l_j \ge \sum_{i:(i,j) \in \mathcal{A}} \Big( \tau_{ij}^T y_{ij} + \big(\max(\tau_{ij}^T, T_{\text{s}}) - \tau_{ij}^T \big)w_{ij} \Big)  - M_1(1-z_j) , \quad \forall j \in \mathcal{H}  \label{nmodel:late_2}\\
& z_{0} = 1, t_{0}^+ = 0 \label{nmodel:v_1}\\
&x_{ij} \in \{0,1\}, y_{ij} \in \{0,1\}, w_{ij} \in \{0,1\}, \quad \forall (i,j) \in \mathcal{A}\label{nmodel:v_2} \\
&z_i \in \{0,1\}, u_i \in [0, n+1], t_i^-, t_i^+ \in [0, T_{\text{bl}}], l_i \in \mathcal{R}_+, \quad \forall i \in \mathcal{H} \label{nmodel:v_3}
\end{align}

To make the proof self-contained, we introduce the notations that are applied in TDTL model and reproduce the TDTL model from \cite{gonzalez2020truck}.

\begin{table}
	\caption{Notations used in the TDTL MIP formulation. \strut } 
	\centering
	\setlength{\tabcolsep}{4pt}
	\renewcommand{\arraystretch}{1.3}
	\begin{tabular}{ll} 
		\hline 
		\multicolumn{2}{l}{\textbf{Sets}}\\
		\hline
        $ \mathcal{N}$ & Set of nodes of graph $G = (\mathcal{N}, \mathcal{A})$. \\
        $\mathcal{A}$ & Set of directed links in $G = (\mathcal{N}, \mathcal{A})$.\\
		$o/e$ & The origin/ending node of the mission, $o,e \in \mathcal{N}$.\\
		$\delta^+(i)$ & Nodes that can be reached from $i$, $i  \in \mathcal{N}$.\\
		$\delta^-(i)$ & Nodes that can reach to node $i$, $i  \in \mathcal{N}$.\\		
		\hline 
		\multicolumn{2}{l}{\textbf{Parameters}} \\
		\hline
		Q & Battery capacity expressed in time units. \\
		$t_{ij}^T/$ &  Truck travel time at link $(i,j) \in \mathcal{A}$.\\
		$t_{ij}^D$ &  Drone travel time at link $(i,j) \in \mathcal{A}$.\\
		M & A big enough constant.\\
		\hline   
	    \multicolumn{2}{l}{\textbf{Variables}} \\
		\hline
	    $u_{ij}$  & If $u_{ij} = 1$, link $(i,j)$ is traversed by the truck. \\
	    $v_{ij}$ & If $v_{ij} = 1$, link $(i,j)$ is traversed by the drone.\\
        $s_i$ & The earliest departure time from node $i\in \mathcal{N}$.\\
        $b_i^-$ & Drone battery level when if drone is just coming to the node $i \in \mathcal{N}$.\\
        $b_i^+$ & Drone battery level when the drone is just departing from node $i \in \mathcal{N}$.\\
		\hline
	\end{tabular}
	\label{tab:g_notation}
\end{table}

\begin{align}
(\text{TDTL})\quad  \min_{} \quad  &  s_e \nonumber\\
\textrm{s.t.} \quad 
&\sum_{j \in \delta^-(i)} u_{ji} \le 1, \quad \forall i \in \mathcal{N} \setminus \{o,e\} \label{g:1}\\
&\sum_{j \in \delta^+(i) } u_{ij} - \sum_{j \in \delta^-(i)} u_{ji}=0, \quad \forall i \in \mathcal{N} \setminus \{o,e\} \label{g:2}\\
&\sum_{j\in \delta^+(o)} u_{oj} = 1 \label{g:3}\\
&\sum_{i \in \delta^-(e)} u_{ie} = 1 \label{g:4}\\
&\sum_{j \in \delta^-(i)} v_{ji} \le 1,  \quad \forall i \in \mathcal{N} \setminus \{o,e\} \label{g:5}\\
& \sum_{j \in \delta^+(i)} v_{ij} - \sum_{j\in \delta^-(i)} v_{ji} = 0,  \quad \forall i \in \mathcal{N} \setminus \{o,e\} \label{g:6}\\
& \sum_{j \in \delta^+(o)} v_{oj} = 1\label{g:7}\\
& \sum_{i \in \delta^-(e)} v_{ie} = 1 \label{g:8}\\
&\sum_{i \in \delta^-(j)} u_{ij} + \sum_{i \in \delta^-(j)}v_{ij} \ge 1, \quad \forall j \in \mathcal{N} \setminus \{o\}\label{g:9}\\
&s_j \ge s_i + t_{ij}^Tu_{ij} - M\Big(1-u_{ij}\Big), \quad (i,j) \in \mathcal{A}\label{g:10}\\
&s_j \ge s_i + t_{ij}^Dv_{ij} - M\Big(1 - v_{ij} + u_{ij}\Big), \quad (i,j) \in \mathcal{A}\label{g:11}\\
&s_o = 0\label{g:12}\\
& b_j^- \le Q + M\Big(2 - v_{ij} - u_{ij}\Big), \quad (i,j) \in \mathcal{A}\label{g:13}\\
& b_j^- \ge Q - M\Big(2 - v_{ij} - u_{ij}\Big), \quad (i,j) \in \mathcal{A}\label{g:14}\\
& b_j^+ \le Q + M\Big(2 - v_{ij} - u_{ij}\Big), \quad (i,j) \in \mathcal{A}\label{g:15}\\
& b_j^+ \ge Q - M\Big(2 - v_{ij} - u_{ij}\Big), \quad (i,j) \in \mathcal{A}\label{g:16}\\
& b_j^- \le  b_i^+ - t_{ij}^D + M\Big(1-v_{ij} + u_{ij} + \sum_{k \neq i} u_{kj}\Big), \quad (i,j) \in \mathcal{A}\label{g:17}\\
& b_j^- \ge  b_i^+ - t_{ij}^D - M\Big(1-v_{ij} + u_{ij} + \sum_{k \neq i} u_{kj}\Big), \quad (i,j) \in \mathcal{A}\label{g:18}\\
&b_j^+ \le b_j^- + M\Big(1 - v_{ij} + u_{ij} + \sum_{k \neq i}u_{kj}\Big), \quad (i,j) \in \mathcal{A}\label{g:19}\\
&b_j^+ \ge b_j^- - M\Big(1 - v_{ij} + u_{ij} + \sum_{k \neq i}u_{kj}\Big), \quad (i,j) \in \mathcal{A}\label{g:20}\\
& b_j^- \le b_i^+ - t_{ij}^D + M\Big(1 - v_{ij} + u_{ij} + 1 - \sum_{k \neq i} u_{kj}\Big), \quad (i,j) \in \mathcal{A}\label{g:21}\\
& b_j^- \ge b_i^+ - t_{ij}^D - M\Big(1 - v_{ij} + u_{ij} + 1 - \sum_{k \neq i} u_{kj}\Big), \quad (i,j) \in \mathcal{A}\label{g:22}\\
& b_j^+ \le Q + M\Big(1 - v_{ij} + u_{ij} + 1 - \sum_{k\neq i} u_{kj}\Big), \quad (i,j) \in \mathcal{A}\label{g:23}\\
& b_j^+ \ge Q - M\Big(1 - v_{ij} + u_{ij} + 1 - \sum_{k\neq i} u_{kj}\Big), \quad (i,j) \in \mathcal{A}\label{g:24}\\
& b_o^+ =  Q \label{g:25}
\end{align}

Let the polyhedron of the linear relaxation of models $\text{ZONVRP}$ and $\text{TDTL}$ be defined by 
\begin{align}
&P(\text{ZONVRP}) = \Big\{(l,t^-, t^+,z,u,w,x,y) \in \mathcal{R}^{5(n+1)}_{\ge 0} \times [0,1]^{3n^2} \big| \text{constraints: (\ref{nmodel:drone_1})--(\ref{nmodel:v_3})} \Big\} \nonumber\\
&P(\text{TDTL}) = \Big\{(s, b^-, b^+, u, v) \in \mathcal{R}^{3(n+1)}_{\ge 0}\times [0,1]^{2n^2}\big|\ \text{constraints:  (\ref{g:1})--(\ref{g:25})} \Big\} \nonumber
\end{align}

Given the linear transformation $\Phi$ in (\ref{tran:x})--(\ref{tran:tp}) below, we will show that each constraint in $P(\text{TDTL})$ is implied by that in $P(\text{ZONVRP})$. We use the notation TDTL.(k) to refer to the constraint (k) in the TDTL model.
	\begin{align}
	& x_{ij} = v_{ij}, \quad \forall (i,j) \in \mathcal{A} \label{tran:x} \\
	&y_{ij} = u_{ij}, \quad \forall (i,j) \in \mathcal{A} \label{tran:y}\\
	& z_j = \sum_{i: (i,j) \in \mathcal{A}} u_{ij} \label{tran:z}, \quad \forall j \in \mathcal{N}  \\
	& t_i^- = Q - b_i^-,\quad \forall i \in \mathcal{N} \label{tran:tm}\\
	& t_i^+ = Q - b_i^+,\quad \forall i \in \mathcal{N}\label{tran:tp}
	\end{align}
	
\vspace{.1 in}

\noindent \text{TDTL}.(\ref{g:1})
	\begin{equation}
	\sum_{j \in \delta^-(i) } u_{ji} = \sum_{j:(j,i) \in \mathcal{A}}  y_{ji} = z_i \le 1 
	\nonumber\end{equation}
	\text{TDTL}.(\ref{g:2})--(\ref{g:4}) are implied by constraint (\ref{nmodel:truck_1})--(\ref{nmodel:y_order}).
	\newline
	\text{TDTL}.(\ref{g:5})--(\ref{g:8}) are implied by constraints (\ref{nmodel:drone_1})--(\ref{nmodel:drone_4}) directly.
	\newline
	\text{TDTL}.(\ref{g:9})
	\begin{equation}
	\sum_{i \in \delta^-(j)} u_{ij} + \sum_{i\in \delta^-(j)} v_{ij} = z_j + \sum_{i:(i,j)\in \mathcal{A}} x_{ij} = z_j + 1 \ge 1 \nonumber
	 \end{equation}
	\text{TDTL}.(\ref{g:10}) is in charge of setting the departure time at node $j$ given that arc $(i,j)$ is traversed by the truck.
	\begin{align*}
	& s_j - s_i - t_{ij}^Tu_{ij} + M(1-u_{ij})\\
	& = s_j - s_i - \tau_{ij}^Ty_{ij} + M(1-y_{ij})\\
	& \ge s_j - s_i  -\tau^T_{ij},  \quad \text{implied by\ }y_{ij}=1 \\
	& = \max \Big\{\tau_{ij}^T, \max \{\tau_{ij}^T, T_s \} \Big\} -\tau^T_{ij}, \text{where $s_j - s_i$ is no less than the truck travel time from location i to j}  \\
	&\ge 0 
	\end{align*}
	\text{TDTL}.(\ref{g:11}) regulates the departure time at node $j$ given that arc $(i,j)$ is traversed by the drone.
	\begin{align*}
	& s_j - s_i - t_{ij}^Dv_{ij} + M(1-v_{ij} + u_{ij})\\
	& = s_j - s_i - \tau_{ij}^Dx_{ij} + M(1-x_{ij} + y_{ij})\\
	& \ge s_j - s_i - \tau_{ij}^D,  \quad \text{implied by\ } x_{ij}=1, y_{ij}=0\\
	& =\tau_{ij}^D - \tau_{ij}^D  \\
	&\ge 0
	\end{align*}
	 \text{TDTL}.(\ref{g:12}) is implied by constraint (\ref{nmodel:v_1}).
	 
	 \text{TDTL}.(\ref{g:13})--(\ref{g:16}) describe the drone's battery level at the time it just arrives at or departs from a node $j$. 
	\begin{align*}
	&b_j^- - Q - M(2 - v_{ij} - u_{ij}) = -t_j^- - M(2-x_{ij} - y_{ij}) \le 0\\
	&b_j^- - Q + M(2 - v_{ij} - u_{ij}) = -t_j^- + M(2-x_{ij} - y_{ij}) \ge 0,\quad \text{implied by constraints  (\ref{nmodel:w_1})-- (\ref{nmodel:time_4})}\\
	&b_j^+ - Q - M(2 - v_{ij} - u_{ij}) = -t_j^+ - M(2 - x_{ij} - y_{ij}) \le 0\\
	&b_j^+ - Q + M(2 - v_{ij} - u_{ij}) = -t_j^+ + M(2 - x_{ij} - y_{ij}) \ge 0,   \quad \text{implied by\ }\text{constraints (\ref{nmodel:time_5})-- (\ref{nmodel:time_4})}
	\end{align*}
	 \text{TDTL}.(\ref{g:17}), $\forall (i,j) \in \mathcal{A}$
	\begin{align*}
	&b_j^- - b_i^+ + \tau_{ij}^D - M\Big(1 - v_{ij} + u_{ij} + \sum_{k \neq i} u_{kj}\Big) \\
	&= t_i^+  - t_j^- + \tau_{ij}^D - M(1 - x_{ij}  + z_j) \\
	&\le t_i^+  - t_j^- + \tau_{ij}^D ,  \quad \text{implied by\ } x_{ij} = 1,   z_j = 0\\
	& = -\tau_{ij}^D + \tau_{ij}^D\\
	& \le 0
	\end{align*}
	
	\text{TDTL}.(\ref{g:18}), $\forall (i,j) \in \mathcal{A}$
	\begin{align*}
	&b_j^- - b_i^+ + \tau_{ij}^D + M\Big(1 - v_{ij} + u_{ij} + \sum_{k \neq i} u_{kj}\Big) \\
	&= t_i^+  - t_j^- + \tau_{ij}^D + M(1 - x_{ij}  + z_j) \\
	&\ge t_i^+  - t_j^- + \tau_{ij}^D,\quad \text{implied by\ } x_{ij} = 1,   z_j = 0\\
	& = - \tau_{ij}^D  + \tau_{ij}^D  \\
	& \ge 0
	\end{align*}
	\text{TDTL}.(\ref{g:19}), $\forall (i,j) \in \mathcal{A}$
	\begin{align*}
	& b_j^+ - b_j^- - M\Big(1 - v_{ij} + u_{ij} + \sum_{k\ne i } u_{kj}\Big) \\
	& =  t_j^- - t_j^+  - M(1 - x_{ij} + z_j)\\
	& \le t_j^- - t_j^-(1-z_j), \quad \text{implied by\ } x_{ij} = 1,   z_j = 0, \text{constraint (\ref{nmodel:time_5})}\\
	&\le 0
	\end{align*}
	\text{TDTL}.(\ref{g:20}), $\forall (i,j) \in \mathcal{A}$
	\begin{align*}
	& b_j^+ - b_j^- + M\Big(1 - v_{ij} + u_{ij} + \sum_{k\ne i } u_{kj}\Big) \\ 
	& =  t_j^- - t_j^+  + M(1 - x_{ij} + z_j)\\
	& \ge t_j^- - t_j^-(1-z_j), \quad \text{implied by\ } x_{ij} = 1,   z_j = 0, \text{constraint (\ref{nmodel:time_5})}\\
	&\ge 0
	\end{align*}
	\text{TDTL}.(\ref{g:21})
	\begin{align*}
	&b_j^- - b_i^+ + \tau_{ij}^D -  M\Big(1 - v_{ij}+u_{ij} + 1 - \sum_{k \neq i}u_{kj}\Big)\\
	& = -t_j^- + t_i^+ + t_{ij}^D - M\Big(1 - x_{ij}+ y_{ij} + 1 - \sum_{k \neq i}y_{kj}\Big) \\
	&\le -t_j^- + t_i^+ + t_{ij}^D ,  \quad \text{implied by\ } x_{ij}=1, y_{ij} = 0, z_j = 1\\
	& = -t_{ij}^D + t_{ij}^D\\
	&\le 0     
	\end{align*}
	\text{TDTL}.(\ref{g:22}), $\forall (i,j) \in \mathcal{A}$
	\begin{align*}
	&b_j^- - b_i^+ + \tau_{ij}^D +  M\Big(1 - v_{ij}+u_{ij} + 1 - \sum_{k\ne i} u_{kj}\Big)\\
	& = -t_j^- + t_i^+ + t_{ij}^D + M\Big(1 - x_{ij}+ y_{ij} + 1 - \sum_{k\ne i} y_{kj}\Big) \\
	&\ge -t_j^- + t_i^+ + t_{ij}^D ,  \quad \text{implied by\ } x_{ij}=1, y_{ij} = 0, z_j = 1\\
	& = -t_{ij}^D + t_{ij}^D\\
	&\ge 0 
	\end{align*}
	\text{TDTL}.(\ref{g:23}), $\forall (i,j) \in \mathcal{A}$
	\begin{align*}
	& b_j^+ - Q - M\Big(1 - v_{ij} + u_{ij} + 1 - \sum_{k\neq i} u_{kj}\Big)\\
	& = -t_j^+ - M\Big(1 - x_{ij} + y_{ij} + 1 - \sum_{k\neq i} y_{kj}\Big)\\
	& \le -t_j^-(1-z_j),  \quad \text{implied by\ } x_{ij}=1, y_{ij} = 0, z_j = 1 \\
	& \le 0
	\end{align*}
	\text{TDTL}.(\ref{g:24}), $\forall (i,j) \in \mathcal{A}$
	\begin{align*}
	& b_j^+ - Q + M\Big(1 - v_{ij} + u_{ij} + 1 - \sum_{k\neq i} u_{kj}\Big)\\
	& = -t_j^+ - M\Big(1 - x_{ij} + y_{ij} + 1 - \sum_{k\neq i} y_{kj}\Big)\\
	& \ge -t_j^-(1-z_j),  \quad \text{implied by\ } x_{ij}=1, y_{ij} = 0, z_j = 1 \\
	& \ge 0
	\end{align*}
	\text{TDTL}.(\ref{g:25}) is implied by constraint (\ref{nmodel:v_1}).
	
	\sloppy
	We have finally shown that $\Phi(P(\text{ZONVRP})) \subseteq P(\text{TDTL})$. To further establish that the feasible region of $\Phi(P(\text{ZONVRP}))$ is strictly contained in $P(\text{TDTL})$, it is sufficient to give a solution that is valid in TDTL model but infeasible in ZONVRP model. 
	
	\begin{figure}[h!]
		\centering
		\includegraphics[width=0.6\linewidth]{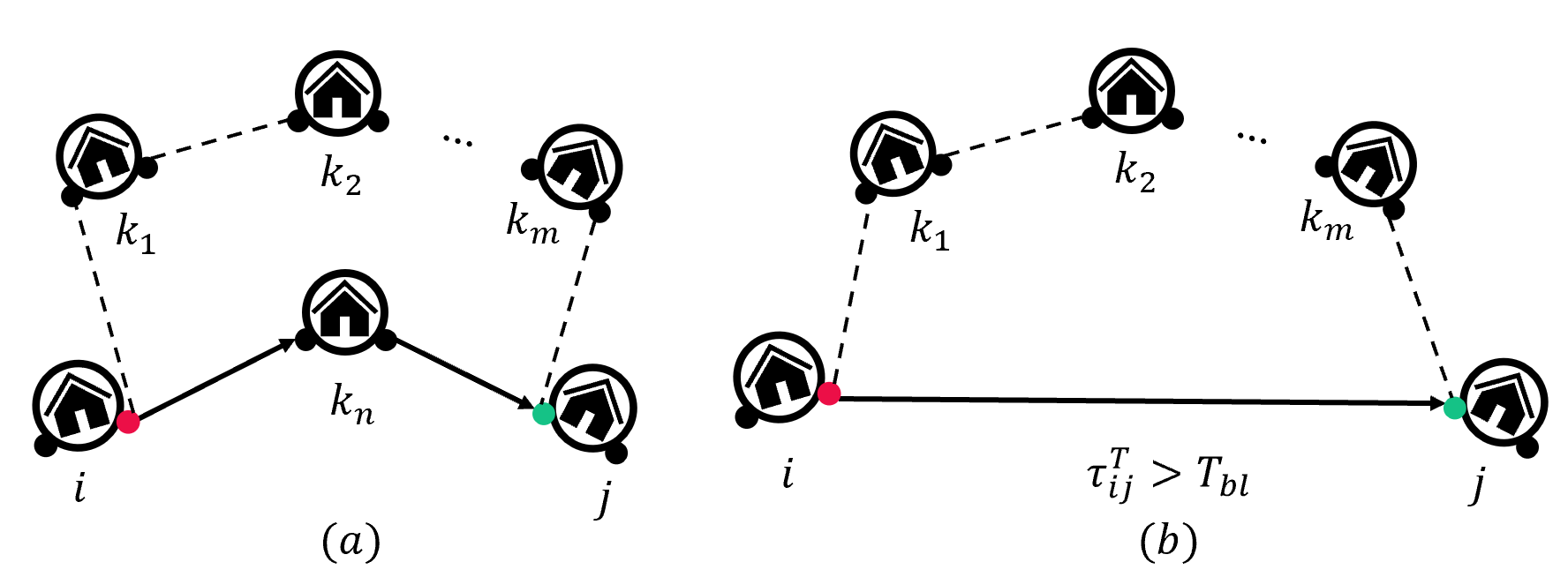}
		\caption{Partial solutions that are avoided in ZONVRP\@.\label{fig:outlier}}
	\end{figure}
	
	In Figure \ref{fig:outlier} (a), we depict a nested unit in which the truck visits location $k_n$ before drives to rendezvous location $j$. This nested unit could be part of any TDTL solution. However, such a solution is infeasible to the ZONVRP model since constraints (\ref{model:drone_1})--(\ref{model:drone_2}) require that each location has to be visited by the drone exactly once. In this case shown in Figure \ref{fig:outlier} (b), the truck travels for more than $T_{\text{bl}}$ time units to the designated destination $j$; and here the drone has depleted its battery and has been forced to land on the ground. But the ZONVRP excludes this situation via the constraint (\ref{model:waiting}).  \quad $\Box$
\end{pf}

\section{Proof for proposition 1 }\label{app:prop1}

\begin{pf}
The following analysis applies to general cases where the total number of locations $n$ is more than two. 

Consider the left hand side of constraints (\ref{dl1}) with the coefficient of term $x_{j,n+1}$ being replaced by $a_{j,n+1}$, we compute the largest possible value for $a_{j,n+1}$ such that the inequality remains valid. 
\begin{equation}
    1 + (1 - x_{0j}) + a_{j,n+1}x_{j,n+1} \le u_j \label{dl_left}
\end{equation}
\textit{case 1.} $x_{0j}=1$. Since location $j$ is right after location $0$, we have $u_j = 1$. Meanwhile, location $j$ should not be connected to the designated location $n+1$ when $n>2$ (i.e., $x_{j, n+1}=0$), otherwise, the drone misses out locations before returning back to the depot. In this case, constraint (\ref{dl_left}) becomes $0 \le 1$ which is valid.

\textit{Case 2.} $x_{0j}=0$. Since the location $j$ is not the one right after $0$, the order $u_j$ of location $j$ must satisfy $2 \le u_j \le n+1$. Furthermore, if $x_{j, n+1} = 0$, we have $2 \le u_j$ which is valid. However, if $x_{j, n+1} = 1$, we have $u_j=n$ and the constraint (\ref{dl_left}) becomes $ a_{j, n+1} \le u_j - 2 =n-2$. Therefore, we set $a_{j, n+1} = n - 2$.

Next, we consider the right hand side of constraint (\ref{dl1}) with the coefficient of term $x_{0j}$ being replaced by $b_{0j}$. Similarly, we compute the largest possible value of $b_{0j}$.

\begin{equation}
    u_j \le (n+1) - b_{0j}x_{0j} - (1 - x_{j,n+1}) \label{dl_right}
\end{equation}

\textit{Case 1. } $x_{j,n+1} = 1$. Since location $j$ is the last location visited before the drone coming back to depot $n+1$, we know the rank of location $j$ is $n$ (i.e., $u_j = n$). Moreover, location $j$ must not be right after depot $0$ Thus, the inequalities (\ref{dl_right}) become $n \le n+1$ which are satisfied at all times.

\textit{Case 2. } $x_{j,n+1} = 0$. If $x_{0j} = 0$, since location $j$ could be anywhere not right after $0$ and right before $n+1$, the rank of location $j$ must satisfies $ 2 \le u_j \le n$ inequalities (\ref{dl_right}) become $u_j \le n $ which is satisfied at all times. However, if $x_{0j} = 1$, we have $u_j=1$ and $b_{0j}  \le n-1$. Therefore, we choose $b_{0j} = n - 1$.

Consider the constraint (\ref{dl2}) with the coefficient of term $x_{ji}$ being replaced by $c_{ji}$, if $c_{ji}=0$, the constraint is the same with constraint (\ref{model:drone_4})

\begin{equation}
 u_i - u_j + (n+1)x_{ij} + c_{ji}x_{ji} \le n  
\end{equation}

\textit{Case 1.} $x_{ji} = 0$. The constraint is the same with constraint (\ref{model:drone_4}).

\textit{Case 2.} $x_{ji} = 1$. This means that $x_{ij} = 0$ and $u_i\ge u_j +1$ in the drone order. Thus, constraint (\ref{dl2}) becomes $u_i - u_j + c_{ji} \le n$. We have $c_{ji} \le n -1$. 
\end{pf}

\section{Proof for proposition 2} \label{app:prop2}

\begin{pf}
We would like to show that constraints (\ref{sd_1}) -- (\ref{sd_5}) are valid restatement of constraints (\ref{model:ubound}) and (\ref{model:drone_4}). To achieve the goal, we follow the RTL technique proposed in \cite{sherali2002tightening} which consists of a reformulation and linearization steps.

We restate the MTZ subtour elimination constraints as follows:
\begin{align}
    \quad & u_j x_{ij} = (u_i + 1)x_{ij}, \quad \forall (i,j) \in A, i \neq 0 \label{Q1}\\
    & u_j x_{0j} = x_{0j}, \quad \forall j \in H \setminus \{ 0\} \label{Q2}\\
    & u_j x_{j, n+1} = nx_{j, n+1}, \quad \forall j \in H \setminus \{n +1\} \label{Q3}\\
    & 1 \le u_j \le n+1, \quad \forall j \in H \setminus \{ 0 \} \label{Q4}
\end{align}

Constraint (\ref{Q1}) -- (\ref{Q4}) excludes any subtour in the drone route. To see that, if $x_{ij} = 1$, location $j$ increases rank by 1 as compared to that of the location $i$. Specifically, if $x_{0j} =1$, location $j$ must have rank 1. If $x_{j, n+1} = 1$, location $j$ has rank $n$. Note that  Constraint (\ref{Q1}) -- (\ref{Q3}) contain quadratic terms, we therefore perform linearization process by introducing new variables.

\textbf{Reformulation:}
We reformulate the constraints (\ref{Q1}) -- (\ref{Q4}) by generating additional implied constraints:

\begin{align*}
    & (\text{r1}): u_i \big( \sum_{j: (i,j) \in A} x_{ij} -1\big) = 0, \quad \forall i \in H \setminus \{n+1\} \\
    & (\text{r2}): u_j \big( \sum_{i: (i,j) \in A} x_{ij} - 1 \big) = 0, \quad \forall j \in H \setminus \{0\}\\
    & (\text{r3}):(u_i-1)x_{ij} \ge 0, \quad \forall (i,j) \in A, i \neq 0\\
    & (\text{r4}): (n+1 - u_i)x_{ij} \ge 0, \quad \forall (i,j) \in A, i \neq 0  \\
    & (\text{r5}):(u_i - 1)(1 - x_{ij} - x_{ji}) \ge 0, \quad \forall (i,j) \in A, i \neq 0\\
    & (\text{r6}): (n+1 - u_i)(1 - x_{ij} - x_{ji}) \ge 0, \quad \forall (i,j) \in A, i \neq 0   \\
    & (\text{r7}): (u_i - 2)(1 - x_{0i} - x_{i,n+1}) \ge 0, \quad \forall i \in H \setminus \{0, n+1\}  \\
    & (\text{r8}): (n-1 - u_i)(1 - x_{0i} - x_{i,n+1}) \ge 0, \quad \forall i \in H \setminus \{0, n+1\}  \\
\end{align*}

\textbf{Linearization:}

Let $y_{ij} = u_i x_{ij}, \forall (i,j) \in A, i \neq 0$ and $z_{ij} = u_jx_{ij}, \forall (i,j) \in A, i \neq 0$. Constraints (\ref{Q1})--(\ref{Q3}) become:

\begin{align} \label{eqn}
    \begin{cases}
    & z_{ij} = y_{ij} + x_{ij}, \quad \forall (i,j) \in A, i \neq 0 \\
    & u_jx_{0j} = x_{0j}, \quad \forall j \in H \setminus \{0\} \\
    & u_j x_{j, n+1} = n x_{j, n+1}, \quad \forall j \in H \setminus \{n+1\} 
     \end{cases}
\end{align}

We then linearize constraints (r1)--(r8) one by one by leveraging the equations in (\ref{eqn}).  

\textbf{r1:}

$\forall i \in H \setminus \{n+1\}$:
\begin{align*}
    & \qquad u_i \big( \sum_{j: (i,j) \in A} x_{ij} -1\big) = 0 \\
    & \Longleftrightarrow u_i x_{i1} + u_i x_{i2} + \cdots + u_i x_{i, n+1} - u_i = 0 \\
    & \Longleftrightarrow \sum_{j: (i,j) \in A, j \neq n+1} y_{ij} + nx_{i,n+1} - u_i = 0\\
    & \implies \text{constraint\  (\ref{sd_1})}
\end{align*}

\textbf{r2:}

$\forall j \in H \setminus \{0\}$:

\begin{align*}
    & \qquad u_j \big( \sum_{i: (i,j) \in A} x_{ij} - 1 \big) = 0\\
    & \Longleftrightarrow \sum_{i: (i,j) \in A} (x_{ij} + y_{ij}) - u_j = 0\\
    & \Longleftrightarrow \sum_{i: (i,j) \in A} y_{ij} + 1 - u_j = 0 \\
    & \implies \text{constraint\  (\ref{sd_2})}
\end{align*}

\textbf{r3:}

$\forall (i,j) \in A, i \neq 0$:
\begin{align*}
    & \qquad (u_i-1)x_{ij} \ge 0\\
    & \Longleftrightarrow x_{ij} \le y_{ij}\\
    & \implies \text{constraint\  (\ref{sd_3})} \text{\ lower bound}
\end{align*}

\textbf{r4:}

$\forall (i,j) \in A, i \neq 0$:

\begin{align*}
    & \qquad (n+1 - u_i)x_{ij} \ge 0\\
    & \Longleftrightarrow (n+1) x_{ij} - y_{ij} \ge 0 \\
    & \Longleftrightarrow  y_{ij} \le (n+1) x_{ij}\\
    & \implies \text{constraint\  (\ref{sd_3})} \text{\ upper bound}
\end{align*}

\textbf{r5:}

$\forall (i,j) \in A, i \neq 0$:
\begin{align*}
    & \qquad (u_i - 1)(1 - x_{ij} - x_{ji}) \ge 0\\
    & \Longleftrightarrow u_i - y_{ij} - z_{ij} - 1 + x_{ij} +x_{ji} \ge 0\\
    & \Longleftrightarrow y_{ij} + y_{ji} \le  u_i - (1 - x_{ij})\\
    & \implies \text{constraint\  (\ref{sd_4})} \text{\ upper bound}
\end{align*}

\textbf{r6:}

$\forall (i,j) \in A, i \neq 0$:

\begin{align*}
    & \qquad (n+1 - u_i)(1 - x_{ij} - x_{ji}) \ge 0\\
    & \Longleftrightarrow  (n+1) - (n+1)x_{ij} - (n+1)x_{ji} - u_i + y_{ij} + x_{ji} + y_{ji} \ge 0\\
    & y_{ij} + y_{ji} \ge u_i + (n+1)(x_{ij}-1) + nx_{ji}\\
    & \implies \text{constraint\  (\ref{sd_4})} \text{\ lower bound}
\end{align*}

\textbf{r7:}

$\forall i \in H \setminus \{0, n+1\}$:

\begin{align*}
    & \qquad (u_i - 2)(1 - x_{0i} - x_{i,n+1}) \ge 0\\
    & \Longleftrightarrow u_i \ge 2 + x_{0i} + n x_{i, n+1} - 2x_{0i} - 2x_{i,n+1}\\
    & \Longleftrightarrow u_i \ge 1 + (1 - x_{0i}) + (n-2)x_{i, n+1}\\
     & \implies \text{constraint\  (\ref{sd_5})} \text{\ lower bound}
\end{align*}

\textbf{r8:}

$\forall i \in H \setminus \{0, n+1\}$:

\begin{align*}
    & \qquad (n-1 - u_i)(1 - x_{0i} - x_{i,n+1}) \ge 0\\
    & \Longleftrightarrow n-1 - (n-1)x_{0i} - (n-1)x_{i, n+1} - u_i + x_{0i} + nx_{i,n+1}\\
    & \Longleftrightarrow u_i \le n - (n-2)x_{0i} - (1- x_{i,n+1})\\
     & \implies \text{constraint (\ref{sd_5})} \text{\ upper bound}
\end{align*}
\end{pf}

\section{Proof of Theorem \ref{thm:LB}} \label{app:proof_lb}
\begin{pf}

Given a Nested-VRP described in graph $\mathcal{G} = (\mathcal{H}, \mathcal{A})$, the optimal solution to the problem consists of the drone route $\mathcal{X}=\{(i,j)\ |\  (i,j) \in \mathcal{A}, \ \text{and}\  x_{ij}=1 \}$ and the truck route $\mathcal{Y}=\{(i,j)\ |\  (i,j) \in \mathcal{A}, \ \text{and}\  y_{ij}=1 \}$. The optimal solution can also be described as a set of non-overlapping nested units $U$. Mathematically, a nested unit $u \in U$ can be viewed as a sub-graph of $\mathcal{G}$. The sub graph contains a set of locations $V(u)$ to be observed by the drone  and the corresponding drone path $E(u)$. Most importantly, one can see that $\mathcal{X} \equiv \{(i,j)\ |\ (i,j) \in E(u), u \in U\}$ and $\mathcal{H}  \equiv \{k \ | \ k \in V(u), u \in U \} $.

In a nested unit $u$, when the drone is about to meet with the truck, the drone arrives at the rendezvous location either earlier than the truck and idles for $\Delta_u$ time units or later than the truck without idling. We define an indicator function $\mathbbm{1}_u$ for each nested unit $u$ that forms the optimal Nested-VRP solution. In a nested unit $u$, if the drone arrives earlier than the truck at rendezvous, $\mathbbm{1}_u=1$, otherwise $\mathbbm{1}_u=0$. After two vehicles meet up successfully, the drone relinquishes all remaining battery life before it obtains a new battery. This portion of unused battery life is denoted as $\delta_u$. 

A battery is either used for drone surveillance (e.g., routing between locations, surveying locations, and possibly waiting for the truck) or wasted after the two vehicles meet up. To derive the lower bound of the mission makespan of a Nested-VRP solution, we first investigate the battery consumption associated with a Nested-VRP solution. Denote the IBR $l_u$ of a nested unit as in Equation (\ref{lu}). We summarize the battery consumption breakdowns in Equation (\ref{bc}).  
\begin{eqnarray}
l_u & = &\sum_{(i,j) \in E(u)} \tau_{ij}^D x_{ij} + \sum_{k \in V(u)} o_k + \mathbbm{1}_u \Delta_u
\label{lu} \\
T_{\text{bl}} & = &l_u + \delta_u \label{bc} 
\end{eqnarray}

Due to the conservation of energy, the amount of energy extracted from all battery replacements scheduled en route should be able to balance off the total battery consumption needed for the mission. Therefore, we can derive the necessary number of battery swaps $N_{\text{s}}$ as follows:
\begin{align*}
N_{\text{s}} T_{\text{bl}} &= \sum_{u \in U} (l_u + \delta_u) = \sum_{u \in U} \Big(\sum_{(i,j) \in E(u)} \tau_{ij}^D x_{ij} + \sum_{k \in V(u)} o_k + \mathbbm{1}_u \Delta_u + \delta_u \Big)
\end{align*}
\begin{equation}
N_{\text{s}} = \frac{1}{T_{\text{bl}}} \sum_{u \in U} \Big(\sum_{(i,j) \in E(u)} \tau_{ij}^D x_{ij} + \sum_{k \in V(u)} o_k + \mathbbm{1}_u \Delta_u + \delta_u \Big) \nonumber
\end{equation}
Since the mission makespan is the sum of the IBRs $l_u, \forall u \in U$ and battery swap service times:
\begin{align*}
\text{makespan} &= \sum_{u\in U} l_u + N_{\text{s}}T_{\text{s}} \quad \\
&=\sum_{u \in U} \Big(\sum_{(i,j) \in E(u)} \tau_{ij}^D x_{ij} + \sum_{k \in V(u)} o_k + \mathbbm{1}_u \Delta_u \Big) +  \frac{T_{\text{s}}}{T_{\text{bl}}}
\sum_{u \in U} \Big(\sum_{(i,j) \in E(u)} \tau_{ij}^D x_{ij} + \sum_{k \in V(u)} o_k + \mathbbm{1}_u \Delta_u + \delta_u \Big)
\end{align*}

We relax constraints imposed on how battery swap could happen and allow the drone to charge itself at any time when its battery has depleted. Therefore, the mission makespan of a relaxed version of Nested-VRP instance is given by:
\begin{align}
\text{makespan} &=\sum_{u \in U} \Big(\sum_{(i,j) \in E(u)} \tau_{ij}^D x_{ij} + \sum_{k \in V(u)} o_k \Big) +  \frac{T_{\text{s}}}{T_{\text{bl}}}
\sum_{u \in U} \Big(\sum_{(i,j) \in E(u)} \tau_{ij}^D x_{ij} + \sum_{k \in V(u)} o_k \Big) \nonumber \\
& = \sum_{(i,j) \in \mathcal{X}} \tau_{ij}^D x_{ij} + \sum_{k \in \mathcal{H}}o_k + \frac{T_{\text{s}}}{T_{\text{bl}}} \Big(\sum_{(i,j) \in \mathcal{X}} \tau_{ij}^D x_{ij} + \sum_{k \in \mathcal{H}}o_k \Big) \label{RP-t}\\
&\ge \sum_{(i,j) \in \mathcal{S}} \tau_{ij}^D + \sum_{k \in \mathcal{H}} o_k + \left \lfloor\frac{1}{T_{\text{bl}}} \Big( \sum_{(i,j) \in \mathcal{S}} \tau_{ij}^D  + \sum_{k \in \mathcal{H}} o_k \Big) \right \rfloor  T_{\text{s}}, \label{LB_TSP}
\end{align}

Recall that $\mathcal{S}$ is the collection of arcs that are in the TSP route. Since the union of the drone paths in all nested units will produce a Hamiltonian cycle containing all of the locations and whose total length is no shorter than the TSP route, Equation (\ref{RP-t}) will be lower bounded by specifying the drone route as the TSP route as well as rounding down the total number of battery swaps to an integer. The result is that, the objective function value of the original Nested-VRP model is lower bounded by term (\ref{LB_TSP}). \quad $\Box$
\end{pf}

\section{Computational results on small data set}\label{app:small_data}
\renewcommand\thetable{\thesection.\arabic{table}}

\begin{table*}[!htbp]	
\centering
	\caption{Results from solving instances from the uniform pattern in the small data set.\label{tab:s_uniform}}
	\ra{1}
	\resizebox{\textwidth}{!}{%
	\begin{tabular}{@{}rrrrrrrrrrrrr@{}}
		\toprule
		\textbf{uniform} & \multicolumn{3}{c}{\textbf{MIQCP}} & \phantom{abc}& \multicolumn{3}{c}{\textbf{MILP+SD}} & \phantom{abc} & \textbf{LB}& 
		\phantom{abc} & \multicolumn{2}{c}{\textbf{NS Heuristic}} \\
		\cmidrule{2-4} \cmidrule{6-8}  \cmidrule{10-10}  \cmidrule{12-13}  
		$N$ & $C_{\text{MIQCP}}(s) $  & $\gamma_{\text{MIQCP}}(\%)$ & $T_{\text{MIQCP}}(s)$ && $C_{\text{MILP+SD}}(s) $  & $\gamma_{\text{MILP+SD}}(\%)$ & $T_{\text{MILP+SD}}(s)$ &&$\gamma_{\text{lb}}(\%) $  &&  $ N^\star $  & $T_{\text{NS}}(s)$\\ 
		\hline
		$\alpha =1$\\
		\hline
5  & 1615.5 & 0.0  & 0.4   &  & 1615.5 & 0.0  & 0.4   &  & -0.8 &  & 10/10 & 0.7 \\
6  & 1795.1 & 0.0  & 2.2   &  & 1795.1 & 0.0  & 2.7   &  & -0.8 &  & 10/10 & 1.6 \\
7  & 2313.9 & 0.0  & 26.4  &  & 2313.9 & 0.0  & 16.3  &  & -1.2 &  & 10/10 & 2.0 \\
8  & 2484.2 & 0.0  & 246.4 &  & 2484.2 & 0.0  & 59.8  &  & -1.5 &  & 10/10 & 3.5 \\
9  & 2901.8 & 36.6 & 901.7 &  & 2900.7 & 1.6  & 446   &  & -1.4 &  & 10/10 & 4.1 \\
10 & 3004.9 & 54.8 & 901.8 &  & 3002.5 & 54.4 & 900.4 &  & -1.6 &  & 10/10 & 5.9 \\
		\hline
		$\alpha =2$\\
		\hline
5  & 1431.9 & 0.0  & 1.0   &  & 1431.9 & 0.0  & 0.5   &  & -3.7 &  & 10/10 & 0.7 \\
6  & 1577.6 & 0.0  & 4.5   &  & 1577.6 & 0.0  & 2.1   &  & -2.2 &  & 10/10 & 1.4 \\
7  & 1915.1 & 0.0  & 43.8  &  & 1915.1 & 0.0  & 11.1  &  & -2.9 &  & 10/10 & 2.5 \\
8  & 2305.1 & 0.0  & 374.8 &  & 2305.1 & 0.0  & 72.2  &  & -2.8 &  & 10/10 & 3.5 \\
9  & 2596.9 & 36.8 & 902.2 &  & 2591.8 & 5.1  & 475.4 &  & -3.0 &  & 10/10 & 4.9 \\
10 & 2364.1 & 60.2 & 901.9 &  & 2361.1 & 52.8 & 900.4 &  & -2.9 &  & 10/10 & 7.6\\				           \hline
		$\alpha =3$\\
		\hline
5  & 1474.8 & 0.0  & 1.1   &  & 1474.8 & 0.0  & 0.4   &  & -4.2 &  & 10/10 & 0.9 \\
6  & 1414.8 & 0.0  & 4.6   &  & 1414.8 & 0.0  & 0.8   &  & -2.4 &  & 10/10 & 1.4 \\
7  & 1880   & 0.0  & 30.5  &  & 1880   & 0.0  & 5.5   &  & -4.1 &  & 10/10 & 2.3 \\
8  & 2227.1 & 0.0  & 231.1 &  & 2227.1 & 0.0  & 62.8  &  & -1.9 &  & 10/10 & 3.1 \\
9  & 2562.5 & 40.4 & 902.0 &  & 2562.4 & 3.5  & 500.9 &  & -3.2 &  & 9/10  & 4.7 \\
10 & 2449   & 62.6 & 901.8 &  & 2429.1 & 47.0 & 900.3 &  & -3.0 &  & 9/10  & 6.5\\	
\hline
 Summary & \multicolumn{3}{c}{120/180 of instances reach optimality}& & \multicolumn{3}{c}{145/180 instances reach optimality} &  &  &  &  \\
		\bottomrule
	\end{tabular}
	}
\end{table*}

\begin{table*}[!htbp]	
\centering
	\caption{Results from solving instances from the single-center pattern in the small data set.\label{tab:s_singlecenter}}
	\ra{1}
	\resizebox{\textwidth}{!}{%
	\begin{tabular}{@{}rrrrrrrrrrrrr@{}}
		\toprule
		\textbf{single-center} & \multicolumn{3}{c}{\textbf{MIQCP}} & \phantom{abc}& \multicolumn{3}{c}{\textbf{MILP+SD}} & \phantom{abc} & \textbf{LB}& 
		\phantom{abc} & \multicolumn{2}{c}{\textbf{NS Heuristic}} \\
		\cmidrule{2-4} \cmidrule{6-8}  \cmidrule{10-10}  \cmidrule{12-13}  
		$N$ & $C_{\text{MIQCP}}(s) $  & $\gamma_{\text{MIQCP}}(\%)$ & $T_{\text{MIQCP}}(s)$ && $C_{\text{MILP+SD}}(s) $  & $\gamma_{\text{MILP+SD}}(\%)$ & $T_{\text{MILP+SD}}(s)$ &&$\gamma_{\text{lb}}(\%) $  &&  $ N^\star $  & $T_{\text{NS}}(s)$\\ 
		\hline
		$\alpha =1$\\
		\hline
5  & 1546   & 0.0  & 0.4   &  & 1546   & 0.0  & 0.3   &  & -1.2 &  & 10/10 & 0.7 \\
6  & 2156.7 & 0.0  & 2.0   &  & 2156.7 & 0.0  & 2.3   &  & -1.6 &  & 10/10 & 1.2 \\
7  & 2233.6 & 0.0  & 19.4  &  & 2233.6 & 0.0  & 19.0    &  & -1.3 &  & 10/10 & 1.8 \\
8  & 2562.4 & 0.0  & 194.4 &  & 2562.4 & 0.0  & 178.5 &  & -1.3 &  & 10/10 & 2.9 \\
9  & 2964.2 & 34.9 & 901.7 &  & 2960.3 & 15.0 & 778.5 &  & -1.6 &  & 10/10 & 4.0 \\
10 & 3471.4 & 51.0 & 901.8 &  & 3461.2 & 32.3 & 900.5 &  & -2.3 &  & 9/10  & 5.8\\
		\hline
		$\alpha =2$\\
		\hline
5  & 1340.6 & 0.0  & 0.9   &  & 1340.6 & 0.0 & 0.5   &  & -3.1 &  & 10/10 & 1.2 \\
6  & 1741.3 & 0.0  & 5.0   &  & 1741.3 & 0.0 & 1.2   &  & -4.9 &  & 10/10 & 1.6 \\
7  & 1917.1 & 0.0  & 32.5  &  & 1917.1 & 0.0 & 9.1   &  & -2.3 &  & 10/10 & 2.3 \\
8  & 2241.6 & 0.0  & 363.3 &  & 2241.6 & 0.0 & 67.3  &  & -4.0 &  & 9/10  & 3.2 \\
9  & 2348.9 & 37.7 & 902.4 &  & 2338.8 & 6.4 & 301.9 &  & -3.3 &  & 8/10  & 3.8 \\
10 & 3213.4 & 57.5 & 901.6 &  & 3191.1 & 1.8 & 406.2 &  & -3.3 &  & 7/10  & 6.0\\
		\hline
		$\alpha =3$\\
		\hline
5  & 1299.6 & 0.0  & 1.1   &  & 1299.6 & 0.0 & 0.4   &  & -2.4 &  & 10/10 & 1.3 \\
6  & 1484.9 & 0.0  & 4.5   &  & 1484.9 & 0.0 & 0.7   &  & -3.8 &  & 10/10 & 1.3 \\
7  & 1874.1 & 0.0  & 25.8  &  & 1874.1 & 0.0 & 3.0   &  & -3.0 &  & 10/10 & 2.2 \\
8  & 2418   & 0.0  & 268.8 &  & 2418   & 0.0 & 27.0  &  & -2.8 &  & 9/10  & 2.2 \\
9  & 2475.5 & 37.5 & 902.0 &  & 2468.8 & 1.8 & 104.7 &  & -2.8 &  & 10/10 & 5.2 \\
10 & 2713.9 & 61.8 & 901.8 &  & 2673.5 & 0.9 & 245.8 &  & -1.6 &  & 4/10  & 6.6 \\	
\hline
 Summary & \multicolumn{3}{c}{120/180 of instances reach optimality}& & \multicolumn{3}{c}{156/180 instances reach optimality} &  &  &  &  \\
		\bottomrule
	\end{tabular}
	}
\end{table*}

\begin{table*}[!htbp]	
\centering
	\caption{Results from solving instances from the double-center pattern in the small data set.\label{tab:s_doublecenter}}
	\ra{1}
	\resizebox{\textwidth}{!}{%
	\begin{tabular}{@{}rrrrrrrrrrrrr@{}}
		\toprule
		\textbf{double-center} & \multicolumn{3}{c}{\textbf{MIQCP}} & \phantom{abc}& \multicolumn{3}{c}{\textbf{MILP+SD}} & \phantom{abc} & \textbf{LB}& 
		\phantom{abc} & \multicolumn{2}{c}{\textbf{NS Heuristic}} \\
		\cmidrule{2-4} \cmidrule{6-8}  \cmidrule{10-10}  \cmidrule{12-13}  
		$N$ & $C_{\text{MIQCP}}(s) $  & $\gamma_{\text{MIQCP}}(\%)$ & $T_{\text{MIQCP}}(s)$ && $C_{\text{MILP+SD}}(s) $  & $\gamma_{\text{MILP+SD}}(\%)$ & $T_{\text{MILP+SD}}(s)$ &&$\gamma_{\text{lb}}(\%) $  &&  $ N^\star $  & $T_{\text{NS}}(s)$\\ 
		\hline
		$\alpha =1$\\
		\hline
5  & 2645.8 & 0.0  & 0.4   &  & 2645.8 & 0.0 & 0.2   &  & -1.0 &  & 10/10 & 0.8 \\
6  & 2889.1 & 0.0  & 1.2   &  & 2889.1 & 0.0 & 0.9   &  & -0.6 &  & 10/10 & 1.1 \\
7  & 3254   & 0.0  & 9.8   &  & 3254   & 0.0 & 6.2   &  & -0.8 &  & 10/10 & 1.9 \\
8  & 3687.1 & 0.0  & 78.0  &  & 3687.1 & 0.0 & 41.3  &  & -1.2 &  & 9/10  & 2.6 \\
9  & 3568.4 & 15.5 & 746.6 &  & 3567.3 & 5.0 & 394.9 &  & -0.9 &  & 10/10 & 4.7 \\
10 & 4179.2 & 31.4 & 901.6 &  & 4176.4 & 7.6 & 900.5 &  & -1.4 &  & 10/10 & 5.0 \\
		\hline
		$\alpha =2$\\
		\hline
5  & 1855.7 & 0.0  & 0.5   &  & 1855.7 & 0.0 & 0.2   &  & -3.9 &  & 10/10 & 1.1 \\
6  & 2429.1 & 0.0  & 3.5   &  & 2429.1 & 0.0 & 0.3   &  & -4.8 &  & 10/10 & 1.4 \\
7  & 2354.9 & 0.0  & 23.4  &  & 2354.9 & 0.0 & 0.9   &  & -3.2 &  & 10/10 & 2.0 \\
8  & 3027.8 & 0.0  & 372.5 &  & 3027.8 & 0.0 & 6.7   &  & -4.5 &  & 10/10 & 3.0 \\
9  & 3296.2 & 31.2 & 901.8 &  & 3289.1 & 0.0 & 73.6  &  & -4.5 &  & 10/10 & 4.5 \\
10 & 3362.6 & 50.5 & 901.6 &  & 3353.2 & 0.0 & 417.9 &  & -4.6 &  & 8/10  & 6.4 \\
		\hline
		$\alpha =3$\\
		\hline
5  & 1612.6 & 0.0  & 0.4   &  & 1612.6 & 0.0 & 0.1  &  & -2.6 &  & 9/10  & 1.0 \\
6  & 1711.9 & 0.0  & 1.8   &  & 1711.9 & 0.0 & 0.3  &  & -2.7 &  & 10/10 & 1.4 \\
7  & 2350.4 & 0.0  & 20.8  &  & 2350.4 & 0.0 & 0.6  &  & -2.2 &  & 9/10  & 2.2 \\
8  & 2429.4 & 0.0  & 142.7 &  & 2429.4 & 0.0 & 2.6  &  & -2.3 &  & 10/10 & 3.3 \\
9  & 2806.4 & 35.7 & 901.8 &  & 2794.9 & 0.0 & 23.3 &  & -3.0 &  & 9/10  & 4.8 \\
10 & 3273.3 & 53.8 & 902.3 &  & 3251.7 & 0.0 & 93.9 &  & -2.6 &  & 6/10  & 6.3\\
\hline
 Summary & \multicolumn{3}{c}{123/180 of instances reach optimality}& & \multicolumn{3}{c}{173/180 instances reach optimality} &  &  &  &  \\
		\bottomrule
	\end{tabular}
	}
\end{table*}

\end{document}